\documentclass{WileyMSP-template}
\usepackage{textcomp}
\usepackage{graphicx}
\usepackage{xcolor}
\usepackage{subcaption}
\usepackage[version=4]{mhchem}
\usepackage[linktocpage=true,
  colorlinks=true, 
  pdfborder={0 0 0},
  linkcolor=blue,
  citecolor=blue,
  filecolor=yellow,
  urlcolor=blue,
  bookmarks,
  pdfauthor={},
]{hyperref}

\definecolor{GOrange}{HTML}{F36D21}
\definecolor{GGreen}{HTML}{77B661}
\definecolor{GBlue}{HTML}{3B83C5}
\definecolor{GPurple}{HTML}{842F8A}

\begin{document}

\pagestyle{fancy}
\rhead{}

\title{Tackling Disorder in $\gamma$-\ce{Ga2O3}}

\maketitle

\author{Laura~E.~Ratcliff*}
\author{Takayoshi~Oshima}
\author{Felix~Nippert}
\author{Benjamin~M.~Janzen}
\author{Elias~Kluth}
\author{R\"udiger~Goldhahn}
\author{Martin~Feneberg}
\author{Piero~Mazzolini}
\author{Oliver~Bierwagen}
\author{Charlotte Wouters}
\author{Musbah Nofal}
\author{Martin~Albrecht}
\author{Jack~E.~N.~Swallow}
\author{Leanne~A.~H.~Jones}
\author{Pardeep~K.~Thakur}
\author{Tien-Lin~Lee}
\author{Curran~Kalha}
\author{Christoph~Schlueter}
\author{Tim~D.~Veal}
\author{Joel~B.~Varley}
\author{Markus~R.~Wagner}
\author{Anna~Regoutz*}

\begin{affiliations}
{\small
Laura~E.~Ratcliff\\
Department of Materials, Imperial College London, London SW7 2AZ, United Kingdom.\\
Centre for Computational Chemistry,
School of Chemistry, University of Bristol, Bristol BS8 1TS, United Kingdom.\\
Email Address: laura.ratcliff@bristol.ac.uk

Takayoshi~Oshima\\
Department of Electrical and Electronic Engineering, Saga University, Saga 840-8502, Japan.

Elias~Kluth, R\"udiger~Goldhahn, Martin~Feneberg\\
Institut f\"ur Physik, Otto-von-Guericke-Universit\"at Magdeburg, Universit\"atsplatz 2, 39106 Magdeburg, Germany.

Piero~Mazzolini\\
Paul-Drude-Institut f\"ur Festk\"{o}rperelektronik, Leibniz-Institut im Forschungsverbund Berlin e.V., Hausvogteiplatz 5-7, 10117 Berlin, Germany.\\
Present address: Department of Mathematical, Physical and Computer Sciences, University of Parma, Viale delle Scienze 7/A, 43124 Parma, Italy.

Oliver~Bierwagen\\
Paul-Drude-Institut fur Festk\"{o}rperelektronik, Leibniz-Institut im Forschungsverbund Berlin e.V., Hausvogteiplatz 5-7, 10117 Berlin, Germany.

Charlotte Wouters, Musbah Nofal, Martin Albrecht\\
Leibniz-Institut für Kristallz\"{u}chtung, Max-Born-Str. 2, 12489 Berlin, Germany

Jack~E.~N.~Swallow\\
Department of Materials, University of Oxford, Parks Road, Oxford OX1 3PH, United Kingdom.

Pardeep~K.~Thakur, Tien-Lin~Lee\\
Diamond Light Source Ltd., Diamond House, Harwell Science and Innovation Campus, Didcot OX11 0DE, United Kingdom.

Christoph~Schlueter\\
Deutsches Elektronen-Synchrotron DESY, Notkestrasse 85, 22607 Hamburg, Germany.

Leanne~A.~H.~Jones, Tim~D.~Veal\\
Stephenson Institute for Renewable Energy and Department of Physics, University of Liverpool, Liverpool L69 7ZF, United Kingdom.

Joel~B.~Varley\\
Lawrence Livermore National Laboratory, Livermore, CA 94550, United States of America.

Benjamin~M.~Janzen, Felix~Nippert, Markus~R.~Wagner\\
Technische Universit\"{a}t Berlin, Institute of Solid State Physics, Hardenbergstrasse 36, 10623 Berlin, Germany.

Curran~Kalha, Anna~Regoutz\\
Department of Chemistry, University College London, 20 Gordon Street, London WC1H 0AJ, United Kingdom.\\
Email: a.regoutz@ucl.ac.uk}
\end{affiliations}



\begin{abstract}
\ce{Ga2O3} and its polymorphs are attracting increasing attention. The rich structural space of polymorphic oxide systems such as \ce{Ga2O3} offers potential for electronic structure engineering, which is of particular interest for a range of applications, such as power electronics. $\gamma$-\ce{Ga2O3} presents a particular challenge across synthesis, characterisation, and theory due to its inherent disorder and resulting complex structure -- electronic structure relationship. Here, density functional theory is used in combination with a machine learning approach to screen nearly one million potential structures, thereby developing a robust atomistic model of the $\gamma$-phase. Theoretical results are compared with surface and bulk sensitive soft and hard X-ray photoelectron spectroscopy, X-ray absorption spectroscopy, spectroscopic ellipsometry, and photoluminescence excitation spectroscopy experiments representative of the occupied and unoccupied states of $\gamma$-\ce{Ga2O3}. The first onset of strong absorption at room temperature is found at 5.1~eV from spectroscopic ellipsometry, which agrees well with the excitation maximum at 5.17~eV obtained by PLE spectroscopy, where the latter shifts to 5.33~eV at 5~K. This work presents a leap forward in the treatment of complex, disordered oxides and is a crucial step towards exploring how their electronic structure can be understood in terms of local coordination and overall structure.  
\end{abstract}

\section{Introduction}
\label{sec:intro}

Gallium oxide is an ultra-wide band gap (UWBG) semiconductor that promises to extend the capabilities and application limits in areas such as power electronics, solar blind UV photodetectors, gas-sensing devices, and solar cells.~\cite{Pearton2018, Shi2021} It is already successfully used in some areas, including phosphors and electroluminescent [EL] devices~\cite{Guo2019}, solar-blind photodetectors~\cite{Teng2014,Hou2021}, photocatalysis~\cite{Akatsuka2020}, and power electronics.~\cite{Masataka2016,Xue2018} \ce{Ga2O3} is similar to many other polymorphic oxide systems, such as \ce{Al2O3}, \ce{In2O3}, and \ce{Sb2O3}, in that beyond the thermodynamically stable monoclinic $\beta$-phase ($C2/m$) at least four further phases exist. These include the rhombohedral $\alpha$-\ce{Ga2O3} ($R\bar{3}c$), cubic $\gamma$-\ce{Ga2O3} ($Fd\bar{3}m$), orthorhombic $\varepsilon$/$\kappa$-\ce{Ga2O3} ($Pna2_1$), and cubic $\delta$-\ce{Ga2O3} ($Ia\bar{3}$) phases. It should be noted that the existence of the $\delta$-phase is still subject to some discussion and it has been suggested that it may be formed by a mixture of the $\beta$- and $\varepsilon$-phases~\cite{Playford2013}.

\begin{figure}[ht!]
\centering
\includegraphics[width=0.95\linewidth]{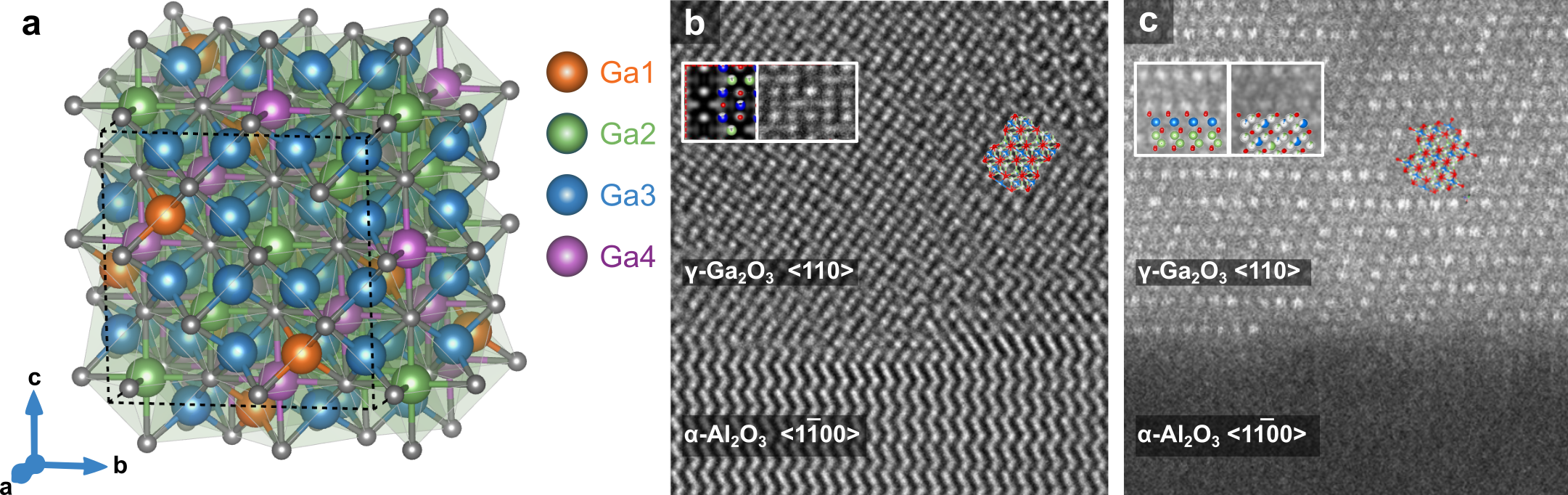}
\caption{Atomic structure of $\gamma$-\ce{Ga2O3}. (a) Schematic representation of the crystal structure with inequivalent Ga positions given numbers (1,3) for tetrahedral $T_d$ and (2,4) for octahedral $O_h$ ordination. (b)-(c) Atomic resolution image of $\gamma$-\ce{Ga2O3} crystallised on a sapphire substrate. (b) shows a high resolution phase contrast image along the [110] projection. The inset shows an expanded view as well as an image simulation. The latter is overlaid with an atomic model (red atoms are oxygen, green and blue are four-fold and six-fold coordinated Ga atoms. (c) shows a STEM high angle annular darkfield image of the same area. Bright atoms correspond to Ga. An atomic model is overlaid to the image. The image pattern fluctuates between a single periodicity and a double periodicity along the (111) planes of the structure. The inset shows details of the micrograph that correspond to an occupation resembling that of the $\beta$-structure in the $<$132$>$ projection (double periodicity, left inset) and to an occupation of the $\gamma$-structure along the $<$110$>$ projection (single periodicity, right inset).}
\label{fig:structure}
\end{figure}

The existence of $\gamma$-\ce{Ga2O3} was first suspected by B\"{o}hm in 1939,~\cite{Boehm1940} and subsequent works by Roy~\textit{et al.} and Pohl led to the conclusion that it has a spinel-type structure similar to $\gamma$-\ce{Al2O3}.~\cite{Roy1952,Pohl1968} Although these initial observations of the $\gamma$-phase took place in the first half of the 20th century, it took until 2013 for detailed structural investigations to be performed by Playford~\textit{et al.}\ using total neutron diffraction.~\cite{Playford2013,Playford2014} Whilst the analogy to $\gamma$-\ce{Al2O3} still holds in that $\gamma$-\ce{Ga2O3} can be considered a cubic, cation-deficient spinel with only partial occupancy of its tetrahedral and octahedral sites, Playford~\textit{et al.}\ conclusively showed that the distribution of occupied sites results in an inherently disordered structure. Figure~\ref{fig:structure}(a) shows a schematic representation of the crystal structure of $\gamma$-\ce{Ga2O3}. In addition to the expected ideal spinel sites, tetrahedral ($T_d$) $8a$ (Ga1) and octahedral ($O_h$) $16d$ sites (Ga2), $T_d$ $48f$ (Ga3) and $O_h$ $16c$ (Ga4) sites are also partially occupied, with a refined tetrahedral to octahedral ratio from Playford~\textit{et al.}\ of 1:1.35. Furthermore, the local structure of $\gamma$-\ce{Ga2O3} is distorted, with the $O_h$ $16d$ sites having distinct long and short Ga--O distances and showing the most significant degree of local distortion.

Previous first principles calculations of $\gamma$-\ce{Ga2O3} based on density functional theory (DFT)~\cite{Hohenberg1964,Kohn1965} have focused on systematically exploring all possible structures arising due to different combinations of vacant Ga sites, both for pure $\gamma$-\ce{Ga2O3}~\cite{Yoshioka2007} and Mn-doped $\gamma$-\ce{Ga2O3}.~\cite{Hayashi2012}
Similar approaches have been employed for $\gamma$-\ce{Al2O3}.~\cite{Gutierrez2001,Pinto2004,Taniike2006} While the exhaustive search approach has led to some interesting insights regarding the preferred vacancy sites in $\gamma$-\ce{Ga2O3}, it is made tractable due to the assumption of a 2-site defective spinel structure.  However, the proposed 4-site model of Playford~\emph{et al.}\ leads to a number of possible configurations that makes such an approach prohibitively expensive.  Similarly, the need to impose stoichiometry, combined with the low occupancies for two of the four sites, increases the size of the unit cell needed to effectively define the structure. Other approaches for structure searching and optimisation of \ce{Ga2O3} have been employed, however, these have focused on other phases~\cite{Wang2020} or 2D structures~\cite{Meng2020}, while the large number of required calculations in such approaches poses challenges for systems containing many atoms. One way to overcome such size limitations is \emph{via} the use of interatomic potentials, which have been previously employed to explore a very large number of potential structures for $\gamma$-\ce{Al2O3}~\cite{Paglia2005}. However, this relies on the availability of a potential which is accurate enough to distinguish between structures which are close in energy.

In a previous study by some of the present authors, a combination of X-ray spectroscopy and theory was successfully used to explore the influence of local Ga coordination on the electronic structure across the $\alpha$, $\beta$ and $\varepsilon$ polymorphs.\cite{Swallow2020} However, the $\gamma$-phase was not included at the time as its inherent disorder proved challenging. The present work extends our previous efforts by combining first principles calculations with a machine learning (ML) approach to accelerate the screening of possible structures in 160 atom cells. The resulting low energy configurations are then explored in greater detail, and put in relation to experimental results from soft and hard X-ray photoelectron spectroscopy (XPS and HAXPES), X-ray absorption spectroscopy (XAS), spectroscopic ellipsometry, and photoluminescence excitation spectroscopy (PLE).

\section{Results}

\subsection{Atomic Level Disorder}

Whilst the neutron diffraction experiments and analysis by Playford~\emph{et al.}~\cite{Playford2013,Playford2014} provide crucial insights into the disorder present in $\gamma$-\ce{Ga2O3}, electron microscopy can probe this on much shorter length scales. Figure~\ref{fig:structure}(b) shows an atomic resolution phase contrast image of $\gamma$-\ce{Ga2O3} thin films obtained under imaging conditions, where atoms appear as bright dots. The image shows the interface between the sapphire substrate and the crystalline \ce{Ga2O3} layer along the $<$1$\bar{1}$00$>$ projection of the sapphire substrate. The layer is single crystalline, and the image pattern fits well to that of the $\gamma$-phase of \ce{Ga2O3}. A ball and stick model is overlaid to indicate the positions of the oxygen (red) and the octahedrally coordinated (blue) and tetrahedrally coordinated (green) Ga atoms. Image simulations were done considering only partial occupation of the various Ga sites according to the model by Playford~\emph{et al.}~\cite{Playford2013,Playford2014} The best fit between the image pattern and the simulations was obtained for projected sample thickness of 10~nm. The inset shows the simulation with the atoms overlaid and a detail of the micrograph. Two key observations can be made in this image. While the oxygen sublattice is periodic and characterised by intense bright spots under the imaging conditions used here, translational symmetry in the Ga sublattice is not present and the image pattern fluctuates at the nanometre scale. This indicates a strong local fluctuation in the occupancy of the various Ga sites, as expected for the defective spinel structure. While neutron scattering and X-ray methods are integrating across larger volumes, TEM data such as those presented here can resolve these fluctuations on much shorter length scales. This finding is confirmed by scanning transmission electron microscopy images using high angle annular darkfield detector (see Figure~\ref{fig:structure}(c)), where only Ga atoms are visible. As can be seen, the image pattern fluctuates, with the apparent change of the periodicity along the (111) planes between a single periodicity and a double periodicity. The latter corresponds to an occupancy of the Ga sites that is close to the $\beta$-phase in its [132] projection, shown in the inset in Figure~\ref{fig:structure}(c), while single periodicity corresponds to the $\gamma$-structure in the [110] projection. The structures in the projections show a common oxygen lattice with almost identical positions with the $\gamma$-phase having a higher number of Ga sites with stochastic occupation. These TEM results provide further evidence of the inherent disorder of both Ga and O in $\gamma$-\ce{Ga2O3}, emphasising the need to build atomistic models that take this aspect of its intrinsic structure into account.

\subsection{High Throughput Structure Screening}

\begin{figure}[ht!]
\centering
\includegraphics[scale=1.0]{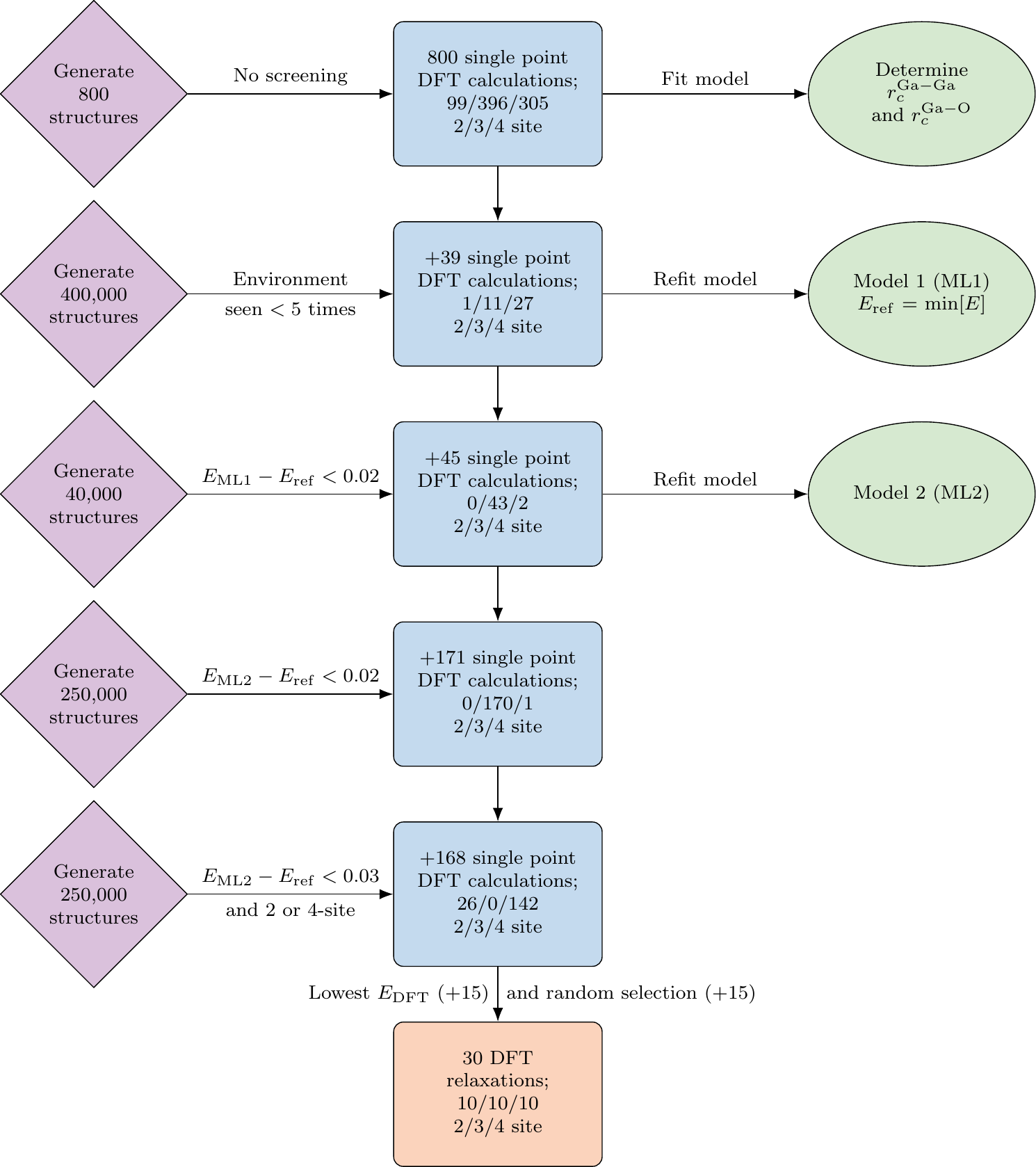}
\caption{Flowchart summarizing the process for generating, screening and calculating structures, as well as fitting the machine learning (ML) model. Energy thresholds are in eV/atom. \label{fig:workflow}}
\end{figure}

\begin{figure}[ht]
\centering
\includegraphics[scale=0.85]{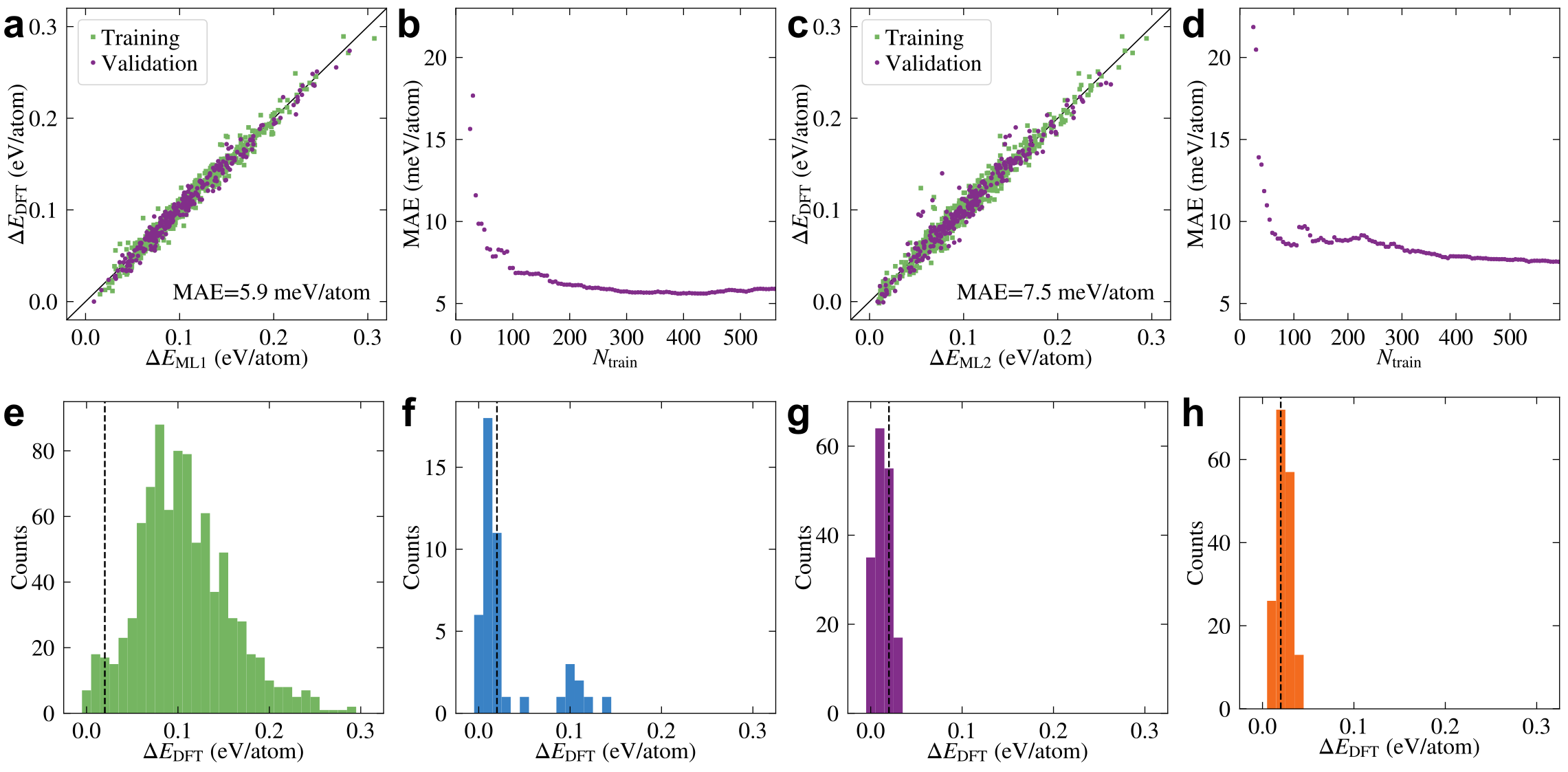}
\caption{Performance of the different models, including a comparison to the DFT results and ability to predict low energy structures. (a) and (b) show the model performance for the initial 800 structures plus those containing rare environments (`ML1'), while (c) and (d) are for the refitted model, which has an additional 45 predicted low energy structures (`ML2'). Also shown in (a) and (c) are the mean absolute errors (MAEs) for the largest training set.
(e)-(h) show the distribution of PBE-calculated energies for (e) the initial structures plus those containing rare environments, (f) structures where the ML1 predicted energy was less than 0.02~eV/atom above the reference energy, (g) structures where the ML2 predicted energy ML2 was less than 0.02~eV/atom above the reference energy, and (h) structures where the ML2 predicted energy was less than 0.03~eV/atom above the reference energy, and for which only 2 or 4 distinct Ga sites were occupied.\label{fig:model}}
\end{figure}

The initial random structure generation, PBE-based DFT calculations, and process for augmenting the data set, which are described in Section~\ref{sec:ml_approach}, are summarised in the first two rows of Figure~\ref{fig:workflow}.  The performance of the model coming from the resulting 839 structures and denoted `ML1', is depicted in Figures~\ref{fig:model}(a)-(b).  As can be seen, the mean absolute error (MAE) across the validation set converges at around 300 structures in the training set, while the MAE for the largest training set size is 5.9~meV/atom. However, the DFT energies across the 839 structures (depicted in Figure~\ref{fig:model}(e)) are spread out across a range of around 0.3~eV/atom, with the majority distributed around the centre of that range. Therefore, a more important measure of the success of the model is whether or not it is able to find low energy structures, which are not well represented in the data set.  

To this end, an additional 40,000 structures were generated, using the four site probabilities.  
All structures with a predicted energy lower than $E_{\mathrm{ref}}+0.02$~eV/atom were calculated, where  $E_{\mathrm{ref}}$ is the lowest energy seen thus far. The majority of the 45 new structures (69~\%) were indeed found to have DFT energies within 0.02~eV/atom of $E_{\mathrm{ref}}$ (see Table~S1 in the Supporting Information).  However, as shown in Figure~\ref{fig:model}(f), a number of structures were 0.1~eV/atom or higher in energy, demonstrating that the initial model did not adequately cover the low energy region of interest.  Therefore, the model was refit by splitting the full set of data into 590 (294) training (validation) structures, as summarised in row three of Figure~\ref{fig:workflow}.

The performance of the new model, denoted `ML2', is depicted in Figures~\ref{fig:model}(c)-(d).  The convergence with training set size is slower than for ML1, while the MAE for the largest training set size is \newline 7.5~meV/atom, slightly higher than for ML1, but nonetheless reasonable. In order to verify whether or not the new model had improved predictivity for low energy structures, a further 250,000 structures were generated, this time including 10\% of structures with explicit two site occupancies.  The same energy criterion was used to identify potential low energy structures, i.e.\ $E_{\mathrm{ML2}}-E_{\mathrm{ref}}<0.02$~eV/atom. As shown in row four of Figure~\ref{fig:workflow}, this resulted in an additional 171 structures, all but one of which had three of the Ga sites occupied.  As can be seen from Figure~\ref{fig:model}(g) and Table~S1 in the Supporting Information, the majority of the structures were within the targeted range (79~\%), with the remainder being only slightly higher in energy. More importantly, there were no high-energy outliers.

As a final step, another 250,000 structures were generated, as in the previous step, however the energy threshold was increased to 0.03~eV/atom, and all 3-site structures were rejected.  This resulted in an additional 168 DFT structures, of which 26 (142) had two (four) sites occupied, as summarised in row five of Figure~\ref{fig:workflow}.  As shown in Figure~\ref{fig:model}(h) all of these structures were higher in energy than the lowest identified 3-site structures.

\subsection{Atomic Structure}

Out of the full set of 1223 DFT-calculated structures, 30 structures were relaxed.  The structures were selected first by taking the five lowest energy 2-, 3- and 4-site structures, respectively.  An additional 15 structures were then randomly selected to give a total of 10 structures each with two, three and four occupied Ga sites.  Particularly for the higher energy structures, some Ga atoms moved considerably during relaxation, and thus the type of Ga site was redetermined for each Ga atom by identifying the closest corresponding Ga site in the pristine structure.  The structures were then reclassified by the number of occupied Ga sites.  The unrelaxed and relaxed energies, and the site occupancies for the 30 structures are given in Table~S2 in the Supporting Information.

For the unrelaxed structures, the lowest energy structures all have three occupied Ga sites, with the 4-site structures being next lowest in energy, at around 8~meV/atom higher in energy. The 2-site structures are slightly higher in energy. This is further evident in Figure~S1 in the Supporting Information, where the distribution of all unrelaxed energies is given by number of occupied Ga sites. After relaxation, the lower energy structures remain lower in energy than those selected randomly, but with some rearrangements within the group (see Figure~S2 in the Supporting Information). Notably, the 3- and 4-site relaxed structures are at similar energies, with the lowest energy 3-site structure being less than 1~meV/atom lower in energy than the lowest energy 4-site structure. The 2-site relaxed structures remain slightly higher in energy, but are also closer in energy than before relaxation, with the lowest 2-site being 7~meV/atom higher than the overall lowest energy structure.  However, such small energy differences are relatively insignificant given the sensitivity to simulation parameters including the basis set and pseudopotential.  Indeed, when comparing relative energies of the relaxed structures calculated using CASTEP with those calculated with BigDFT (depicted in Figure~S2 in the Supporting Information), the ordering changes and the relative energies differ on average by 4.0~meV/atom.  Therefore, in the following, structures within 10~meV/atom of the lowest energy structure are grouped together, and are all considered to be low in energy.  We note that the energies calculated using HSE after further relaxation differ more significantly from the BigDFT PBE energies, however, despite a smaller range of values, the trend is well preserved (see Figure~S2 in the Supporting Information), and thus unless otherwise stated, relative energies are those calculated with BigDFT. For the randomly selected structures, a considerable number of structures contain Ga atoms which change site following relaxation, while the relative energies also change considerably.  As a result, only two randomly selected structures have two occupied Ga sites, both of which are relatively low in energy, while there remain some 3- and 4-site structures which are relatively high in energy.

\begin{figure}[ht!]
\centering
\includegraphics[scale=0.85]{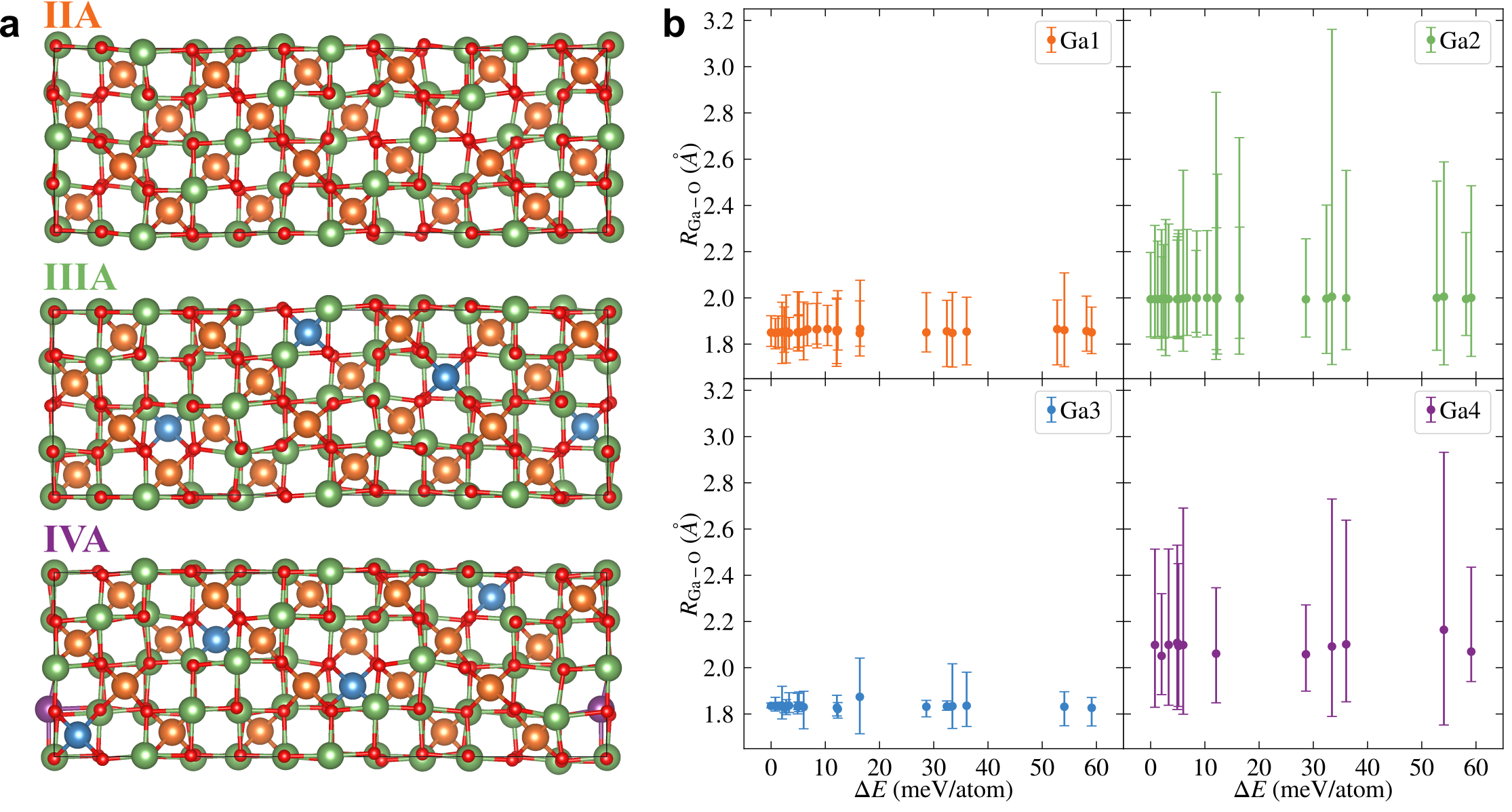}
\caption{Relaxed structures of $\gamma$-\ce{Ga2O3}. (a) Lowest energy relaxed 2- (IIA), 3- (IIIA), and 4- (IVA) site structures, with Ga sites 1/2/3/4 depicted in orange/green/blue/purple, and O atoms in red, as viewed along the $b$-axis. (b) Ga-O bond lengths for the 30 relaxed structures \emph{vs.}\ the relative energy of that structure, for each type of Ga site. Shown is both the average bond length (points) and the minimum and maximum bond lengths (error bars) in a given structure.  \label{fig:bonds}}
\end{figure}

The lowest energy PBE-relaxed 2-, 3- and 4-site structures are depicted in Figure~\ref{fig:bonds}(a), where the distortions away from ideal octa- and tetrahedra are clearly visible, as expected based on the neutron diffraction results~\cite{Playford2013,Playford2014}.  This is also evident in the smearing of the correlation function (depicted for the lowest energy structure in Figure~S3 in the Supporting Information) and in the spreading out of both Ga and O atoms around their undistorted positions (depicted in Figure~S4 in the Supporting Information for the low energy structures).  We note that the structures were further relaxed using HSE for the band gap calculations, however, this did not have a significant effect on the atomic positions (0.01~\AA\ average displacement). To further investigate the distortions, the Ga-O bond lengths were analysed for all 30 relaxed structures.  These were determined by identifying the 4 (6) closest O atoms for each tetrahedral (octahedral) Ga atom, giving rise to the average, maximum and minimum bond length for each type of Ga site in a given structure.  The results are depicted in Figure~\ref{fig:bonds}(b). Disorder is present in the bonds associated with all four Ga sites, but to a much greater extent for the $O_h$ sites. The two $O_h$ sites also show a larger difference in average bond lengths -- taking the lowest energy 4-site structure (IVA), the average bond lengths are 1.85, 1.99, 1.83 and 2.10~\AA\ for Ga sites 1 to 4, respectively, which are in good agreement with the values from Playford \emph{et al.}~\cite{Playford2014}.

\subsection{Occupied and Unoccupied States}

\begin{figure}[ht!]
\centering
\includegraphics[scale=0.85]{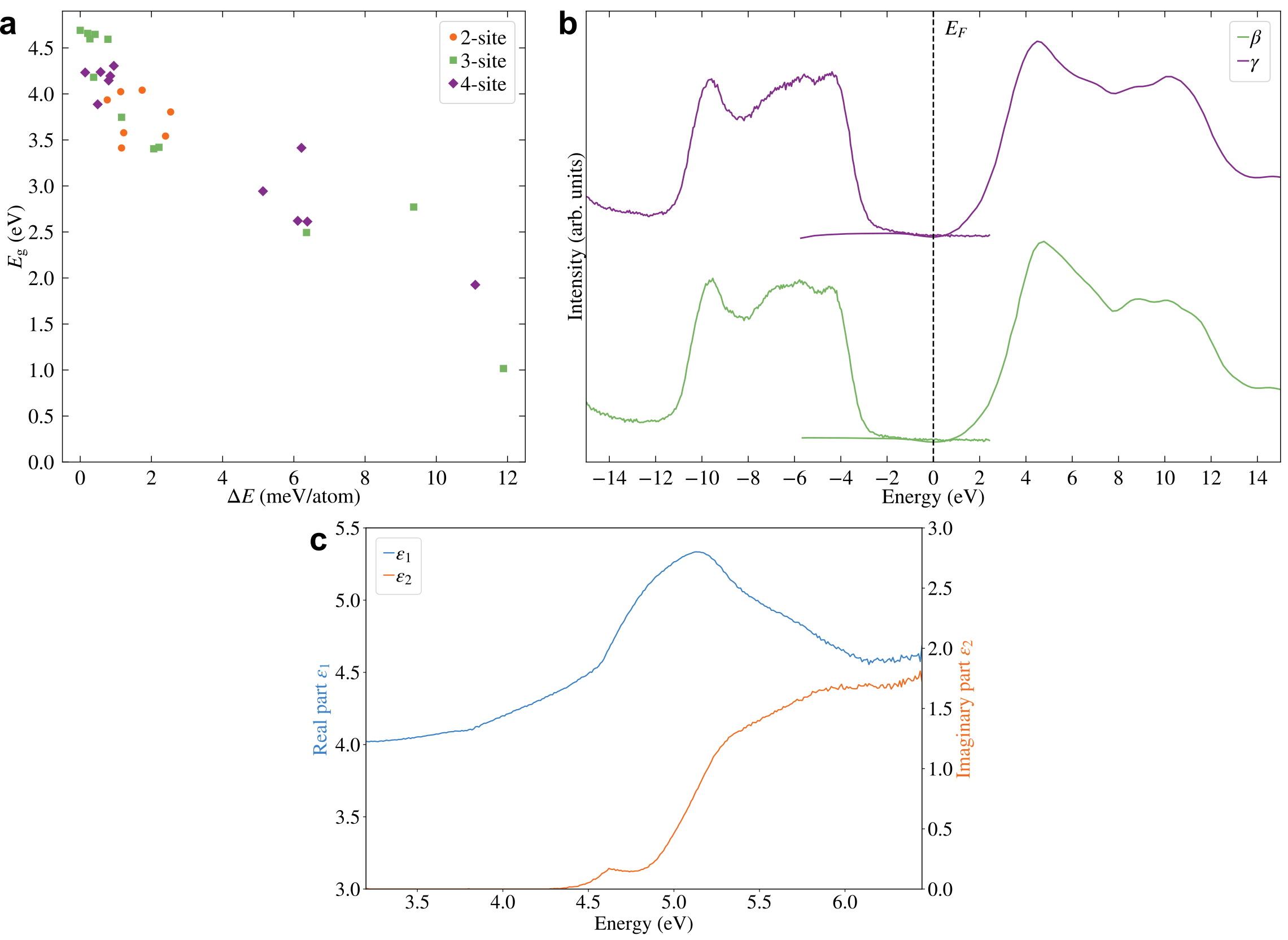}
\caption{Band gap and band alignment of $\gamma$-\ce{Ga2O3}. (a) Relative energy, $\Delta E$ \emph{vs.}\ band gap, $E_{\mathrm{g}}$ for all relaxed structures, where both quantities have been calculated using HSE. (b) Soft X-ray photoelectron spectra of the valence band states and O X-ray absorption spectra of the conduction states of both $\gamma$- and $\beta$-\ce{Ga2O3} plotted on a common energy axis.~\cite{Swallow2020} (c) Point-by-point fitted dielectric function of $\gamma$-\ce{Ga2O3} in the visible and ultraviolet spectral range.}
\label{fig:gap}
\end{figure}

A detailed understanding of the influence of structural disorder on the electronic structure of $\gamma$-\ce{Ga2O3}, including the nature of the occupied and unoccupied states as well as the band gap and alignment, is crucial to provide a firm knowledge base for this and other disordered materials as well as for furthering optimisation and implementation across different applications. HSE was used to calculate the band gaps for the 30 relaxed structures (see Figure~\ref{fig:gap}(a)) with a clear trend of lowest energy structures having the largest band gaps. The band gap value for the lowest energy structure, which is a 3-site structure, is 4.69~eV, which is smaller than the 4.87~eV band gap of the $\beta$-\ce{Ga2O3} from HSE calculations.~\cite{Varley2010} This is in agreement with the narrowing of the band gap observed in the experimental SXPS and XAS data, which are plotted on a common energy scale for both the $\gamma$ and $\beta$ polymorphs in Figure~\ref{fig:gap}(b) (see Figure~S5 in the Supporting Information for the full range O K-edge XAS spectrum). The 0~eV point of the common energy scale is aligned to the experimental Fermi energy $E_F$ position as determined from the SXPS experiments. In line with our previous results on the other polymorphs, the $E_F$ appears within the band gap towards the conduction band minimum (CBM), indicating that $\gamma$-\ce{Ga2O3} is non-degenerate n-type. However, due to limitations in the common energy scale alignment as well as possible small influences from surface band bending, this approach cannot be used to extract reliable band gap values. Therefore, a combination of spectroscopic ellipsometry and photoluminscence excitation spectroscopy (PLE) was used to further explore the band gap experimentally.

The experimentally determined point-by-point fitted dielectric function is shown in Figure~\ref{fig:gap}(c). The onset of strong absorption is visible by the increase of $\varepsilon_2$ at an energy of $\approx$5.1~eV, determined by the characteristic energy of a model dielectric function used to describe the point-by-point result. This value is related to the lowest allowed direct band-to-band transition in the material, but lowered by Coulomb interaction, i.e.\ excitonic effects. This value is very similar to that of other polymorphs of \ce{Ga2O3}, namely an averaged optical response over $x$ and $y$ directions of the dielectric tensor of stable $\beta$-\ce{Ga2O3}\cite{Sturm2015} and the ordinary dielectric function of $\epsilon$-\ce{Ga2O3}\cite{unpublished}, but lower than that of corundum $\alpha$-\ce{Ga2O3}.~\cite{Feneberg2018} The tiny contribution to $\varepsilon_2$ at $\approx$4.6eV is most likely an artifact due to an imperfect model description of the sample. From $\varepsilon_1(\hbar \omega \longrightarrow 0)$,  the dielectric limit of the material, usually referred to as $\varepsilon_\infty$, can be estimated to be $3.9 \pm 0.1$.

\begin{figure}[ht]
\centering
\includegraphics[scale=0.85]{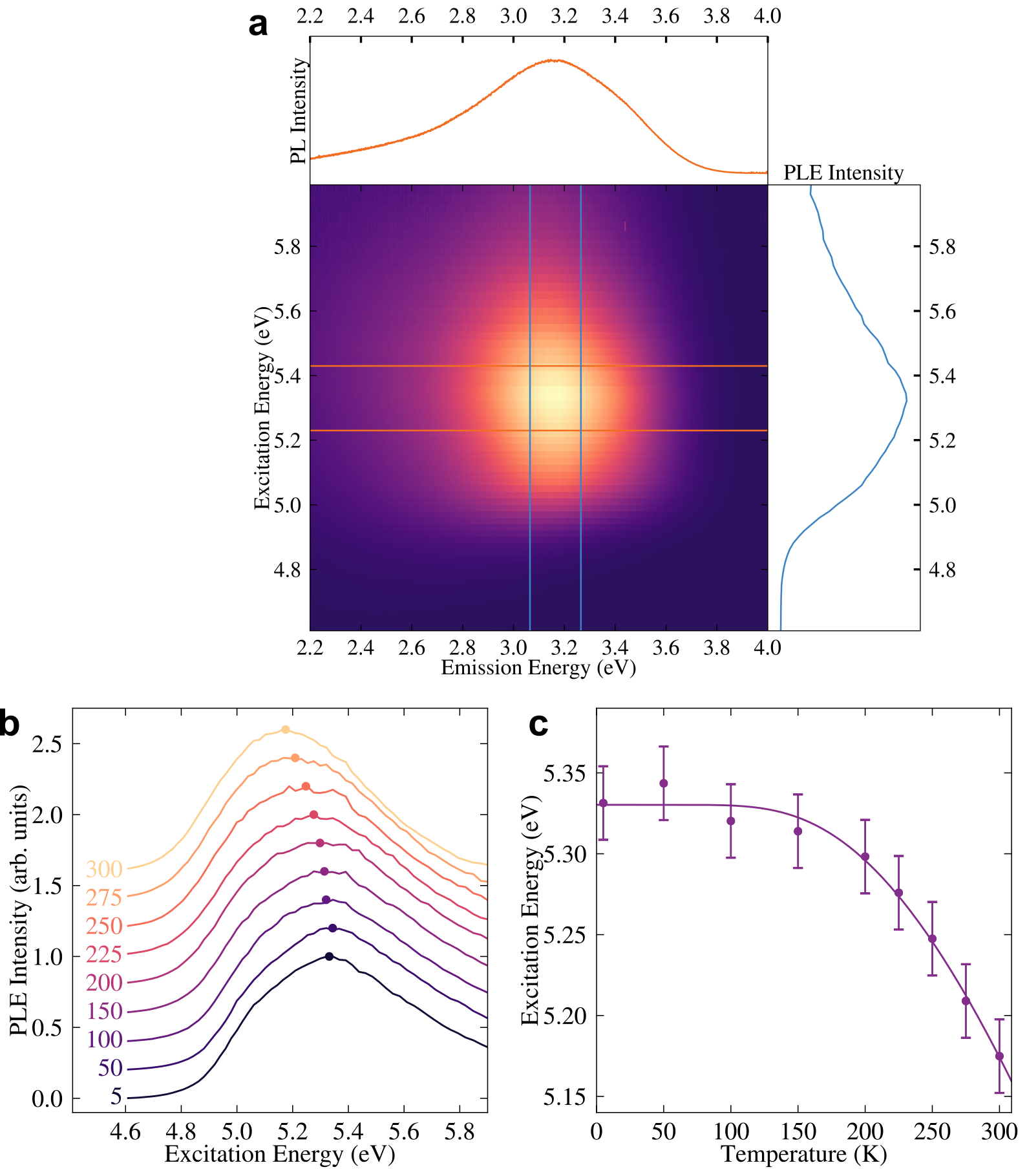}
\caption{PLE spectra of $\gamma$-\ce{Ga2O3}. (a) 2D map of excitation energy as function of emission energy as obtained by PLE measurement of the $\gamma$-\ce{Ga2O3} film at a temperature of 5~K. The top and right graphs display corresponding PL and PLE spectra as obtained within $\pm 100$~meV energy intervals around the intensity maxima of 3.17~eV and 5.33~eV, respectively. (b) Integrated PLE spectra as a function of temperature between 5~K and 300~K. The spectra have been normalised and shifted vertically for clarity. The dots mark approximate excitation energy corresponding to highest PLE signal. (c) Energy of the excitation maximum in the PLE spectra as function of temperature between 5~K and 300~K (dots and corresponding error bars). The solid line represents a semi-empirical fit of the temperature dependence of the excitation energy.~\cite{ODonnell1991,Paessler1997}
\label{fig:ple}}
\end{figure}

Figure~\ref{fig:ple}(a) shows a response- and substrate-corrected PLE map of $\gamma$-\ce{Ga2O3} at 5~K. It is dominated by a strong and broad excitation channel centered around 5.33~eV feeding a broad luminescence with an intensity maximum at 3.17~eV. The corresponding excitation spectra for selected temperatures between 5 and 300~K are displayed in Figure~\ref{fig:ple}(b). The maxima of PLE intensity (dots) shift to lower excitation energies starting at around 150~K. This shift can be well approximated by phenomenological expressions commonly used to describe the temperature dependence of band gaps in semiconductors~\cite{ODonnell1991,Paessler1997} as shown by the solid line in Figure~\ref{fig:ple}(c). The derived values for the high-temperature slope of $\mathrm{3.2(9)}$~meV/K and the low temperature energy gap $\mathrm{E(T=0)=5.330(3)}$~eV do not vary significantly depending on the specific model chosen (Ref.~\cite{ODonnell1991} was used here). At room temperature the PLE maximum is observed at 5.17~eV, thus down-shifted by 160~meV compared to $T=0$~K, which indicates that the highest excitation efficiency coincides with the onset of strong absorption (5.1~eV) observed in spectroscopic ellipsometry (Figure~\ref{fig:gap}(c)). It should be noted that the PLE spectrum is not generally expected to be identical to the corresponding absorption spectrum as only absorptive processes that lead to the occupation of electronic states which participate in the emission of the detected luminescence contribute to the PLE signal. In the present case, this can be observed in the high energy region ($\mathrm{E>5.5}$~eV), where the PLE signal decreases (Figure~\ref{fig:ple}(b)) despite increasing absorption (Fig~\ref{fig:gap}(c)), implying that at higher energies different relaxation paths become available that do not contribute to the measured luminescence band. The band gap values from spectroscopic ellipsometry and room temperature PLE are in good agreement with each other and somewhat larger than the SXPS/XAS and HSE predicted values, pointing towards a mismatch of electronic and optical band gap. It is worth noting that this effect appears more pronounced for $\gamma$-\ce{Ga2O3}, as calculations with the same methods yield good agreement with experiment for other polymorphs such as the $\beta$ and $\alpha$ phases.~\cite{Swallow2020,Sturm2015,Furthmuller.2016.10.1103/physrevb.93.115204,Bechstedt.2019.10.1063/1.5084324} One potential explanation may be due to localisation effects in the uppermost valence band states, which we describe in more detail below and in the Supporting Information.

\begin{figure}[ht]
\centering
\includegraphics[scale=0.85]{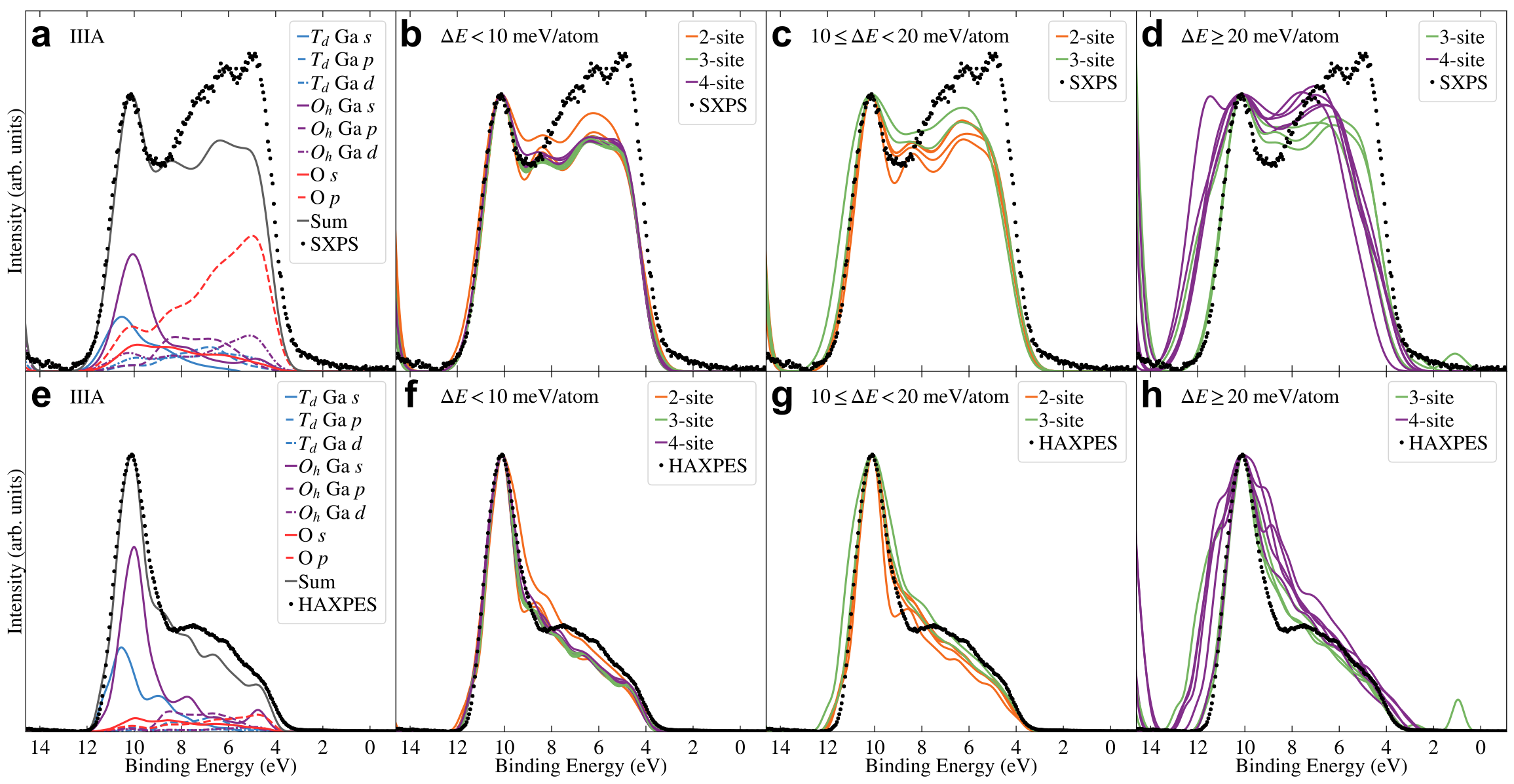}
\caption{Comparison between the calculated and measured valence XPS for (a) - (d) soft and (e) - (h) hard X-rays. The theoretical results are for the 30 relaxed structures listed in Table~S2 in the Supporting Information grouped by energy relative to the lowest energy relaxed structure. (a) and (e) show the weighted PDOS for the lowest energy structure for the SXPS and HAXPES case, respectively. The theoretical spectra have been aligned and normalised to the peak dominated by Ga $s$ states at the bottom of the VB.}
\label{fig:valence}
\end{figure}

In order to explore the electronic structure of the occupied states in more detail, PDOS calculations of the 30 relaxed structures were performed (see Figures~S6 and S7 in the Supporting Information for PDOS results of select structures). After broadening and photoionisation cross section weighting the PDOS can be directly compared to the experimental SXPS and HAXPES valence spectra (see Figures~\ref{fig:valence} (a) and (e), respectively). As is the case of the other \ce{Ga2O3} polymorphs, the valence band (VB) of the $\gamma$ phase is dominated by O~2$p$ states with small contributions from Ga~3$d$ and 4$s$ states at the top and bottom of the VB.
Whilst good agreement is found with the photoelectron spectra for the overall valence band, significant localisation of the highest-lying valence band states of $\gamma$-\ce{Ga2O3} is found, while no such localisation was observed in the conduction band states. Localisation extends to $\sim$0.8~eV below the highest occupied state for the lowest energy structure, as can be seen in Figures~S7 and S8 in the Supporting Information. This behavior in the model $\gamma$-\ce{Ga2O3} structures is similar to that observed in other semiconducting amorphous oxides,~\cite{Kim_Kang_Park_2012} and may account for the discrepancy between the measured optical and electronic band gaps.

Overall good agreement is found between the PDOS of the low energy structures ($\Delta E<10$~meV/atom) and the PES results. The 3- and 4-site structures are almost indistinguishable in their PDOS, with the 2-site structures showing some differences. Variations in PDOS increase when moving to the medium energy structures ($10\leq\Delta E<20$~meV/atom), while the high energy structures ($\Delta E \geq 20$~meV/atom) start to vary extensively, deviating clearly from the experimental spectra. In addition, in the high energy structures in-gap states start to appear which are not present in the experimental data. Therefore, whilst PES cannot distinguish between the lowest energy structures predicted from theory, it demonstrably shows that the higher energy structures are not a realistic description of $\gamma$-\ce{Ga2O3}.

\subsection{Semicore Spectra}

\begin{figure}[ht]
\centering
\includegraphics[scale=0.33]{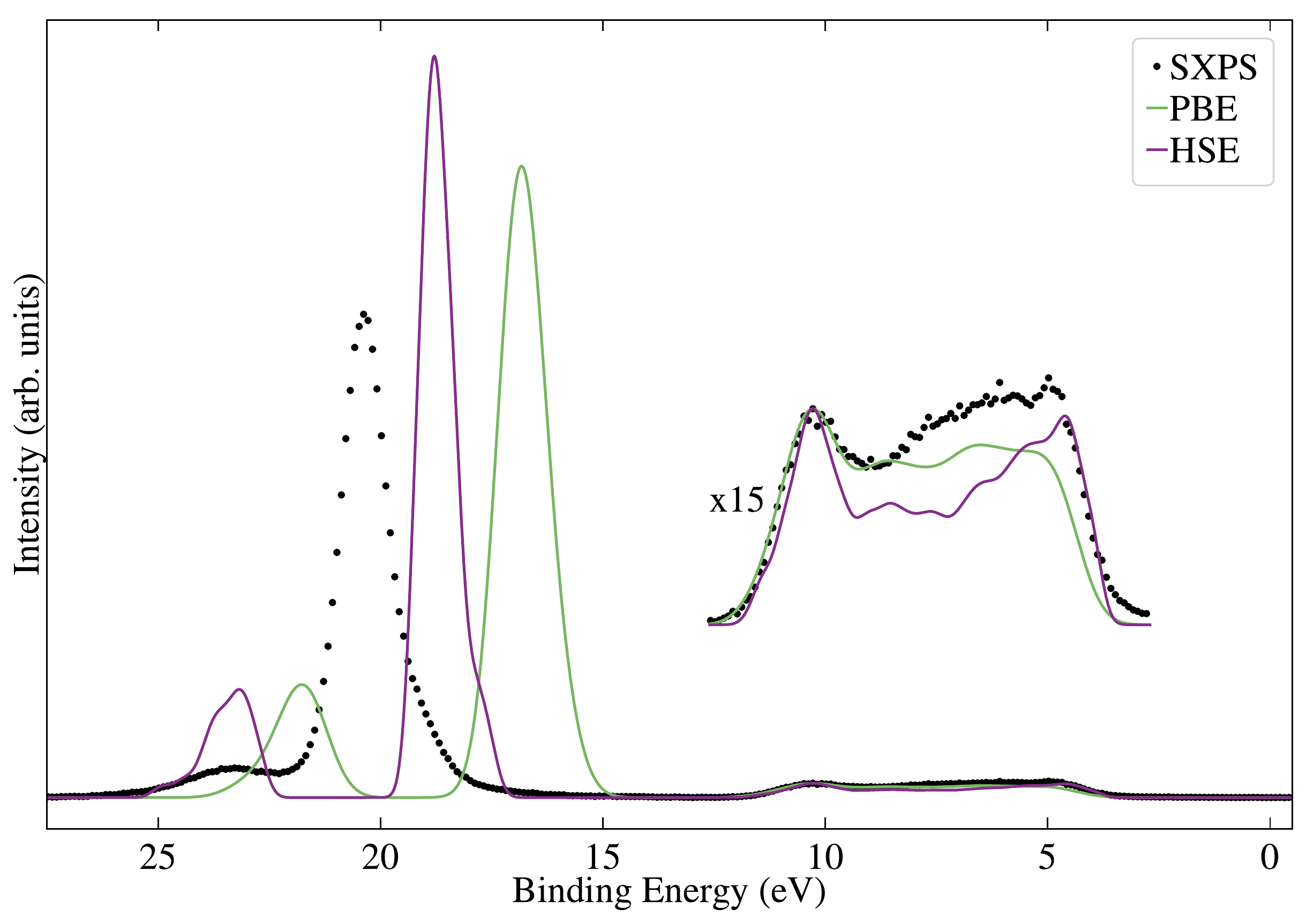}
\caption{Comparison of the weighted occupied PDOS from both PBE and HSE for the lowest energy structure with the experimental SXPS results, including both the valence bands and semicore states. The theoretical spectra are aligned to the dominant Ga $s$ feature at the bottom of the valence band. The detailed PDOS from both PBE and HSE can be found in Figure~S10 in the Supporting Information.}
\label{fig:dos}
\end{figure}

Semicore (shallow core) states can have a significant influence on the final electronic structure of metal oxide semiconductors. Figure~\ref{fig:dos} shows the theoretical spectra from PBE and HSE calculations as well as the experimental SXPS spectra of the semicore states and the valence band, with the calculated spectra coming from the lowest energy structure. HSE shows an improvement in the agreement between its predicted semicore energies and SXPS compared to PBE. As expected from the overestimation of the $p$--$d$-repulsion in the theoretical approach used, which leads to an underestimation of the band gap, the binding energy (BE) positions of the semicore levels relative to the valence band are underestimated compared to experiment. In addition, theory underestimates the level of hybridisation in \ce{Ga2O3}, leading to differences in relative peak intensities of the semicore states as well as the obvious disparity in their BE separation. Nevertheless, their overall shape and orbital character is described well even with PBE, as will be discussed in the following.

\begin{figure}[ht]
\centering
\includegraphics[scale=0.85]{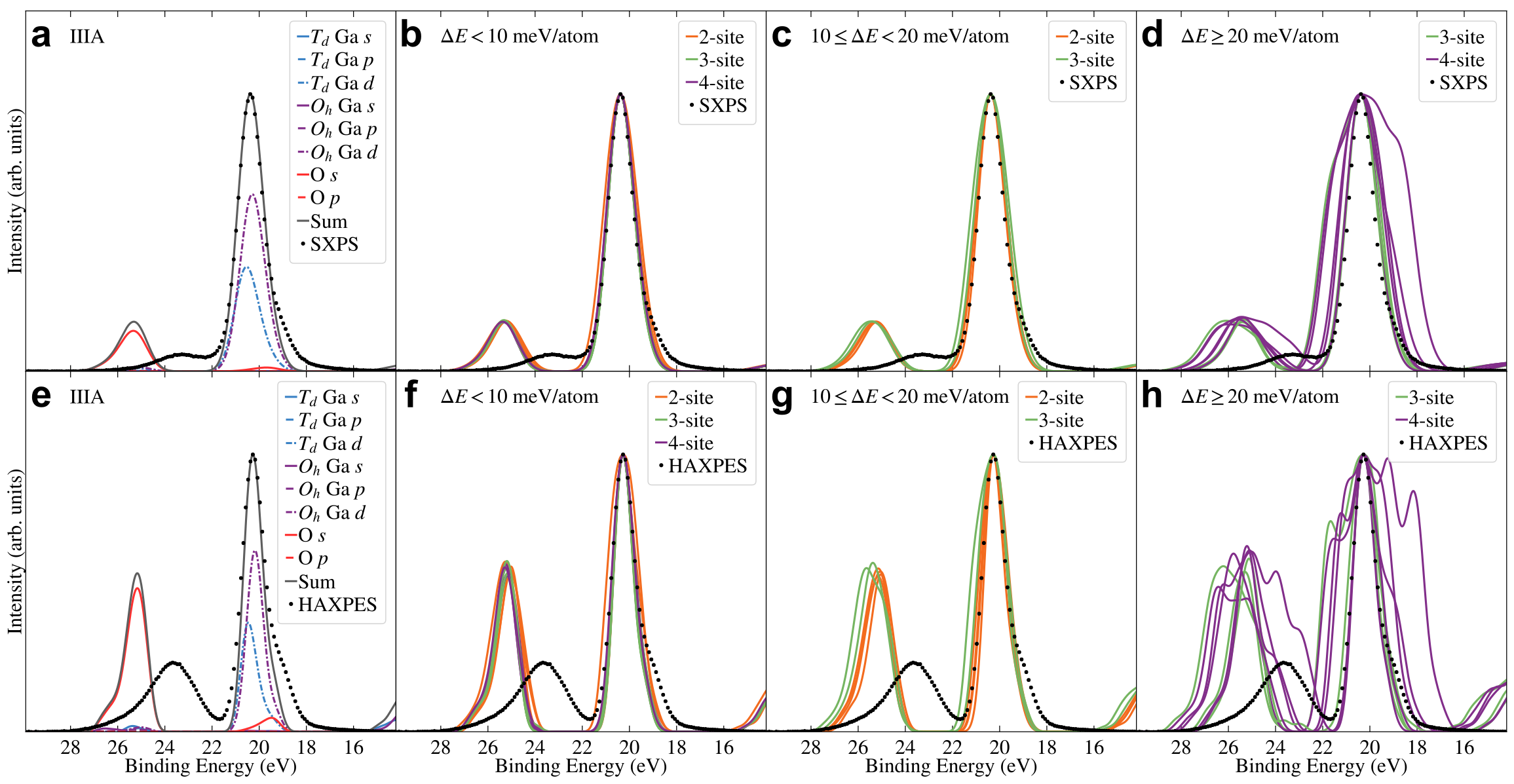}
\caption{Comparison between calculated and measured semicore states for SXPS and HAXPES. Figures (a) and (e) show the weighted PDOS for the lowest energy structure (3-site structure) for SXPS and HAXPES, respectively. Figures (b)-(d) and (f)-(h) show the theoretical results for the 30 relaxed structures listed in Table~S2 in the Supporting Information grouped by energy relative to the lowest energy relaxed structure, and compared to the experimental SXPS and HAXPES spectra, respectively. The theoretical spectra are aligned and normalised to the peak with the maximum height.
\label{fig:semicore}}
\end{figure}

The semicore states in \ce{Ga2O3} show two peaks of O~2$s$ and Ga~3$d$ character, for which the broadened and cross section corrected PDOS for the lowest energy $\gamma$-\ce{Ga2O3} structure, as well as the SXPS and HAXPES data, are shown in Figures~\ref{fig:semicore}(a) and (e). Following the arguments presented for the valence spectra above, combining SXPS and HAXPES can help to identify the orbital character of observed spectral features and verify theoretical projections. The dominant feature in the semicore spectra is the predominantly Ga~3$d$ peak (at 21.1~eV in the SXPS data), with contributions from $T_d$ and $O_h$ sites reflecting the ratio of sites present within the structure, with $O_h$ dominating. In addition, a shoulder towards the lower BE of the main feature is due to hybridisation with O~2$s$ states. The difference in the decay in photoionisation cross sections between O~2$s$ and Ga~3$d$, with the O~2$s$ cross section decreasing at a slower rate compared to the Ga~3$d$ one, leads to this shoulder being more pronounced in the HAXPES spectra compared to the SXPS spectra. This clear cross sectional dependence is also obvious in the second semicore state (at 24.5~eV in the HAXPES data), which is largely dominated by O~2$s$ states and which changes considerably in relative intensity when going from soft to hard X-rays.

Figures~\ref{fig:semicore}(b)-(d) and (f)-(h) show the comparison of the experimental spectra with the calculated spectra for the 30 relaxed structures listed in Table~S2 in the Supporting Information, grouped by energy relative to the lowest energy relaxed structure, for SXPS and HAXPES, respectively. Unweighted PDOS and comparison between the calculated and measured semicore XPS for soft and hard X-rays for select relaxed structures can be found in Figures~S11 and S12 in the Supporting Information. Whilst only very minimal differences exist between the lowest energy ($<10$~meV/atom) structures, changes become more obvious for structures with energies $10\leq\Delta E<20$~meV/atom, particularly in the HAXPES case due to the discussed relative increase in the O~2$s$ cross section. Finally, for the high energy structures with $\Delta E \le 20$~meV/atom the calculated spectra no longer resemble the experiment. The resulting width increase, in particular for the Ga~3$d$ dominated feature, originates from the spreading in energies for both Ga and O states, which is magnified due to hybridisation in the semicore states. As was evident from VB spectra, the higher energy structures are not a realistic representation of $\gamma$-\ce{Ga2O3}.

\subsection{Core Level Spectra}

\begin{figure}[ht!]
\centering
\includegraphics[scale=0.85]{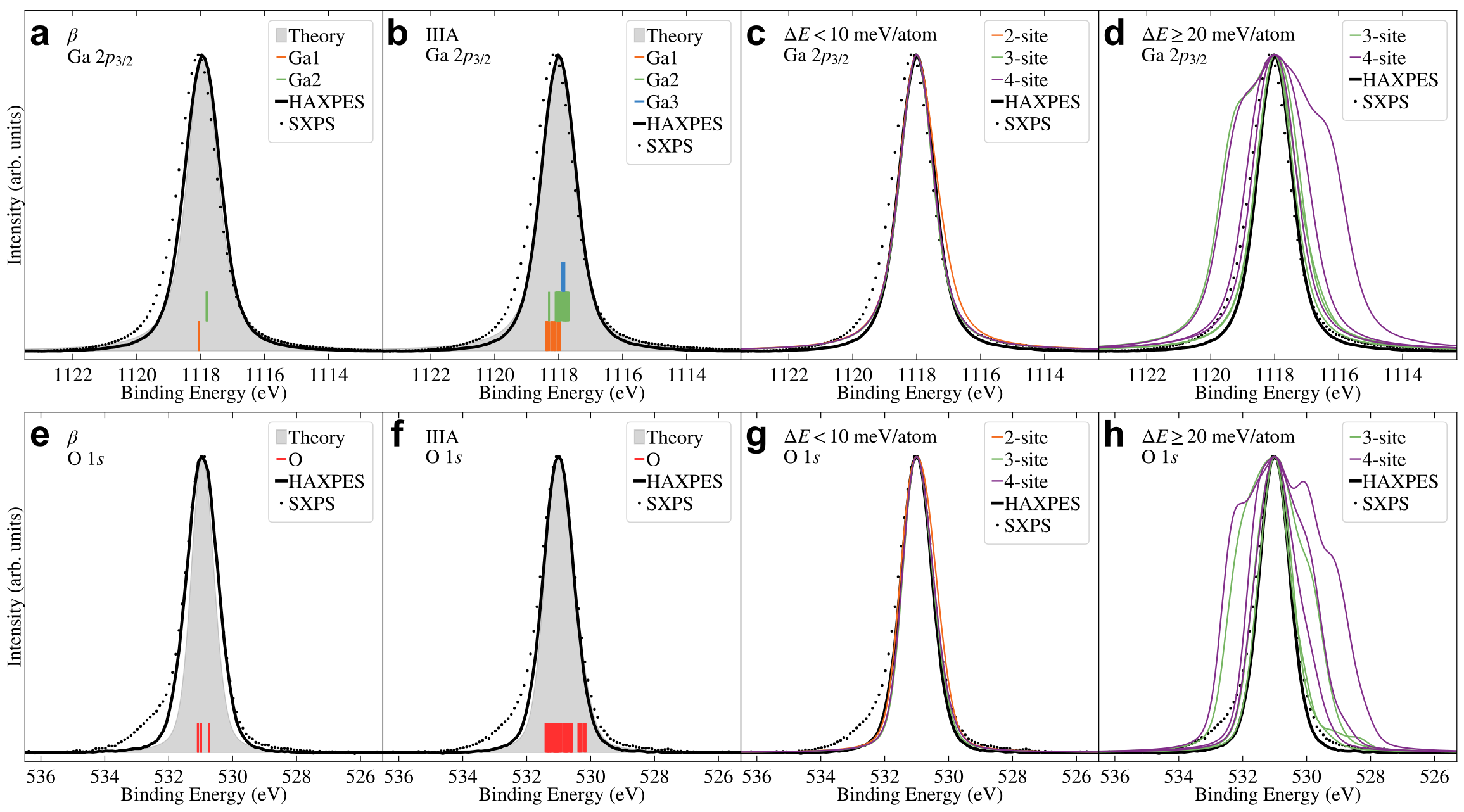}
\caption{Comparison between the calculated and measured core states for Ga~2$p_{3/2}$ [top] and O~1$s$ [bottom], for both SXPS and HAXPES. Figures (b) and (f) show the different site contributions for the lowest energy $\gamma$-\ce{Ga2O3} structure (equivalent plots for the other structures are shown in Figure~S11 in the Supporting Information), and Figures (a) and (e) the reference results for $\beta$-\ce{Ga2O3}. Figures (c), (d), (g), and (h) show the theoretical results for the lowest 2-, 3-, and 4-site relaxed structures, as well as the six highest energy (3- and 4-site) relaxed structures. Calculations are grouped by energy relative to the lowest energy relaxed structure. The theoretical spectra are normalised and aligned with respect to experiment.}
\label{fig:core}
\end{figure}

Although core level spectra are often overlooked as a source of information regarding the relationship between local coordination and electronic structure of a material, recent work has shown that the full width at half maximum (FWHM) of core levels, in particular of O~1$s$, can change significantly depending on local coordination environments in \ce{Ga2O3} polymorphs.~\cite{Swallow2020} Ga~2$p_{3/2}$ and O~1$s$ core level spectra of $\gamma$-\ce{Ga2O3}, as well as $\beta$-\ce{Ga2O3} for comparison, were collected using both SXPS and HAXPES and are shown in Figure~\ref{fig:core}. The advantage of collecting core level spectra with both SXPS and HAXPES is that any surface related effects affecting SXPS data, such as hydroxylation, undercoordination, and band bending, do not influence the HAXPES spectra considerably due to the minute contribution from the sample surface to the overall signal. The Ga~2$p_{3/2}$ spectra are identical in SXPS and HAXPES except for the small difference in the energy resolution of the two measurements. The O~1$s$ spectra also have comparable FWHM and line shapes in SXPS and HAXPES, with the SXPS data showing a small feature on the higher BE side of the main photoionisation peak due to surface species, such as hydroxyl groups.

The FWHM of the HAXPES core level spectra are almost identical at 1.15$\pm$0.01~eV for O~1$s$ and \newline 1.29$\pm$0.01~eV for Ga~2$p_{3/2}$ for both $\gamma$- and $\beta$-\ce{Ga2O3}. To understand this observation, core BE calculations were performed and the results are shown in Figures~\ref{fig:core}(b) and (f) for the lowest energy $\gamma$ and (a) and (e) for the reference $\beta$ structure. In the much simpler $\beta$ case, two clearly separated contributions (with a difference in energy $\Delta$E of 0.25~eV) from $T_d$ Ga1 and $O_h$ Ga2 to the Ga~2$p_{3/2}$ core level are found. In contrast, the relative energies in the $\gamma$ case are spread out due to the differences in coordination environment, with a maximum difference in energy $\Delta E_{\mathrm{max}}=0.71$~eV. O~1$s$ shows similar behaviour albeit with an in general larger $\Delta E_{\mathrm{max}}$ (0.36~eV in $\beta$-\ce{Ga2O3}) and also a greater magnitude of spreading in the $\gamma$ structure (to $\Delta E_{\mathrm{max}}$ = 1.25~eV). This heightened sensitivity of O~1$s$ to changes in local coordination compared to Ga~2$p_{3/2}$ is in agreement with previous observations on \ce{Ga2O3} polymorphs.~\cite{Swallow2020}

Due to the high computational cost of core BE calculations (where each atom in the cell requires a separate DFT calculation), core BEs were only calculated for nine structures: the lowest energy relaxed 2-, 3-, and 4-site structures, and the six highest energy relaxed structures.  Since there are no high energy relaxed 2-site structures, this comprised three 3-site and three 4-site structures. Figures~\ref{fig:core}(c) and (g) show the comparison of the experimental spectra to calculated spectra for the lowest energy structures with two, three and four occupied Ga sites. Similar to the observations made for the semicore states, it is clear that only very minimal differences exist between these structures, which are too small to influence the overall line profile of the experimental core level spectra. In contrast, the spectra from the highest energy structures shown in Figures~\ref{fig:core}(d) and (h) show various levels of deviation, including some extreme cases for the 4-site structures, where there is a much larger spread of BE values for both Ga~2$p_{3/2}$ and O~1$s$.  This is common to all Ga sites, as shown in Figure~S13 and Table~S3 in the Supporting Information. A possible explanation for this greater spread of values comes from the rough trend that the higher the energy of the structure, the greater the number of distinct environments present in the unrelaxed structure, as depicted in Figure~S14 in the Supporting Information.  Although the higher energy structures in particular change significantly during relaxation, it is nonetheless likely that they typically retain a greater diversity in local environment, leading to the wider spread of BE values.

\section{Conclusion}

The present work showcases a successful strategy for theoretically screening and experimentally validating potential atomic structures of the disordered material $\gamma$-\ce{Ga2O3} and assessing their relationship to its electronic structure. By combining the screening of nearly a million structures with more than a thousand DFT calculations, a number of low energy candidate structures were identified. Based on the developed robust atomistic model, band gaps, densities of states, as well as semi core and core state energies are calculated and directly compared to experimental efforts across a range of advanced techniques. The results convincingly show that the predicted low energy structures are a suitable representation of disorder in $\gamma$-\ce{Ga2O3}, whilst higher energy structures result in dramatic deviations from experiments. Of the low energy structures, both 3- and 4-site models provide good descriptions of the experimental data, with 2-site models showing worse agreement with experiment. This work opens up pathways to more extensive theoretical explorations of this material, such as going beyond fixed lattice parameters, and treating larger supercells to enable a more realistic treatment of long range disorder. Furthermore, the complementarity between theoretical and experimental approaches demonstrated here shows the potential for applying such a strategy to many other disordered systems where, whilst technologically interesting, structure -- electronic structure relations have been challenging (or impossible) to explore.

\section{Theoretical and Experimental Approaches} 

\subsection{Theoretical Approach}

\subsubsection{Structure Generation}\label{sec:struc_gen}

Atomic structures were generated using the structure of Playford~\emph{et al.},~\cite{Playford2013} using a fixed lattice constant of 8.2376~\AA. Preliminary calculations in smaller cells showed a strong sensitivity to stoichiometry, therefore all calculations were performed in a stoichiometric $1 \times 1 \times 3$ supercell (160 atoms). The Ga sites were randomly occupied, first by selecting a site with a probability such that each type of Ga site was selected with the same frequency.  The selected site was then occupied using the following probabilities.  For 10\% of the 800 initial $1 \times 1 \times 3$ structures only the first two sites (Ga1 and Ga2) were occupied, with equal probability. For the remainder of the structures, all four Ga sites were randomly occupied, with probabilities of 0.741, 0.741, 0.066 and 0.024 for Ga1 through Ga4 respectively, following Ref.~\cite{Playford2013}.  To avoid unphysically short Ga-Ga distances, a minimum Ga-Ga distance of 2.4~\AA\ was imposed.  Due to the combination of the distance constraint and randomised approach, even in the case where all four sites had a non-zero probability, many structures nonetheless had only two or three sites occupied, as summarised in Table~S1 in the Supporting Information.

\subsubsection{Density Functional Theory Calculations}\label{sec:dft}

Except where stated otherwise, density functional theory (DFT) calculations were performed using the semi-local PBE functional.~\cite{Perdew1996}
Single point energy calculations and geometry optimisations employed the wavelet-based BigDFT code,~\cite{Ratcliff2020} using HGH-GTH pseudopotentials (PSPs),~\cite{Goedecker1996,Hartwigsen1998} also including non-linear core corrections for O.~\cite{Willand2013} The Ga PSP included three valence electrons, with the 3$d$ electrons treated as part of the core. Calculations used a wavelet grid spacing of 0.38~bohr and coarse (fine) radius multipliers of 5 (7), and were performed at the $\Gamma$-point only.  In order to aid convergence for screening calculations, and since a number of the generated structures were found to have a negligible band gap, density mixing was used with a finite temperature of 0.001~Ha.  Structures which did not converge within 500 diagonalisation iterations ($\sim$2\% of initial structures) were discarded.  Geometry optimisations were performed using a direct minimisation approach without finite temperature, since a band gap opened up during relaxation for all structures.  A maximum force threshold of 0.03~eV/\AA\ was employed, while the unit cell was kept fixed.

Projected density of states (PDOS) calculations were performed using the CASTEP plane-wave DFT code,~\cite{Clark2005} employing a kinetic energy cut-off of 700~eV.  Ultrasoft PSPs were employed with the Ga~3$d$ electrons treated as valence states.  The density was obtained at the $\Gamma$-point only, while the PDOS was calculated on a $2\times 2\times 1$ Monkhorst-Pack $k$-point grid.~\cite{Monkhorst1976}
Post-processing was performed using OptaDOS,~\cite{Morris2014} with 0.44~eV (0.25~eV) Gaussian smearing applied for comparison with soft (hard) X-rays to reflect the experimental broadening. To further aid comparison of theory and experiment, Scofield photoionisation cross sections for the experimental soft and hard X-ray energies were applied to the calculated PDOS using Galore.~\cite{Scofield1973, Dig_Sco2020, Jackson2018} Unweighted PDOS were broadened using 0.44~eV Gaussian smearing. 

Relative core binding energies (BE) were also calculated with CASTEP, using the $\Delta$SCF approach, in which a core hole is introduced into the excited atom by means of an on-the-fly generated core hole PSP. Core-hole calculations are then performed with a net charge. For Ga the core-hole PSP was generated using an averaged approach, i.e.\ the electron was not removed from a specific $p$-orbital, and thus the calculated BEs do not specifically correspond to the 2$p_{3/2}$ states.  The same cut-off energy and PSPs were employed as for PDOS calculations. Due to the higher computational cost of core BE calculations, which require one calculation per atom in a given structure, calculations were only performed for select structures, at the $\Gamma$-point only. Core BE and PDOS calculations were also performed for $\beta$-\ce{Ga2O3}, using the same PSPs and cut-off energy, and an $8\times 8 \times 8$ Monkhorst-Pack $k$-point grid. The unit cell was taken from Ref.~\cite{Ahman1996}. The structure was relaxed using CASTEP using a maximum force threshold of 0.03~eV/\AA\ while keeping the cell fixed, otherwise using the same parameters as the core BE calculations.  BEs were calculated for each atom in the cell, regardless of whether they are equivalent by symmetry. For both $\gamma$ and $\beta$ phases, the spectra were generated by applying a combination of 0.25~eV Gaussian smearing to reflect the experimental broadening of the HAXPES data and 0.38~eV (0.133~eV) Lorentzian smearing for Ga (O) to reflect the different intrinsic line widths of the Ga and O lines.

To evaluate the band gaps beyond semilocal functionals, HSE06 screened hybrid functional calculations were employed. These calculations were performed with the VASP code with 32\% mixing of exact exchange, a 400~eV plane-wave cut-off and PAW potentials that include the Ga $3d$ states as valence electrons, which has been previously shown to accurately describe the properties of $\beta$-\ce{Ga2O3}.\cite{Swallow2020}  All atomic coordinates in the 160-atom model $\gamma$-\ce{Ga2O3} structures were allowed to relax and the direct band gap was evaluated at the zone-center with a $\Gamma$-centered $k$-point mesh including $\Gamma$ and another $k$-point at the zone boundary at 0.5,0.5,0.5. Owing to the size of the supercells, this band gap may also include contributions from higher-lying valence band states that are folded to the $\Gamma$-point in the supercells. For the lowest energy structure, the density of states was further resolved with a 5$\times$5$\times$1 $\Gamma$-centered mesh. This larger $k$-point sampling confirmed the indirect band gap nature of $\gamma$-\ce{Ga2O3}, with the indirect valence band maximum (VBM) falling at least 0.05 eV higher than the highest-lying valence band state at the $\Gamma$-point of the supercell.

\subsubsection{Machine Learning Approach}\label{sec:ml_approach}

Since the number of possible structures is too large for an exhaustive search, a machine learning (ML) approach was implemented to nonetheless enable the screening of a large number of structures. The employed model relies on the decomposition of a structure into a set of distinct (`reference') local atomic environments.  Each reference environment, $\alpha$ has an associated energy, $\varepsilon_\alpha$, with the total energy of a given structure, $E$, defined as 
\begin{equation}\label{eq:model_energy}
E =\sum_{\alpha=1}^{N_\mathrm{env}} M_\alpha \varepsilon_\alpha\;,
\end{equation}    
where $M_\alpha$ is the number of instances of $\alpha$ in the structure and $N_\mathrm{env}$ is the total number of reference environments.
The decomposition of the total energy into local atomic energies is well established in the ML community, and has been employed with a range of descriptors.~\cite{Musil2021,Keith2021} In this work we use a simple descriptor based only on the local coordination environment, as described below.

For each atom $i$ of species $s$ in a given structure, all Ga (O) atoms within a radius of $r_c^{s-\mathrm{Ga}}$ ($r_c^{s-\mathrm{O}}$) of atom $i$ are identified.  Atom $i$ and these neighbouring atoms constitute environment $\alpha$.  Given another environment $\beta$, centred on atom $j$ of the same species $s$, the two environments may be said to be equivalent if both $\alpha$ and $\beta$ contain the same number of Ga and O atoms, i.e.\ if $N_{\alpha}^{\mathrm{Ga}} = N_{\beta}^{\mathrm{Ga}}$ and  $N_{\alpha}^{\mathrm{O}} = N_{\beta}^{\mathrm{O}}$.  Otherwise, the two environments are treated as distinct.  In other words, only the number of atoms of each species present within the environment is taken into consideration, irrespective of their actual atomic coordinates. 
Since all O sites are fully occupied, all O-O interactions are excluded, i.e.\ $r_c^{\mathrm{O-O}}=0.0$~\AA. Therefore, the approach relies on only two parameters: $r_c^{\mathrm{Ga-Ga}}$ and $r_c^{\mathrm{Ga-O}}\equiv r_c^{\mathrm{O-Ga}}$.

The decomposition of a given atomic structure then proceeds as follows, as illustrated in Figure~S15 in the Supporting Information. For each atom $i$ in the input structure, its local environment $\alpha$ is obtained, given user-defined cut-off radii.  Environment $\alpha$ is then compared to each of the existing reference environments $\beta$ associated with an atom of the same species as atom $i$.  If none of the existing environments are equivalent to $\alpha$, then $\alpha$ is a new, distinct environment and is added to the set of reference environments. The process is repeated for each atom in the structure, and again for subsequent structures, taking the already existing set of reference structures as an input, adding new reference environments as needed.

Once all structures within a given data set have been decomposed into a set of reference environments, the data is split into training and validation sets, where each structure which contains a previously unseen reference environment must be included in the training set. The energies of each environment, $\varepsilon_\alpha$, are then obtained by fitting to DFT-calculated energies using ridge regression, as implemented in Scikit-learn.~\cite{Pedregosa2011} In order to determine the cut-off distances, $r_c$, single point calculations were performed for 800 initial structures.  The structures were split randomly into 534 (226) training (validation) structures. To account for variations due to the choice of training set, five randomised splits were tested.  The model was then fit using different values of $r_c$, which were chosen so as to include increasing shells of neighbours, as depicted in Figure~S16 in the Supporting Information.  For each pair of $r_c$ values, the regularisation parameter was varied between 0.5 and 50 in intervals of 0.5. For each case, the value giving the lowest mean absolute error (MAE) of the validation set, when averaged across the five splits, was then employed.  As shown in Table~S4 in the Supporting Information, the values $r_c^{\mathrm{Ga-Ga}}=4.3$~\AA\ and $r_c^{\mathrm{Ga-O}}=4.8$~\AA\ gave the lowest MAE and are therefore used in the following.

Having established the model hyperparameters, a number of additional structures were generated using the four site occupancies, in order to improve the coverage of the data set.  A DFT calculation was then performed for each structure which contained either new (i.e.\ not already encountered) reference environments, or rare reference environments, which were present in less than five structures.  The structures were generated in batches of 20,000, discarding unconverged structures following the same criterion as before. The process was repeated until a total of 400,000 additional structures had been generated, giving rise to an extra 39 converged structures.  Although such an approach does not guarantee that all possible environments are represented, it is assumed that any remaining unrepresented environments are either very rare, or only occur in structures which did not converge, and are thus likely to be high in energy and therefore ultimately unfavourable.  The model was then refit by splitting the data into 560 (279) training (validation) structures.  The regularisation value,  $\alpha$, was again varied as described above; a final value of 0.5 was employed.

\subsection{Experimental Approach}

\subsubsection{Growth and Structure}
An epitaxial $\gamma$-\ce{Ga2O3} film was grown on a double polished (001) \ce{MgAl2O4} substrate using conventional plasma-assisted molecular beam epitaxy (MBE). Beam fluxes of Ga and O radicals were supplied to the heated substrate from an effusion cell and an RF-radical cell, respectively. The nominal Ga flux measured in vacuum was 2.2$\times$10$^{-8}$~Torr and an input RF power of 200~W and an oxygen flow rate of 0.50~standard cubic centimeters per minute (sccm) were used to generate oxygen radicals for the growth. In the oxygen background during growth, the Ga effusion cell mainly provided \ce{Ga2O} at a higher flux than the pure Ga flux provided during the beam-flux measurement in vacuum.~\cite{Hoffmann2021} The substrate temperature was 600$^{\circ}$C with a total growth time of 2000~s. The film thickness as estimated from the growth rate \newline(2.8~nm/min) was 92~nm. The surface roughness as determined from atomic force microscopy (AFM) is 0.8~nm. Further details can be found in previous publications.~\cite{Oshima2017, Oshima2019, Kato2019}

The crystal phase of the produced epitaxial film was investigated using X-ray diffraction (XRD). Data collection was performed on a high-resolution X-ray diffractometer (Rigaku SmartLab) using monochromated Cu K$\alpha_1$ radiation. The strong \ce{MgAl2O4} 001 substrate reflection is observed at 44.8$^\circ$ with the $\gamma$-\ce{Ga2O3} 004 film reflection at lower $2\theta$ (42.6$^{\circ}$) indicating successful epitaxial stabilisation (see Figure~S17 in the Supporting Information). The FWHM of the 004 rocking curve is 0.4$^\circ$.

In addition, $\gamma$-\ce{Ga2O3} films were grown by solid-phase epitaxy on c-plane \ce{Al2O3} substrates. For this purpose, an amorphous \ce{Ga2O3} film was grown by plasma-assisted MBE on a 2" c-plane \ce{Al2O3} substrates at a substrate temperature of 125$^\circ$C, a Ga flux of 2$\times$10$^{-7}$~mbar (corresponding to a \ce{Ga2O3} growth rate of 3.6~nm/min),  and an oxygen flux of 3~sccm at an RF plasma power of 300~W. The growth time of 35~min resulted in an estimated film thickness of $\sim$120~nm. Within the first 20~s of growth (i.e., the first 1.2~nm), the spotty reflection high-energy electron diffraction patterns (RHEED) pattern of the substrate changed into a featureless, diffuse one, indicating the formation of an amorphous film. After growth, the substrate was split into smaller pieces for the subsequent crystallisation of the amorphous film.

The amorphous films were annealed in an oven under \ce{O2} at atmospheric pressure isochronal for 30~min increasing the temperature from 400 to 900$^\circ$C in steps of 100$^\circ$C. The structure of the samples was analysed by X-ray diffraction in a Bragg-Brentano geometry with $\omega-2\theta$ scans after each annealing step. Figure~S18 in the Supporting Information shows the XRD $\omega-2\theta$ scans of as grown amorphous \ce{Ga2O3} thin films on sapphire substrate after annealing at the various temperatures. The sample annealed at 400$^\circ$C shows exclusively (0003)$_n$ reflections of the sapphire substrate indicating the sample still to be amorphous within the detection limit of X-ray diffraction. After further annealing to 500$^\circ$C, additional reflections are revealed (indexed with a, b, c, and d). These peaks can be assigned (111)$_n$ reflections of the $\gamma$-phase (111)$_n$, and they are close in position to the (20$\bar{1}$)$_n$ reflections of the $\beta$ phase. The XRD data shows that peaks (a) and (c) match literature data of the (201) and (603) reflections of the $\beta$ phase increasing in intensity with increasing temperature. Peaks (b) and (c) decrease in intensity with increasing temperature and shift to nominal values that correspond to reflections of the $\gamma$ phase. These observations are commensurate with a gradual transition from the $\gamma$-phase to the $\beta$-phase at higher temperatures. The sample annealed at 500$^\circ$C was used for the high resolution transmission electron microscopy experiments presented. 

\subsubsection{X-ray Photoelectron and Absorption Spectroscopy}
$\gamma$-\ce{Ga2O3} samples were investigated using both soft (SXPS) and hard (HAXPES) X-ray photoelectron spectroscopy. Laboratory-based SXPS was performed on a Thermo Scientific K-Alpha+ spectrometer with a monochromated microfocused Al K$\alpha$ X-ray source (h$\nu$ = 1486.7~eV) and a spot size of 400~$\mu$m. The base pressure of the instrument is 2$\times$10\textsuperscript{-9}~mbar and the X-ray source was operated at 6~mA emission current and 12~kV anode bias. Pass energies of 20~eV for core level and 15~eV for valence band spectra were used. HAXPES data were collected at beamline I09 at Diamond Light Source, UK, at a photon energy of 5.9403~keV.~\cite{Lee2018} A double-crystal Si (111) monochromator was combined with a Si (004) channel-cut crystal as a post-monochromator to achieve the final energy resolution. The main end station at beamline I09 is equipped with a VG Scienta EW4000 electron analyzer, which has a wide acceptance angle of $\pm$28$^\circ$. All measurements were performed in grazing incidence geometry at angles below 5$^\circ$ between the incoming X-ray beam and the sample surface and a pass energy of 200~eV was used for all spectra. The resolution of SXPS and HAXPES measurements as determined by the width of the Fermi edge of gold are 0.44~eV and 0.25~eV, respectively. The HAXPES valence band spectrum was also collected at beamline P22 at PETRA III/DESY to confirm the reproducibility under varying beamline conditions.~\cite{Schlueter2019} The I09 and P22 spectra are identical and information on the experimental setup at P22 as well as a comparative plot of the data can be found in Figure~S19 in the Supporting Information. The $\beta$-\ce{Ga2O3} core level reference spectra were collected on a (010) oriented bulk single crystal obtained from Novel Crystal Technology Inc., Tamura Corporation. The semicore and core spectra for $\gamma$- and $\beta$-\ce{Ga2O3} shown in Figures~\ref{fig:semicore} and \ref{fig:core} are all aligned to the O~1$s$ core level of the SXPS experiments. Beamline I09 was also used to collect X-ray absorption spectroscopy (XAS) of the O K-edge in total electron yield mode (TEY) and details on the alignment procedure can be found in a previous publication.~\cite{Swallow2020}

\subsubsection{Spectroscopic Ellipsometry}
Spectroscopic ellipsometry was performed using a scanning variable-angle spectroscopic ellipsometer based on a grating monochromator operational in the spectral range between 0.5 and 6.5~eV. The instrument is equipped with an autoretarder. The $\gamma$-\ce{Ga2O3} film was measured using three different angles of incidence $\Phi$ of 50$^\circ$, 60$^\circ$, and 70$^\circ$. The recorded experimental ellipsometric parameters $\Psi$ and $\Delta$ were analyzed using a multi-layer model to find the complex dielectric function as a function of photon energy $\varepsilon(\hbar \omega)$. For the present case the model consists of the substrate \ce{MgAl2O4}, the $\gamma$-\ce{Ga2O3} layer of interest, and an effective medium approximated surface roughness layer using Bruggeman's formalism.~\cite{Bruggeman1935} Layer thicknesses were found to be 122~nm ($\gamma$-\ce{Ga2O3}) and 3.3~nm (effective medium layer). The dielectric function of \ce{MgAl2O4} has already been determined by Zollner~\textit{et al.}~\cite{Zollner2014} We compared and slightly modified their results to an identical (001) \ce{MgAl2O4} crystal as used as substrate. For the description of the $\gamma$-\ce{Ga2O3} layer, a model independent so-called point-by-point fitted dielectric function was obtained by fitting the calculated optical response of the multi-layer model to the experimental ellipsometric parameters while varying the real and imaginary parts of $\varepsilon (\hbar \omega)=\varepsilon_1 (\hbar \omega)+i \varepsilon_2(\hbar \omega)$ until best agreement was obtained. The cubic crystal structure of $\gamma$-\ce{Ga2O3} has an isotropic optical response, this means the dielectric function is a scalar for this material.

\subsubsection{Photoluminescence Excitation Spectroscopy }
Photoluminescence excitation spectroscopy (PLE) was performed using a 400~W Xe arc lamp, which was monochromatised by a two-stage spectrometer (Acton SP250, $\mathrm{f=250}$~mm, gratings: 1800~l/mm) yielding a spectral FWHM of the excitation beam of around 1~nm. The samples were placed in a He-flow micro-cryostat (Janis ST-500) enabling temperature dependent measurements between 5 and 300~K. The optical excitation and detection of the emitted light was performed in backscattering geometry using a UV fused silica beamsplitter and focusing lens ($\mathrm{NA=0.69}$). The emitted light was dispersed in a single-stage monochromator (Acton SP300, $\mathrm{300}$~mm, grating: 300~l/mm) and detected by a charge-coupled device (Horiba Syncerity). Reference measurements of the \ce{MgAl2O4} substrate were conducted alongside the PLE measurements of the $\gamma$-\ce{Ga2O3} thin film under equal conditions to correct for the non-negligible substrate signal in overlapping spectral regions. The PLE spectra were corrected to account for the spectral power density of the lamp and transmission losses throughout the optical setup by in-situ monitoring of the excitation light using a UV-optimized high sensitivity Si photodiode (Hamamatsu S4349). The excitation spectra were obtained by integrating over the full width of the detected luminescence signal and are proportional to the probability that an exciting photon generates an emitted photon in the observed emission wavelength range.

\medskip
\textbf{Supporting Information} \par 
Supporting Information is available from the Wiley Online Library or from the author.

\medskip
\textbf{Acknowledgements} \par 
L.E.R.\ acknowledges support from an EPSRC Early Career Research Fellowship (EP/P033253/1) and the Thomas Young Centre under grant number TYC-101. Calculations were performed on the Imperial College High Performance Computing Service and the ARCHER UK National Supercomputing Service.
We are also grateful to the UK Materials and Molecular Modelling Hub for computational resources, which is partially funded by EPSRC (EP/P020194/1 and EP/T022213/1).
Work by F.N., E.K., R.G., M.F., P.M., O.B., C.W., M.N., M.A., and M.R.W. was performed in the framework of GraFOx, a Leibniz-ScienceCampus partially funded by the Leibniz association. M. R. W. and B. M. J. acknowledge funding by the German Research Foundation DFG (project number 446185170).
J.E.N.S.\ acknowledges funding through the Engineering and Physical Sciences  Research  Council  (EPSRC)  Centre  for Doctoral Training in New and Sustainable Photovoltaics (EP/L01551X/1). C.K.\ acknowledges the support from the Department of Chemistry at University College London. A.R.\ acknowledges the support from the Analytical Chemistry Trust Fund for her CAMS-UK Fellowship. 
We acknowledge Diamond Light Source for time on Beamline I09 under Proposals SI21430-1 and SI24670-1. The authors would like to thank Dave McCue, I09 beamline technician, for his support of the experiments. We acknowledge DESY (Hamburg, Germany), a member of the Helmholtz Association HGF, for the provision of experimental facilities. Parts of this research were carried out at PETRA III, beamline P22. Beamtime was allocated for proposal H-20010087.
The work of J.B.V.\ was performed under the auspices of the U.S. DOE by Lawrence Livermore National Laboratory under contract DE-AC52-07NA27344. Figures~\ref{fig:structure}(a) and~\ref{fig:bonds}(a) were prepared using the VESTA software package.~\cite{Momma2011}

\medskip
\textbf{Data Availability} \par
Jupyter notebooks and python code used to fit the models, an example notebook demonstrating the structure generation and application of the model, and associated data, including both relaxed and unrelaxed structures and the BigDFT input file used for single point calculations, are available at \url{https://gitlab.com/lratcliff/gamma-ga2o3}.

\medskip

%
\bibliographystyle{MSP}
\bibliography{ga2o3}

\begin{thebibliography}{10}
\providecommand{\url}[1]{\texttt{#1}}
\providecommand{\urlprefix}{URL }

\bibitem{Pearton2018}
S.~J. Pearton, J.~Yang, P.~H. Cary, F.~Ren, J.~Kim, M.~J. Tadjer, M.~A. Mastro,
\newblock \emph{Appl. Phys. Rev.} \textbf{2018}, \emph{5}, 1 011301.

\bibitem{Shi2021}
J.~Shi, J.~Zhang, L.~Yang, M.~Qu, D.-C. Qi, K.~H.~L. Zhang,
\newblock \emph{Advanced Materials} \textbf{2021}, \emph{33}, 50 2006230.

\bibitem{Guo2019}
D.~Guo, Q.~Guo, Z.~Chen, Z.~Wu, P.~Li, W.~Tang,
\newblock \emph{Mater. Today Phys.} \textbf{2019}, \emph{11} 100157.

\bibitem{Teng2014}
Y.~Teng, L.~X. Song, A.~Ponchel, Z.~K. Yang, J.~Xia,
\newblock \emph{Advanced Materials} \textbf{2014}, \emph{26}, 36 6238.

\bibitem{Hou2021}
X.~Hou, Y.~Zou, M.~Ding, Y.~Qin, Z.~Zhang, H.~Sun, S.~Long,
\newblock \emph{J. Phys. D: Appl. Phys.} \textbf{2021}, \emph{54} 043001.

\bibitem{Akatsuka2020}
M.~Akatsuka, Y.~Kawaguchi, R.~Itoh, A.~Ozawa, M.~Yamamoto, T.~Tanabe,
  T.~Yoshida,
\newblock \emph{Appl. Catal. B} \textbf{2020}, \emph{262}, October 2019 118247.

\bibitem{Masataka2016}
H.~Masataka, S.~Kohei, M.~Hisashi, K.~Yoshinao, K.~Akinori, K.~Akito,
  M.~Takekazu, Y.~Shigenobu,
\newblock \emph{Semicond. Sci. Technol.} \textbf{2016}, \emph{31}, 3 34001.

\bibitem{Xue2018}
H.~W. Xue, Q.~M. He, G.~Z. Jian, S.~B. Long, T.~Pang, M.~Liu,
\newblock \emph{Nanoscale Res. Lett.} \textbf{2018}, \emph{13} 1.

\bibitem{Playford2013}
H.~Y. Playford, A.~C. Hannon, E.~R. Barney, R.~I. Walton,
\newblock \emph{Chem. Eur. J.} \textbf{2013}, \emph{19}, 8 2803.

\bibitem{Boehm1940}
J.~B\"{o}hm,
\newblock \emph{Angew. Chem.} \textbf{1940}, \emph{53}, 11-12 131.

\bibitem{Roy1952}
R.~Roy, V.~G. Hill, E.~Osborn,
\newblock \emph{J. Am. Chem. Soc.} \textbf{1952}, \emph{74}, 3 719.

\bibitem{Pohl1968}
K.~Pohl,
\newblock \emph{Naturwissenschaften} \textbf{1968}, \emph{55} 82.

\bibitem{Playford2014}
H.~Y. Playford, A.~C. Hannon, M.~G. Tucker, D.~M. Dawson, S.~E. Ashbrook, R.~J.
  Kastiban, J.~Sloan, R.~I. Walton,
\newblock \emph{J. Phys. Chem. C} \textbf{2014}, \emph{118}, 29 16188.

\bibitem{Hohenberg1964}
P.~Hohenberg, W.~Kohn,
\newblock \emph{Phys. Rev.} \textbf{1964}, \emph{136}, 3B B864.

\bibitem{Kohn1965}
W.~Kohn, L.~J. Sham,
\newblock \emph{Phys. Rev.} \textbf{1965}, \emph{140}, 4A A1133.

\bibitem{Yoshioka2007}
S.~Yoshioka, H.~Hayashi, A.~Kuwabara, F.~Oba, K.~Matsunaga, I.~Tanaka,
\newblock \emph{J. Phys.: Condens. Matter} \textbf{2007}, \emph{19} 346211.

\bibitem{Hayashi2012}
H.~Hayashi, R.~Huang, F.~Oba, T.~Hirayama, I.~Tanaka,
\newblock \emph{Appl. Phys. Lett.} \textbf{2012}, \emph{101}, 24 241906.

\bibitem{Gutierrez2001}
G.~Guti\'errez, A.~Taga, B.~Johansson,
\newblock \emph{Phys. Rev. B} \textbf{2001}, \emph{65} 012101.

\bibitem{Pinto2004}
H.~P. Pinto, R.~M. Nieminen, S.~D. Elliott,
\newblock \emph{Phys. Rev. B} \textbf{2004}, \emph{70} 125402.

\bibitem{Taniike2006}
T.~Taniike, M.~Tada, Y.~Morikawa, T.~Sasaki, Y.~Iwasawa,
\newblock \emph{J. Phys. Chem. B} \textbf{2006}, \emph{110}, 10 4929.

\bibitem{Wang2020}
X.~Wang, M.~Faizan, G.~Na, X.~He, Y.~H. Fu, L.~Zhang,
\newblock \emph{Adv. Electron. Mater.} \textbf{2020}, \emph{6}, 6 2000119.

\bibitem{Meng2020}
R.~Meng, M.~Houssa, K.~Iordanidou, G.~Pourtois, V.~Afanasiev, A.~Stesmans,
\newblock \emph{J. Appl. Phys.} \textbf{2020}, \emph{128}, 3 034304.

\bibitem{Paglia2005}
G.~Paglia, A.~L. Rohl, C.~E. Buckley, J.~D. Gale,
\newblock \emph{Phys. Rev. B} \textbf{2005}, \emph{71} 224115.

\bibitem{Swallow2020}
J.~E.~N. Swallow, C.~Vorwerk, P.~Mazzolini, P.~Vogt, O.~Bierwagen, A.~Karg,
  M.~Eickhoff, J.~Sch\"{o}rmann, M.~R. Wagner, J.~W. Roberts, P.~R. Chalker,
  M.~J. Smiles, P.~Murgatroyd, S.~A. Razek, Z.~W. Lebens-Higgins, L.~F.~J.
  Piper, L.~A.~H. Jones, P.~K. Thakur, T.-L. Lee, J.~B. Varley,
  J.~Furthm\"{u}ller, C.~Draxl, T.~D. Veal, A.~Regoutz,
\newblock \emph{Chem. Mater.} \textbf{2020}, \emph{32}, 19 8460.

\bibitem{Varley2010}
J.~B. Varley, J.~R. Weber, A.~Janotti, C.~G. Van~de Walle,
\newblock \emph{Applied Physics Letters} \textbf{2010}, \emph{97}, 14 142106.

\bibitem{Sturm2015}
C.~Sturm, J.~Furthm\"uller, F.~Bechstedt, R.~Schmidt-Grund, M.~Grundmann,
\newblock \emph{APL Mater.} \textbf{2015}, \emph{3} 106106.

\bibitem{unpublished}
E.~Kluth, P.~Ning, J.~Gr\"umbel, A.~Karg, M.~Eickhoff, R.~Goldhahn,
  M.~Feneberg,
\newblock -,
\newblock Unpublished.

\bibitem{Feneberg2018}
M.~Feneberg, J.~Nixdorf, M.~Neumann, N.~Esser, L.~Art\'us, R.~Cusc\'o,
  T.~Yamaguchi, R.~Goldhahn,
\newblock \emph{Phys. Rev. Mater.} \textbf{2018}, \emph{2} 044601.

\bibitem{ODonnell1991}
K.~O'Donnell, X.~Chen,
\newblock \emph{Appl. Phys. Lett.} \textbf{1991}, \emph{58} 2924.

\bibitem{Paessler1997}
R.~P\"assler,
\newblock \emph{Phys. Status Solidi B} \textbf{1997}, \emph{200} 155.

\bibitem{Furthmuller.2016.10.1103/physrevb.93.115204}
J.~Furthmüller, F.~Bechstedt,
\newblock \emph{Phys. Rev. B} \textbf{2016}, \emph{93}, 11 115204.

\bibitem{Bechstedt.2019.10.1063/1.5084324}
F.~Bechstedt, J.~Furthmüller,
\newblock \emph{Appl. Phys. Lett.} \textbf{2019}, \emph{114}, 12 122101.

\bibitem{Kim_Kang_Park_2012}
M.~Kim, I.~J. Kang, C.~H. Park,
\newblock \emph{Curr. Appl. Phys.} \textbf{2012}, \emph{12} S25–S28.

\bibitem{Perdew1996}
J.~P. Perdew, K.~Burke, M.~Ernzerhof,
\newblock \emph{Phys. Rev. Lett.} \textbf{1996}, \emph{77}, 18 3865.

\bibitem{Ratcliff2020}
L.~E. Ratcliff, W.~Dawson, G.~Fisicaro, D.~Caliste, S.~Mohr, A.~Degomme,
  B.~Videau, V.~Cristiglio, M.~Stella, M.~D'Alessandro, S.~Goedecker,
  T.~Nakajima, T.~Deutsch, L.~Genovese,
\newblock \emph{J. Chem. Phys.} \textbf{2020}, \emph{152}, 19 194110.

\bibitem{Goedecker1996}
S.~Goedecker, M.~Teter, J.~Hutter,
\newblock \emph{Phys. Rev. B} \textbf{1996}, \emph{54}, 3 1703.

\bibitem{Hartwigsen1998}
C.~Hartwigsen, S.~Goedecker, J.~Hutter,
\newblock \emph{Phys. Rev. B} \textbf{1998}, \emph{58} 3641.

\bibitem{Willand2013}
A.~Willand, Y.~O. Kvashnin, L.~Genovese, {\'A}.~V{\'a}zquez-Mayagoitia, A.~K.
  Deb, A.~Sadeghi, T.~Deutsch, S.~Goedecker,
\newblock \emph{J. Chem. Phys.} \textbf{2013}, \emph{138}, 10 104109.

\bibitem{Clark2005}
S.~J. Clark, M.~D. Segall, C.~J. Pickard, P.~J. Hasnip, M.~I. Probert,
  K.~Refson, M.~C. Payne,
\newblock \emph{Z. Kristall.} \textbf{2005}, \emph{220}, 5-6 567.

\bibitem{Monkhorst1976}
H.~J. Monkhorst, J.~D. Pack,
\newblock \emph{Phys. Rev. B} \textbf{1976}, \emph{13} 5188.

\bibitem{Morris2014}
A.~J. Morris, R.~J. Nicholls, C.~J. Pickard, J.~R. Yates,
\newblock \emph{Comput. Phys. Commun.} \textbf{2014}, \emph{185}, 5 1477.

\bibitem{Scofield1973}
J.~H. Scofield,
\newblock {Theoretical Photoionization Cross Sections from 1 to 1500 keV},
\newblock Technical report, U.S. Atomic Energy Commission, Divison of Technical
  Information Extension, \textbf{1973}.

\bibitem{Dig_Sco2020}
C.~Kalha, N.~K. Fernando, A.~Regoutz,
\newblock {Digitisation of Scofield Photoionisation Cross Section Tabulated
  Data}, \textbf{2020}.

\bibitem{Jackson2018}
A.~J. Jackson, A.~M. Ganose, A.~Regoutz, R.~G. Egdell, D.~O. Scanlon,
\newblock \emph{J. Open Source Softw.} \textbf{2018}, \emph{3}, 26 773.

\bibitem{Ahman1996}
J.~{\AA}hman, G.~Svensson, J.~Albertsson,
\newblock \emph{Acta Cryst. C} \textbf{1996}, \emph{52}, 6 1336.

\bibitem{Musil2021}
F.~Musil, A.~Grisafi, A.~P. Bart\'{o}k, C.~Ortner, G.~Cs\'{a}nyi, M.~Ceriotti,
\newblock \emph{Chem. Rev.} \textbf{2021}, \emph{121}, 16 9759.

\bibitem{Keith2021}
J.~A. Keith, V.~Vassilev-Galindo, B.~Cheng, S.~Chmiela, M.~Gastegger, K.-R.
  M\"{u}ller, A.~Tkatchenko,
\newblock \emph{Chem. Rev.} \textbf{2021}, \emph{121}, 16 9816.

\bibitem{Pedregosa2011}
F.~Pedregosa, G.~Varoquaux, A.~Gramfort, V.~Michel, B.~Thirion, O.~Grisel,
  M.~Blondel, P.~Prettenhofer, R.~Weiss, V.~Dubourg, J.~Vanderplas, A.~Passos,
  D.~Cournapeau, M.~Brucher, M.~Perrot, E.~Duchesnay,
\newblock \emph{J. Mach. Learn. Res.} \textbf{2011}, \emph{12} 2825.

\bibitem{Hoffmann2021}
G.~Hoffmann, Z.~Cheng, O.~Brandt, O.~Bierwagen,
\newblock \emph{APL Materials} \textbf{2021}, \emph{9}, 11 111110.

\bibitem{Oshima2017}
T.~Oshima, Y.~Kato, M.~Oda, T.~Hitora, M.~Kasu,
\newblock \emph{Appl. Phys. Express} \textbf{2017}, \emph{10}, 5 051104.

\bibitem{Oshima2019}
T.~Oshima, Y.~Kato, E.~Magome, E.~Kobayashi, K.~Takahashi,
\newblock \emph{Japanese Journal of Applied Physics} \textbf{2019}, \emph{58},
  6.

\bibitem{Kato2019}
Y.~Kato, M.~Imura, Y.~Nakayama, M.~Takeguchi, T.~Oshima,
\newblock \emph{Applied Physics Express} \textbf{2019}, \emph{12}, 6.

\bibitem{Lee2018}
T.-L. Lee, D.~A. Duncan,
\newblock \emph{Synchrotron Radiat. News} \textbf{2018}, \emph{31}, 4 16.

\bibitem{Schlueter2019}
C.~Schlueter, A.~Gloskovskii, K.~Ederer, I.~Schostak, S.~Piec, I.~Sarkar,
  Y.~Matveyev, P.~L\"omker, M.~Sing, R.~Claessen, C.~Wiemann, C.~M. Schneider,
  K.~Medjanik, G.~Sch\"onhense, P.~Amann, A.~Nilsson, W.~Drube,
\newblock \emph{AIP Conf. Proc.} \textbf{2019}, \emph{2054} 040010.

\bibitem{Bruggeman1935}
D.~A.~G. Bruggeman,
\newblock \emph{Ann. Phys.} \textbf{1935}, \emph{24} 636.

\bibitem{Zollner2014}
C.~J. Zollner, T.~I. Willett-Gies, S.~Zollner, S.~Cho,
\newblock \emph{Thin Solid Films} \textbf{2014}, \emph{571} 689.

\bibitem{Momma2011}
K.~Momma, F.~Izumi,
\newblock \emph{J. Appl. Crystallogr.} \textbf{2011}, \emph{44}, 6 1272.

\end{thebibliography}



\newpage

\begin{figure}
\textbf{Table of Contents}\\
\centering
\medskip
  \includegraphics{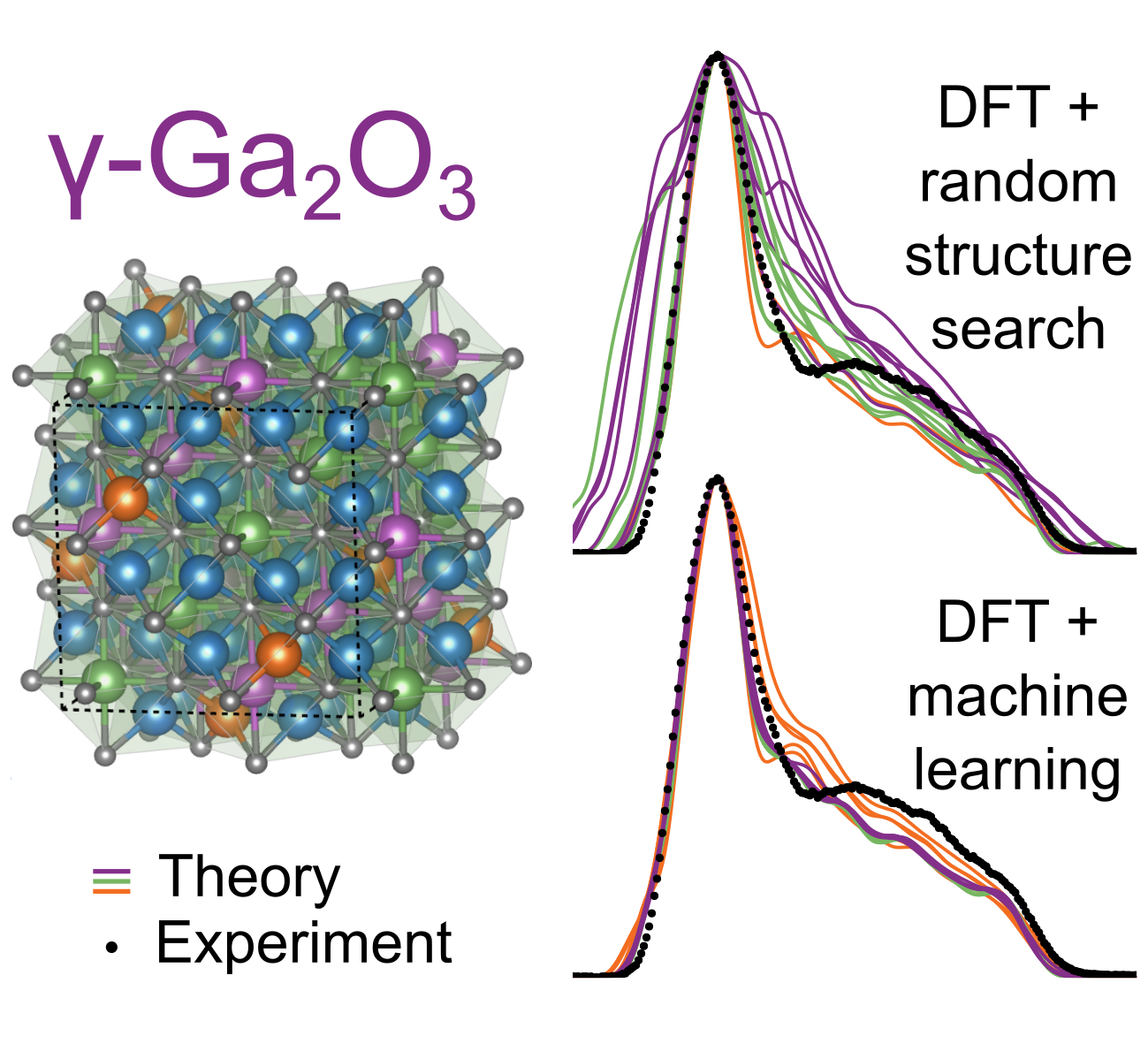}
  \medskip
  \caption*{Gallium oxide (\ce{Ga2O3}) is attracting increasing attention in applications such as power electronics and photodetectors due to its rich structural space and resulting electronic structure. This work explores the influence of structural disorder on a range of crucial material parameters of $\gamma$ phase \ce{Ga2O3} by combining density functional theory and machine learning with a range of experimental probes.}
\end{figure}

\end{document}


\title{{\LARGE Supporting Information}\\
\LARGE Tackling disorder in $\gamma$-\ce{Ga2O3}}
\date{\today}

\author{Laura~E.~Ratcliff}
 \email{laura.ratcliff@bristol.ac.uk}
\affiliation{ 
Department of Materials, Imperial College London, London SW7 2AZ, United Kingdom.}
\affiliation{Centre for Computational Chemistry,
School of Chemistry, University of Bristol, Bristol BS8 1TS, United Kingdom.}

\author{Takayoshi~Oshima}
\affiliation{ 
Department of Electrical and Electronic Engineering, Saga University, Saga 840-8502, Japan.}

\author{Felix~Nippert}%
\affiliation{%
Technische Universit\"{a}t Berlin, Institute of Solid State Physics, Hardenbergstrasse 36, 10623 Berlin, Germany.}

\author{Benjamin~M.~Janzen}
\author{Elias~Kluth}
\author{R\"udiger~Goldhahn}
\author{Martin~Feneberg}
\affiliation{Institut f\"ur Physik, Otto-von-Guericke-Universit\"at Magdeburg, Universit\"atsplatz 2, 39106 Magdeburg, Germany.}

\author{Piero~Mazzolini}
\affiliation{%
Paul-Drude-Institut fur Festk\"{o}rperelektronik, Leibniz-Institut im Forschungsverbund Berlin e.V., Hausvogteiplatz 5-7, 10117 Berlin, Germany.}
\affiliation{Present address: Department of Mathematical, Physical and Computer Sciences, University of Parma, Viale delle Scienze 7/A, 43124 Parma, Italy.}

\author{Oliver~Bierwagen}%
\affiliation{%
Paul-Drude-Institut f\"ur Festk\"{o}rperelektronik, Leibniz-Institut im Forschungsverbund Berlin e.V., Hausvogteiplatz 5-7, 10117 Berlin, Germany.}

\author{Charlotte~Wouters}
\author{Musbah~Nofal}
\author{Martin Albrecht}
\affiliation{Leibniz-Institut für Kristallz\"{u}chtung, Max-Born-Str. 2, 12489 Berlin, Germany}

\author{Jack~E.~N.~Swallow}
 \affiliation{Department of Materials, University of Oxford, Parks Road, Oxford OX1 3PH, United Kingdom.}

\author{Leanne~A.~H.~Jones}%
\affiliation{%
Stephenson Institute for Renewable Energy and Department of Physics, University of Liverpool, Liverpool L69 7ZF, United Kingdom.}

\author{Pardeep~K.~Thakur}%
\affiliation{%
Diamond Light Source Ltd., Diamond House, Harwell Science and Innovation Campus, Didcot OX11 0DE, United Kingdom.}%

\author{Tien-Lin~Lee}%
\affiliation{%
Diamond Light Source Ltd., Harwell Science and Innovation Campus, Didcot OX11 0DE, United Kingdom.}%

\author{Curran~Kalha}%
\affiliation{%
Department of Chemistry, University College London, 20 Gordon Street, London WC1H 0AJ, United Kingdom.}%

\author{Christoph~Schlueter}%
\affiliation{%
Deutsches Elektronen-Synchrotron DESY, Notkestrasse 85, 22607 Hamburg, Germany.}%

\author{Tim~D.~Veal}%
\affiliation{%
Stephenson Institute for Renewable Energy and Department of Physics, University of Liverpool, Liverpool L69 7ZF, United Kingdom.}%

\author{Joel~B.~Varley}%
\affiliation{%
Lawrence Livermore National Laboratory, Livermore, CA 94550, United States of America.}%

\author{Markus~R.~Wagner}%
\affiliation{%
Technische Universit\"{a}t Berlin, Institute of Solid State Physics, Hardenbergstrasse 36, 10623 Berlin, Germany.}

\author{Anna~Regoutz}%
 \email{a.regoutz@ucl.ac.uk}
\affiliation{%
Department of Chemistry, University College London, 20 Gordon Street, London WC1H 0AJ, United Kingdom.}%

\maketitle

\clearpage

\begin{table*}[!t]
\centering
\begin{threeparttable}
\caption{Number of structures, $N$, for which a given number of Ga sites are occupied, total number of structures, and \% which are below $E_{\mathrm{ref}}+\Delta E$ in eV/atom for each data set, where $E_{\mathrm{ref}}$ is the lowest DFT energy out of the first 839 structures. The datasets comprise the initial 800 randomly generated structures, those containing rare environments, those predicted to be less than 0.02~eV/atom above the reference energy using the initial model (`ML1'), those predicted to be less than 0.02~eV/atom above the reference energy using the updated model (`ML2') and those predicted to be less than 0.03~eV/atom above the reference energy using the updated model (`ML1') with only 2 or 4 distinct Ga sites occupied. }
\label{tab:structures}
\begin{tabular*} {0.95\textwidth}{l @{\extracolsep{\fill}} rrrrr}
\hline \hline
 & \multicolumn{2}{c}{No ML}  & \multicolumn{1}{c}{ML1} & \multicolumn{2}{c}{ML2} \\
\cline{2-3} \cline{4-4}\cline{5-6}\\[-2.5ex]
 & \multirow{2}{*}{No Screening} & \multirow{2}{*}{Rare Environments} & \multirow{2}{*}{$E_{\mathrm{ML1}}-E_{\mathrm{ref}}<0.02$} & \multirow{2}{*}{$E_{\mathrm{ML2}}-E_{\mathrm{ref}}<0.02$} & $E_{\mathrm{ML2}}-E_{\mathrm{ref}}<0.03$ \\
&  &  &  &  & and 2 or 4-site\\
\cline{1-1}\cline{2-2} \cline{3-3} \cline{4-4}\cline{5-5}\cline{6-6}\\[-2.5ex]
$N$\\
2 sites & 99  & 1  & 0  & 0   & 26\\
3 sites & 396 & 11 & 43 & 170 & 0 \\
4 sites & 305 & 27 & 2  & 1   & 142\\
Total   & 800 & 39 & 45 & 171 & 168\\
\cline{1-1}\cline{2-2} \cline{3-3} \cline{4-4}\cline{5-5}\cline{6-6}\\[-2.5ex]
$\Delta E$\\
0.02 &  0.8 & 0.0 & 68.9 & 78.9 & 33.3\\
0.03 &  2.0 & 0.0 & 77.8 & 98.8 & 78.0\\ 
 \hline \hline
\end{tabular*}
\end{threeparttable}
\end{table*}

\clearpage

\begin{table*}[!t]
\centering
\begin{threeparttable}
\caption{Information related to the final 30 relaxed structures, which were selected either due to being low in energy, or randomly. Listed are the energies, $\Delta E$, relative to the lowest energy unrelaxed structure, calculated using PBE with BigDFT, the band gaps of the relaxed structures, $E_{\mathrm{g}}$, calculated using HSE with VASP, and the occupancies of the four Ga sites after relaxation.  Where different from the relaxed structure, the Ga site occupancies before relaxation are also given in brackets.}
\label{tab:final_structures}
\begin{tabular*} {0.95\textwidth}{l @{\extracolsep{\fill}} rr r llll}
\hline \hline
 & \multicolumn{2}{c}{$\Delta E$ (meV/atom)} & \multicolumn{1}{c}{$E_{\mathrm{g}}$ (eV)} & \multicolumn{4}{c}{Occupancies}\\
\cline{2-3} \cline{4-4}\cline{5-8}\\[-2.5ex]
 & Unrelaxed & Relaxed & Relaxed & Ga1 & Ga2 & Ga4 & Ga4\\
\cline{1-1}\cline{2-2} \cline{3-3} \cline{4-4}\cline{5-5}\cline{6-6}\cline{7-7}\cline{8-8}\\[-2.5ex]
\textbf{Low Energy}\\
\textbf{2-site}\\
IIA &   14.6 &   4.7 &  3.94  &  1.000 &  0.833 &  0.000 &  0.000 \\
IIB &   16.1 &   6.4 &  4.02  &  1.000 &  0.833 &  0.000 &  0.000 \\
IIC &   17.5 &   6.5 &  3.41  &  1.000 &  0.833 &  0.000 &  0.000 \\
IID &   10.8 &   8.4 &  3.58  &  1.000 &  0.833 &  0.000 &  0.000 \\
IIE &   17.7 &  14.4 &  3.54  &  1.000 &  0.833 &  0.000 &  0.000 \\
\textbf{3-site}\\
IIIA &    0.9 &  -2.0 &  4.69 &  0.833 &  0.833 &  0.028 &  0.000 \\
IIIB &    0.4 &  -0.7 &  4.66 &  0.833 &  0.833 &  0.028 &  0.000 \\
IIIC &    0.0 &   0.0 &  4.60 &  0.833 &  0.833 &  0.028 &  0.000 \\
IIID &    0.8 &   0.4 &  4.18 &  0.833 &  0.833 &  0.028 &  0.000 \\
IIIE &    0.8 &   3.1 &  4.59 &  0.833 &  0.833 &  0.028 &  0.000 \\
\textbf{4-site}\\
IVA &    8.8 &  -1.2 &  4.23 &  0.792 &  0.833 &  0.028 &  0.021 \\
IVB &   11.8 &   0.0 &  3.89 &  0.833 &  0.792 &  0.035 &  0.021 \\
IVC &    9.0 &   1.3 &  4.24 &  0.792 &  0.833 &  0.028 &  0.021 \\
IVD &   12.4 &   2.9 &  4.15 &  0.792 &  0.833 &  0.028 &  0.021 \\
IVE &   12.6 &   3.1 &  4.20 &  0.792 &  0.833 &  0.028 &  0.021 \\
\hline
\textbf{Random}\\
\textbf{2-site}\\
IIa &   20.1 &  10.1 &  4.04  &  0.917 &  0.875 &  0.000 &  0.000 \\
IIb &   27.1 &  14.4 &  3.80  &  1.000 &  0.833 &  0.000 &  0.000 \\
\textbf{3-site}\\
IIIa &    2.3 &   0.8 &  4.65  &  0.833 &  0.833 &  0.028 &  0.000 \\
IIIb &   19.8 &   3.0 &  3.75  &  0.875 &  0.792 &  0.035 &  0.000 \\
IIIc &   36.7 &  10.1 &  3.40  &  0.875 (0.833) &  0.833 &  0.021 &  0.000 (0.021) \\
IIId &   22.7 &  10.3 &  3.42  &  0.875 &  0.833 &  0.021 &  0.000 \\
IIIe &   53.0 &  34.1 &  2.49  &  0.833 &  0.875 &  0.014 &  0.000 \\
IIIf &   91.4 &  50.8 &  2.77  &  0.750 &  0.917 (0.958) &  0.014 (0.000) &  0.000 \\
IIIg &  121.7 &  56.2 &  1.02  &  0.875 (0.917) &  0.875 &  0.000 &  0.021 (0.000) \\
\textbf{4-site}\\
IVa &   12.8 &   4.0 &  4.30  &  0.792 &  0.833 &  0.028 &  0.021 \\
IVb &   64.7 &  26.6 &  2.94  &  0.750 &  0.875 &  0.021 &  0.021 \\
IVc &  102.1 &  30.4 &  2.61  &  0.875 (0.917) &  0.792 (0.812) &  0.028 (0.021) &  0.021 (0.000) \\
IVd &   59.9 &  31.4 &  2.62  &  0.750 &  0.875 &  0.021 &  0.021 \\
IVe &  158.7 &  52.1 &  3.41  &  0.917 (1.000) &  0.833 &  0.007 (0.000) &  0.021 (0.000)  \\
IVf &  162.1 &  57.1 &  1.93  &  0.708 (0.750) &  0.833 (0.875) &  0.028 (0.021) &  0.062 (0.021) \\
 \hline \hline
\end{tabular*}
\end{threeparttable}
\end{table*}

\clearpage

\begin{figure*}[!ht]
\includegraphics[scale=0.33]{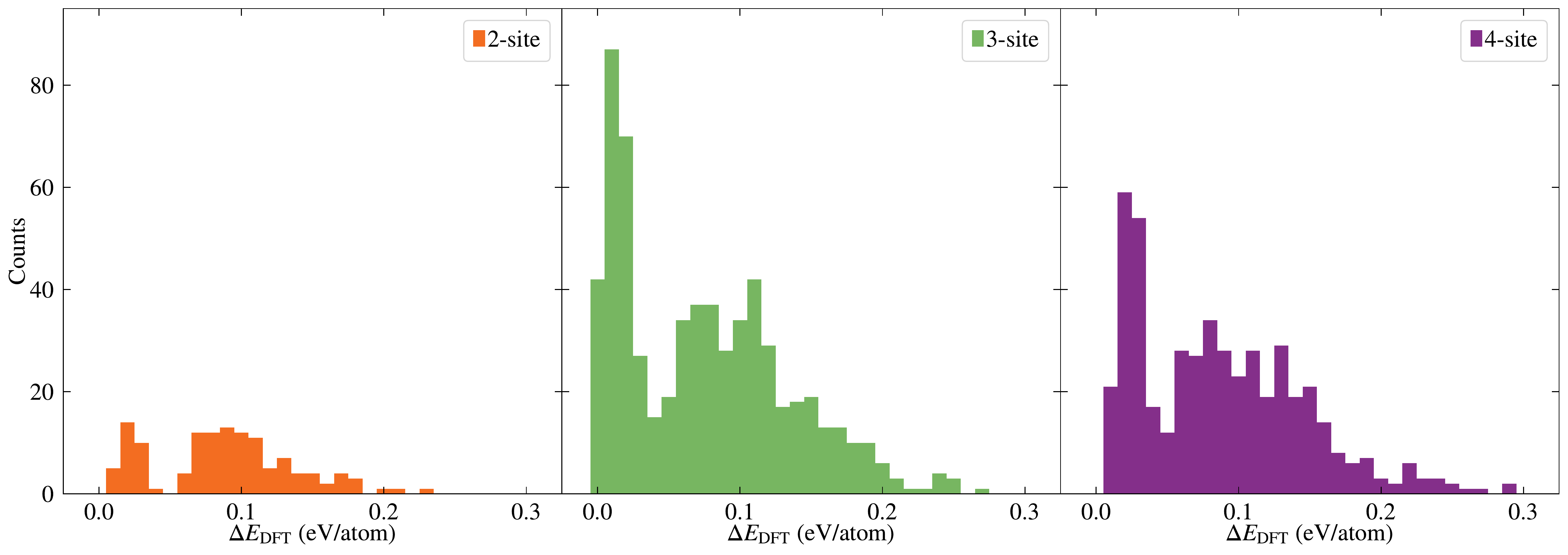}
\caption{Distribution of PBE-calculated energies for the unrelaxed structures.
\label{fig:histos}}
\end{figure*}

\begin{figure}[!ht]
\begin{subfigure}[t]{0.33\linewidth}
\includegraphics[scale=0.33]{{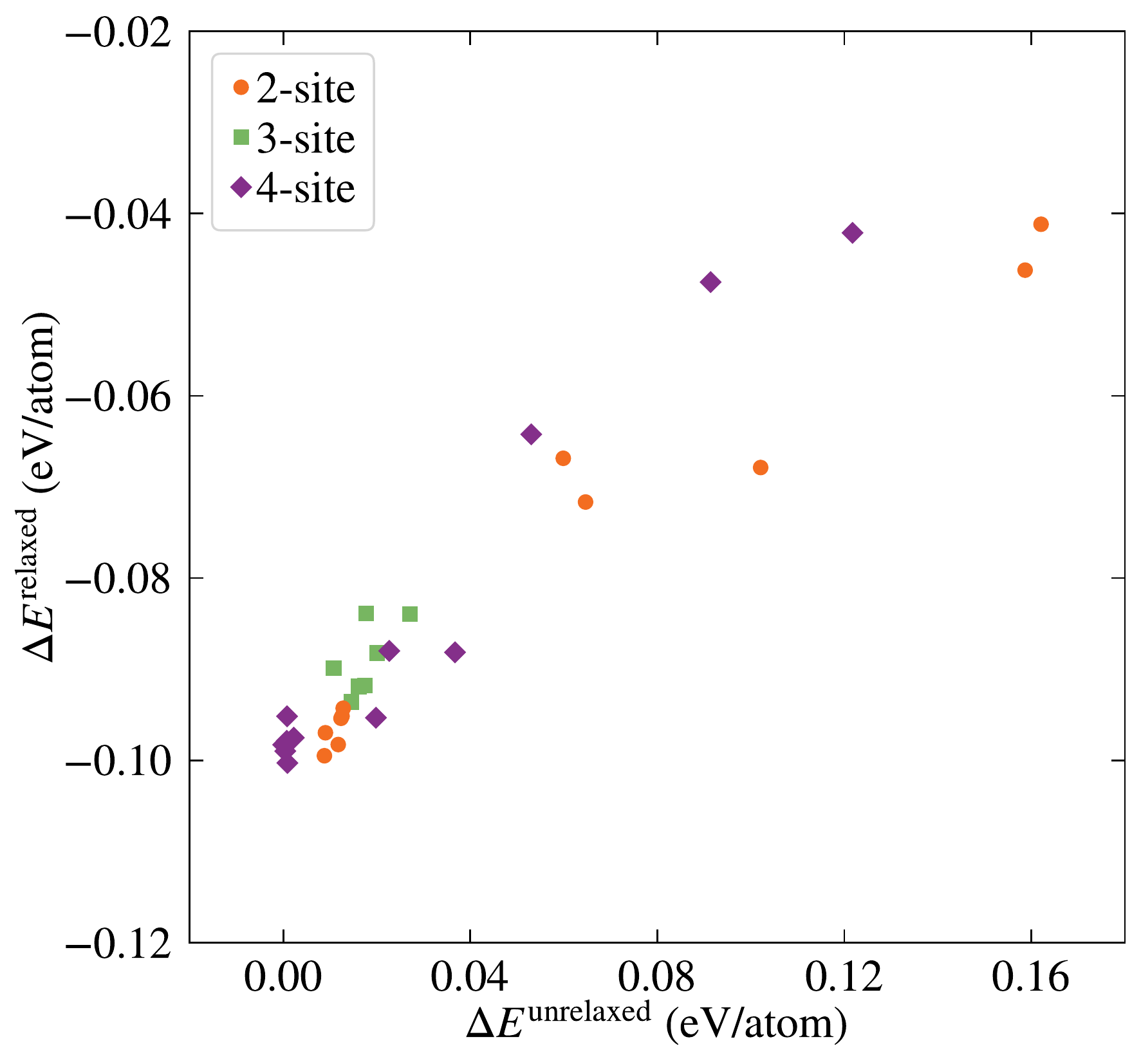}}
\caption{Relaxed and unrelaxed PBE energies, relative to the lowest unrelaxed energy.}
\label{fig:relax}
\end{subfigure}
\begin{subfigure}[t]{0.66\linewidth}
\includegraphics[scale=0.33]{{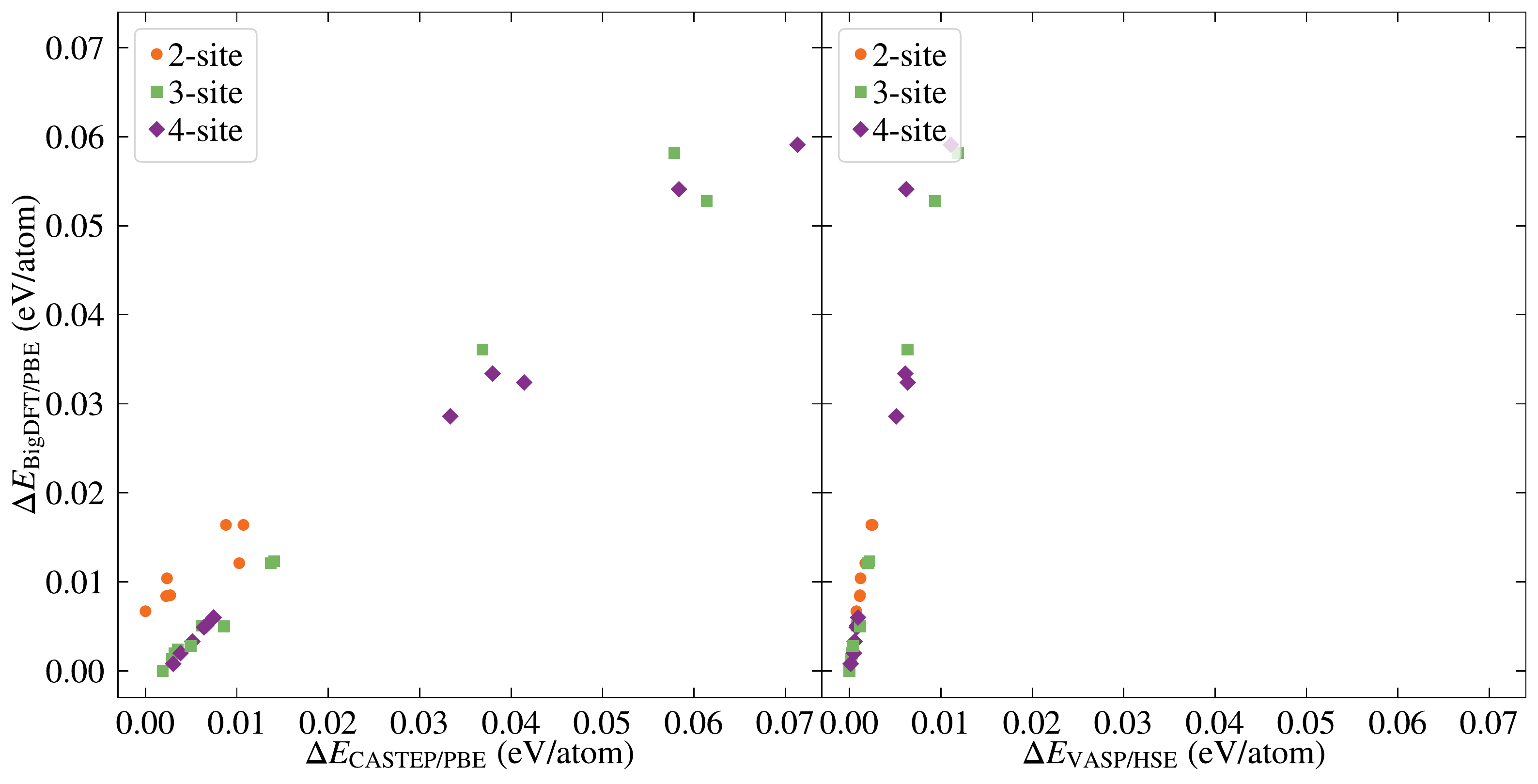}}
\caption{Comparison between relaxed BigDFT calculated PBE energies, and PBE (HSE) energies calculated with CASTEP (VASP).}
\label{fig:code_comp}
\end{subfigure}
\caption{Comparison between relative energies before and after relaxation, as well as across different codes and/or functionals.
\label{fig:energy_correlation}}
\end{figure}

\clearpage

\begin{figure*}[!ht]
\begin{subfigure}[t]{1.0\linewidth}
\includegraphics[scale=0.33]{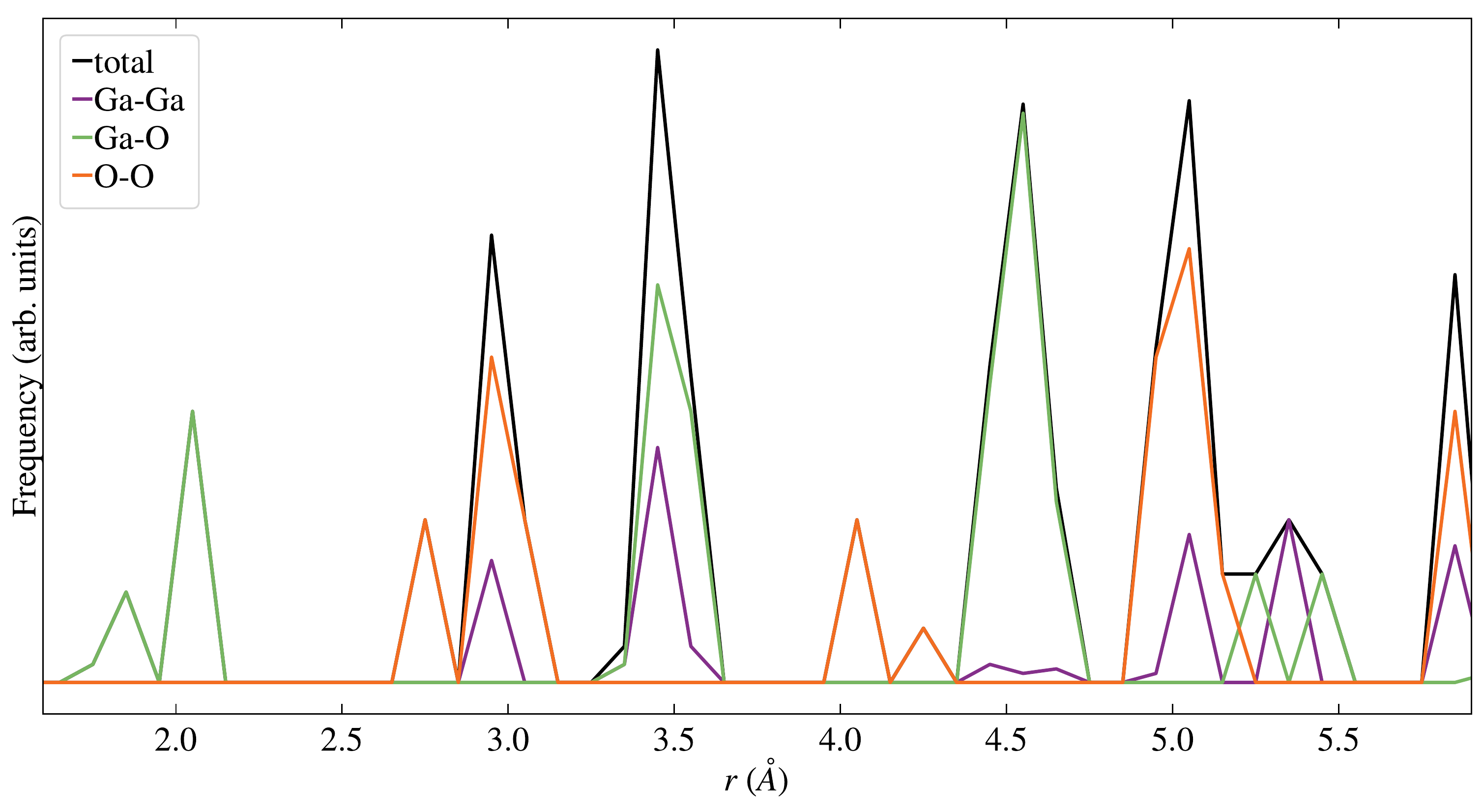}
\caption{Unrelaxed}
\label{fig:corr_1051_unrelaxed}
\end{subfigure}

\begin{subfigure}[t]{1.0\linewidth}
\includegraphics[scale=0.33]{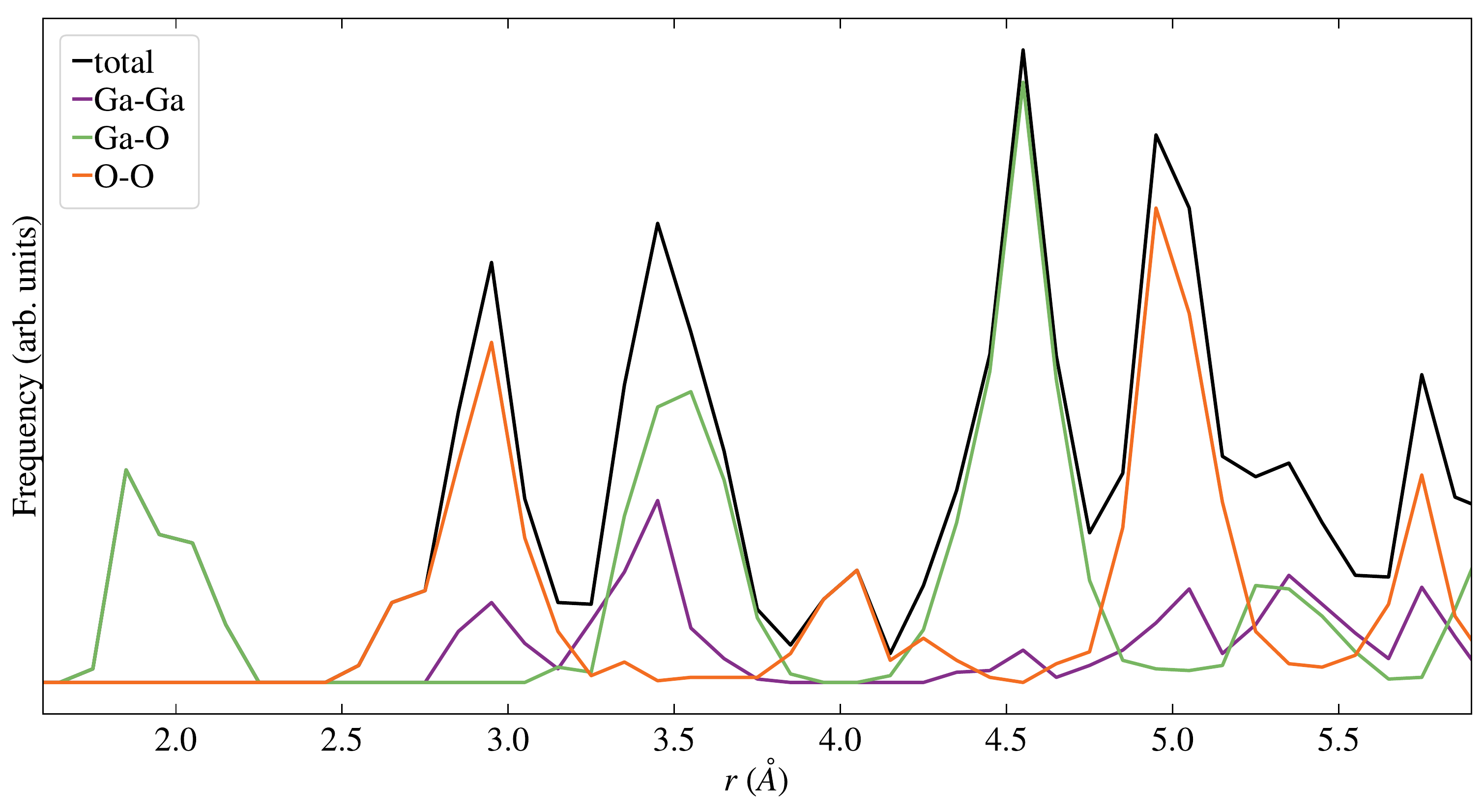}
\caption{Relaxed}
\label{fig:corr_1051_relaxed}
\end{subfigure}

\caption{Correlation function for the lowest energy structure before and after relaxation.
\label{fig:corr_1051}}
\end{figure*}

\begin{figure*}[!ht]
\includegraphics[scale=0.33]{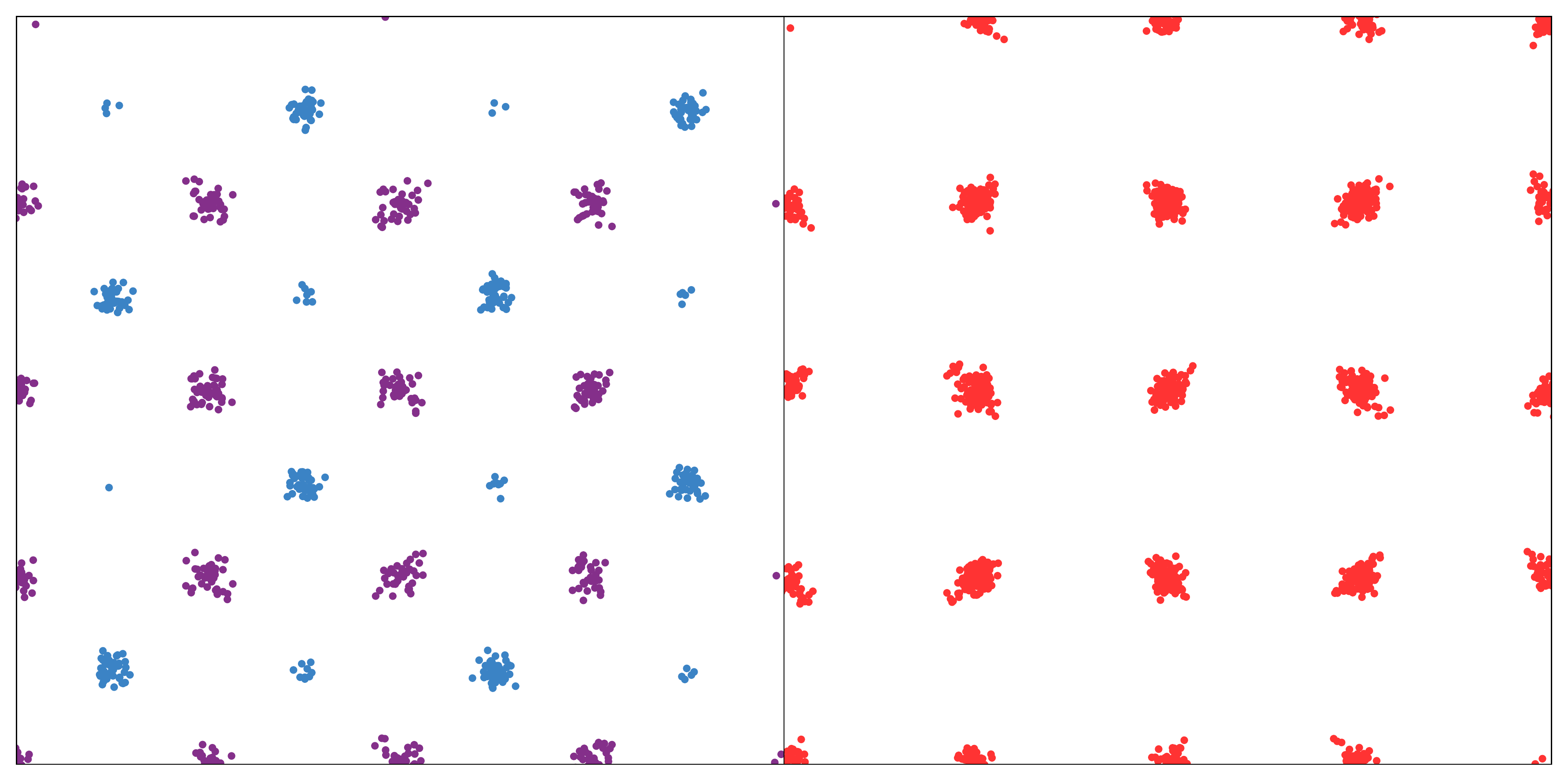}
\caption{Distribution of $T_d$ (blue), and $O_h$ (purple) Ga sites [left] and O sites (red) [right], for the relaxed low energy structures, i.e.\ those within 10~meV/atom of the lowest energy structure, as viewed along the $c$-axis.
\label{fig:projected_sites}}
\end{figure*}

\clearpage

\begin{figure*}[!ht]
\includegraphics[scale=0.33]{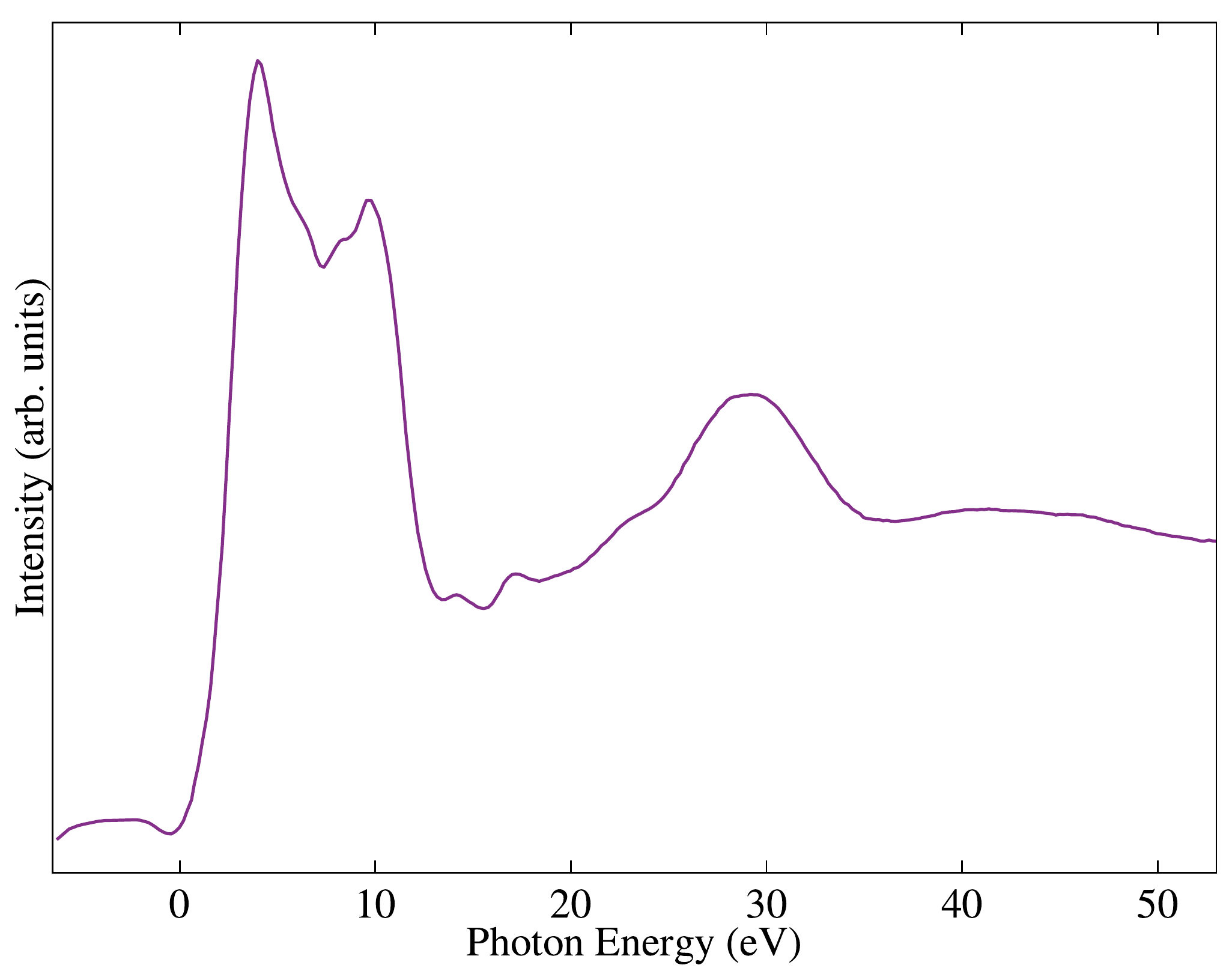}
\caption{O K-edge XAS spectrum of $\gamma$-\ce{Ga2O3}.}
\label{fig:xas_full}
\end{figure*}

\clearpage

\begin{figure*}[htb]
\includegraphics[scale=0.33]{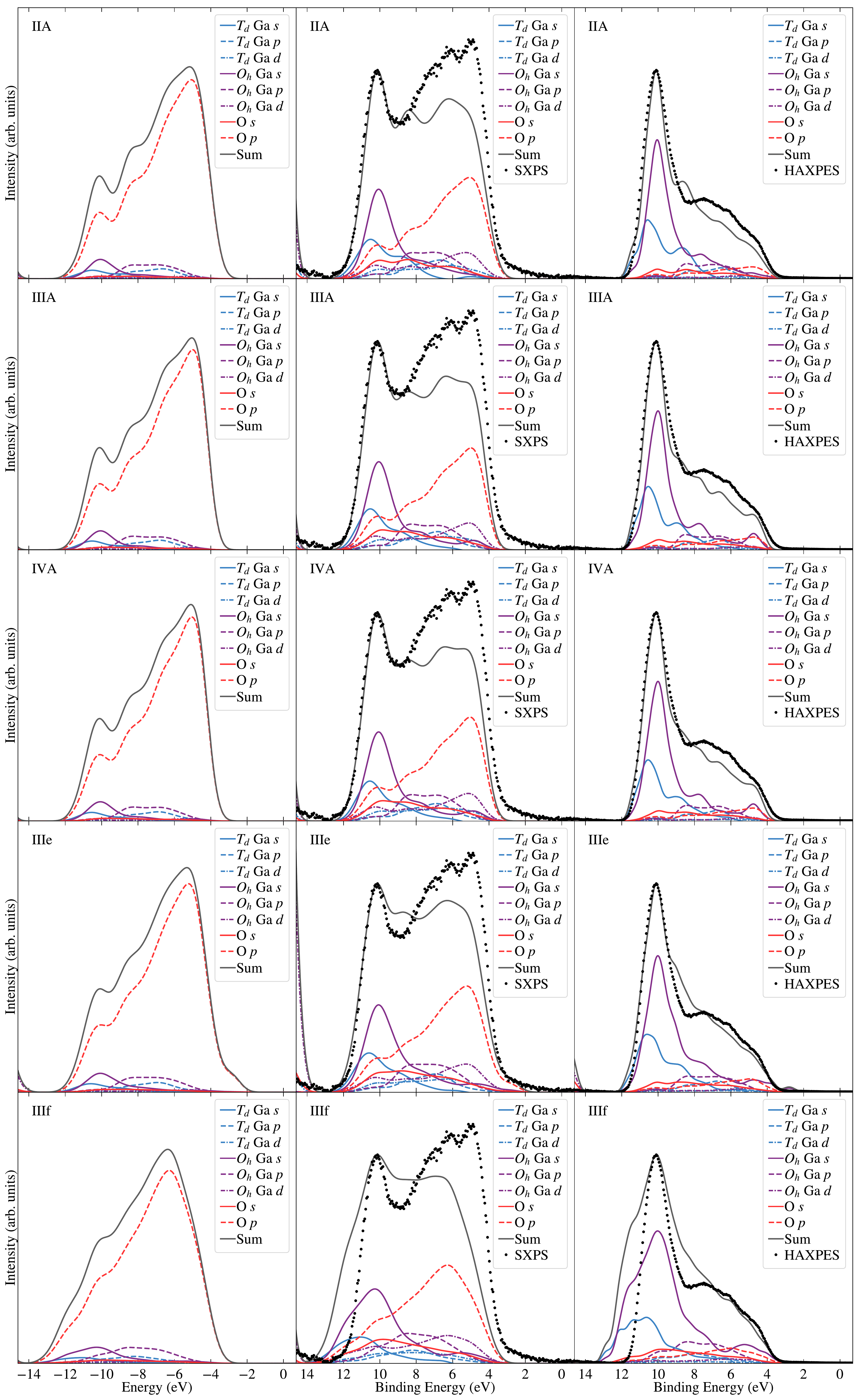}
\caption{Unweighted PDOS and comparison between the calculated and measured valence XPS for soft and hard X-rays for select relaxed structures. Spectra have been aligned and normalised as described in the main text.
\label{fig:valence_si1}}
\end{figure*}

\clearpage

\begin{figure*}[htb]
\includegraphics[scale=0.33]{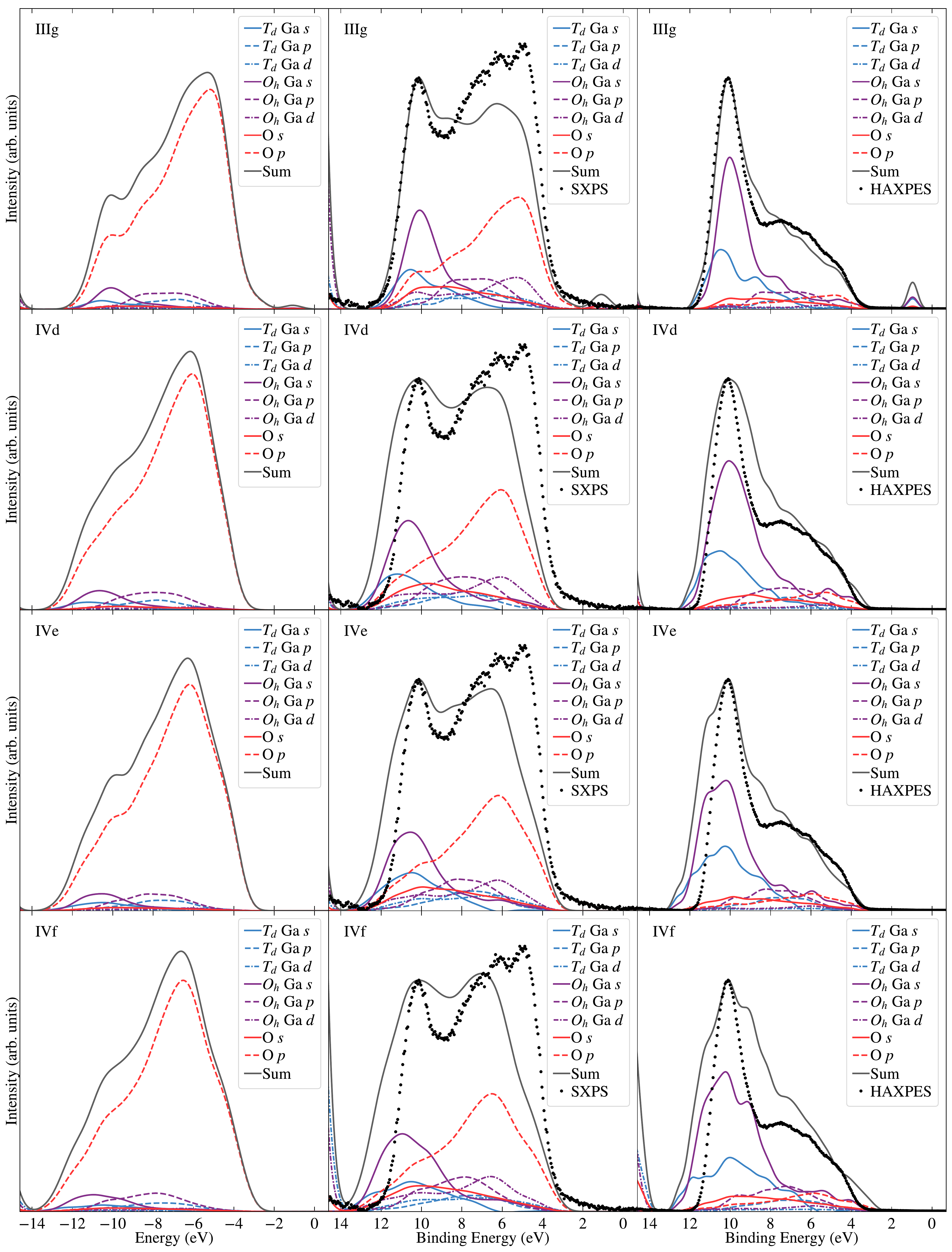}
\caption{Unweighted PDOS and comparison between the calculated and measured valence XPS for soft and hard X-rays for select relaxed structures. Spectra have been aligned and normalised as described in the main text.
\label{fig:valence_si2}}
\end{figure*}

\clearpage

\begin{figure}[!h]
\centering
\includegraphics[scale=0.56]{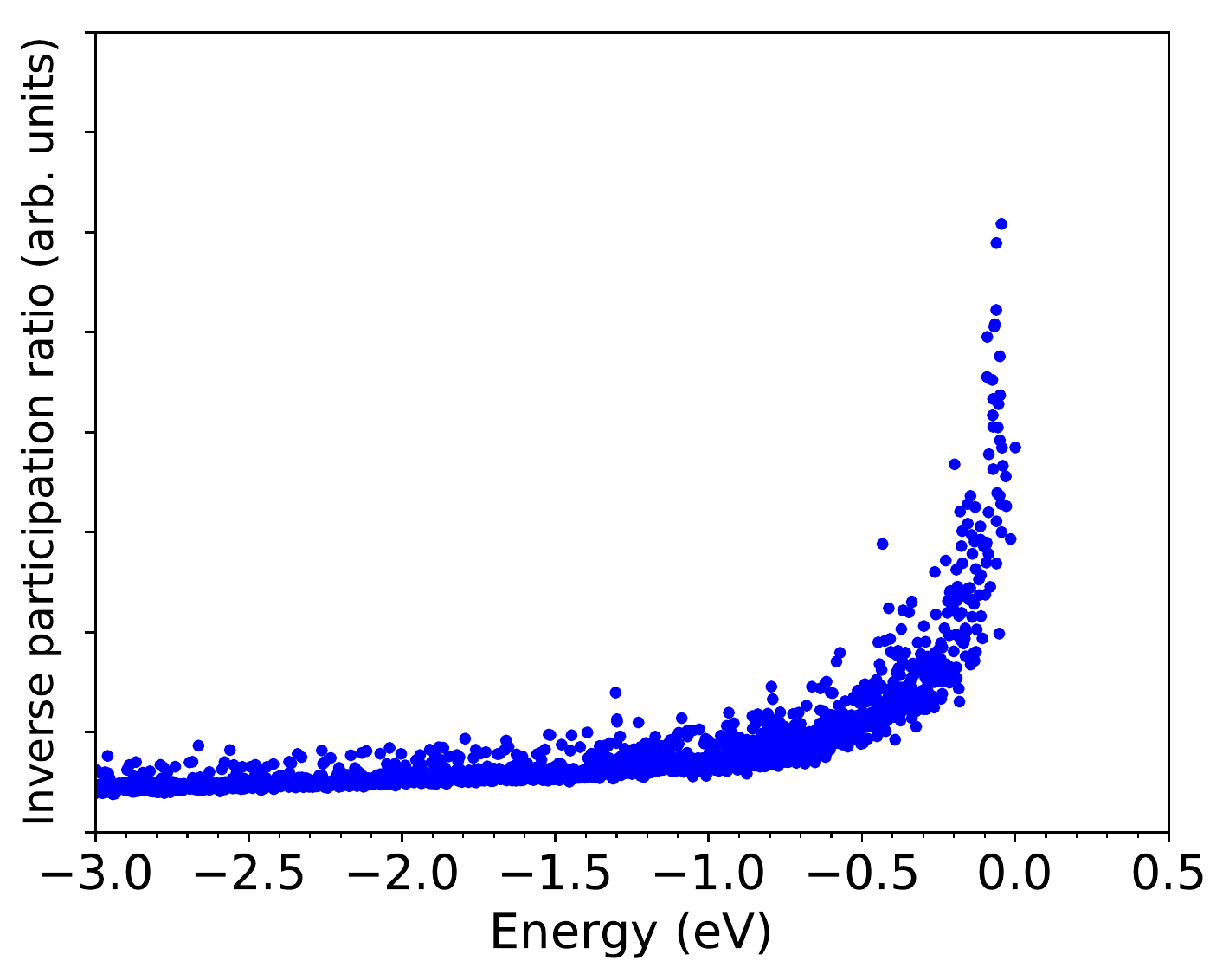}
\caption{Total IPR for the lowest-energy $\gamma$-\ce{Ga2O3} structure (IIIA) calculated with HSE. A significant degree of electronic localization is observed for the highest-lying valence band states up to $\sim$0.5-0.8~eV below the VBM.}
\label{fig:IPR_gamma_lowest}
\end{figure}

\begin{figure}[!ht]
\centering
\includegraphics[scale=0.56]{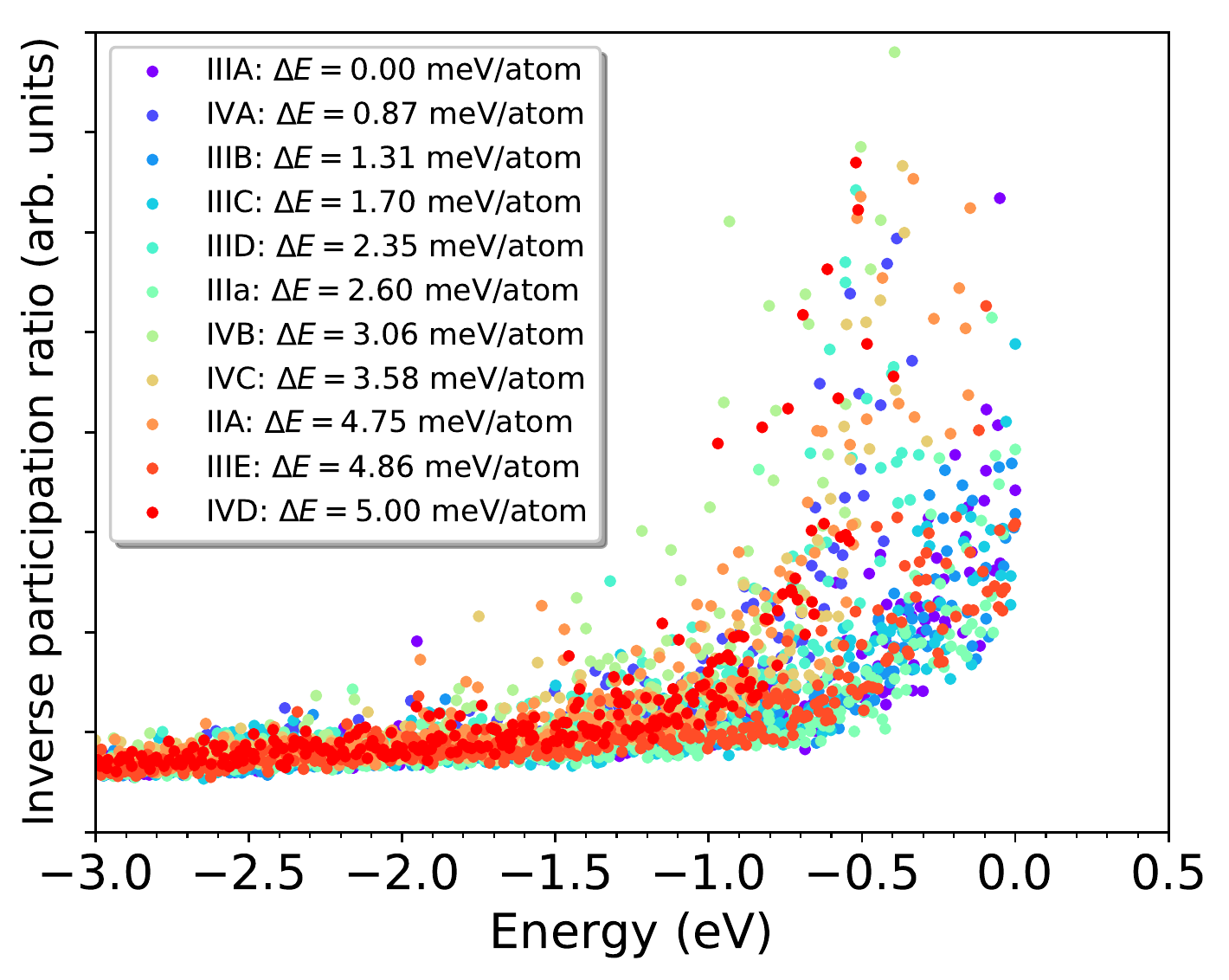}
\caption{Total IPR for several slightly-higher energy $\gamma$-\ce{Ga2O3} structures calculated with HSE. The energy difference of the model structure relative to the lowest-energy structure is included in the legend. All structures are observed to exhibit significant degrees of electronic localization in the upper-most valence band states.}
\label{fig:IPR_gamma_all}
\end{figure}

To assess the extent of localization behavior of the electronic wavefunctions in model structures, we calculated the inverse participation ratio (IPR) according to the following equation:
\begin{equation}
    {\rm IPR}(\phi_{j,k}) = \frac{\sum_n |\phi_{j,k}(n)|^4 }{ |\sum_n |\phi_{j,k}(n)|^2 |^2 }
\end{equation}
where $\psi_{j,k}$ is the orbital of band $j$ at $k$-point $k$, and $n$ spans the grid points that the wavefunction is defined over.
Thus we assess the aggregate IPR over all states as a function of the associated eigenvalues as a measure of the localization of the Kohn-Sham states. In Fig.~\ref{fig:IPR_gamma_lowest} we include the calculated IPR as a function of energy relative to the valence band maximum (VBM) for the lowest-energy structure calculated with the HSE hybrid functional. Our results identify significant localization in the uppermost valence band states that is far larger than in the conduction band states, with the most significant localization extending $\sim$0.8 eV below the VBM. We find the localization of these states is common to the other, lower-energy structures as well, as seen in Fig.~\ref{fig:IPR_gamma_all}. Analysis of the dipole transition matrix elements to the lowest lying conduction band states indicate weaker transitions for these localized states as compared to lower-lying states (e.g.\ relative to states 5.1-5.3~eV below the conduction band minimum) and may account for the discrepancies observed between the electronic and optical band gaps discussed in the main text.

\clearpage

\begin{figure}[!ht]
\includegraphics[scale=0.33]{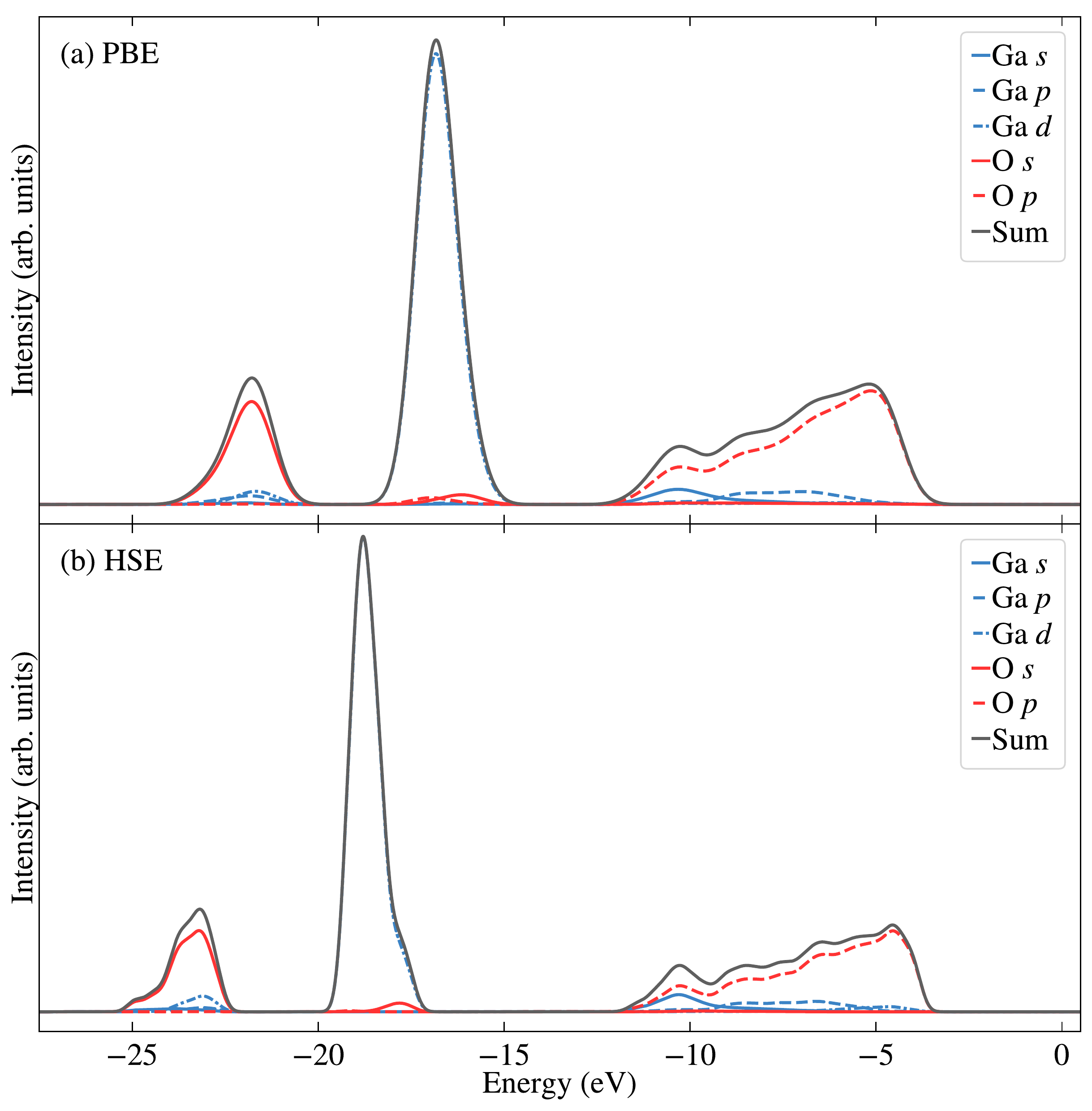}
\caption{Unweighted PDOS for $\gamma$-\ce{Ga2O3} calculated using (a) PBE and (b) HSE.
\label{fig:comp_pdos}}
\end{figure}
 
 \clearpage
 
\begin{figure*}[htb]
\includegraphics[scale=0.33]{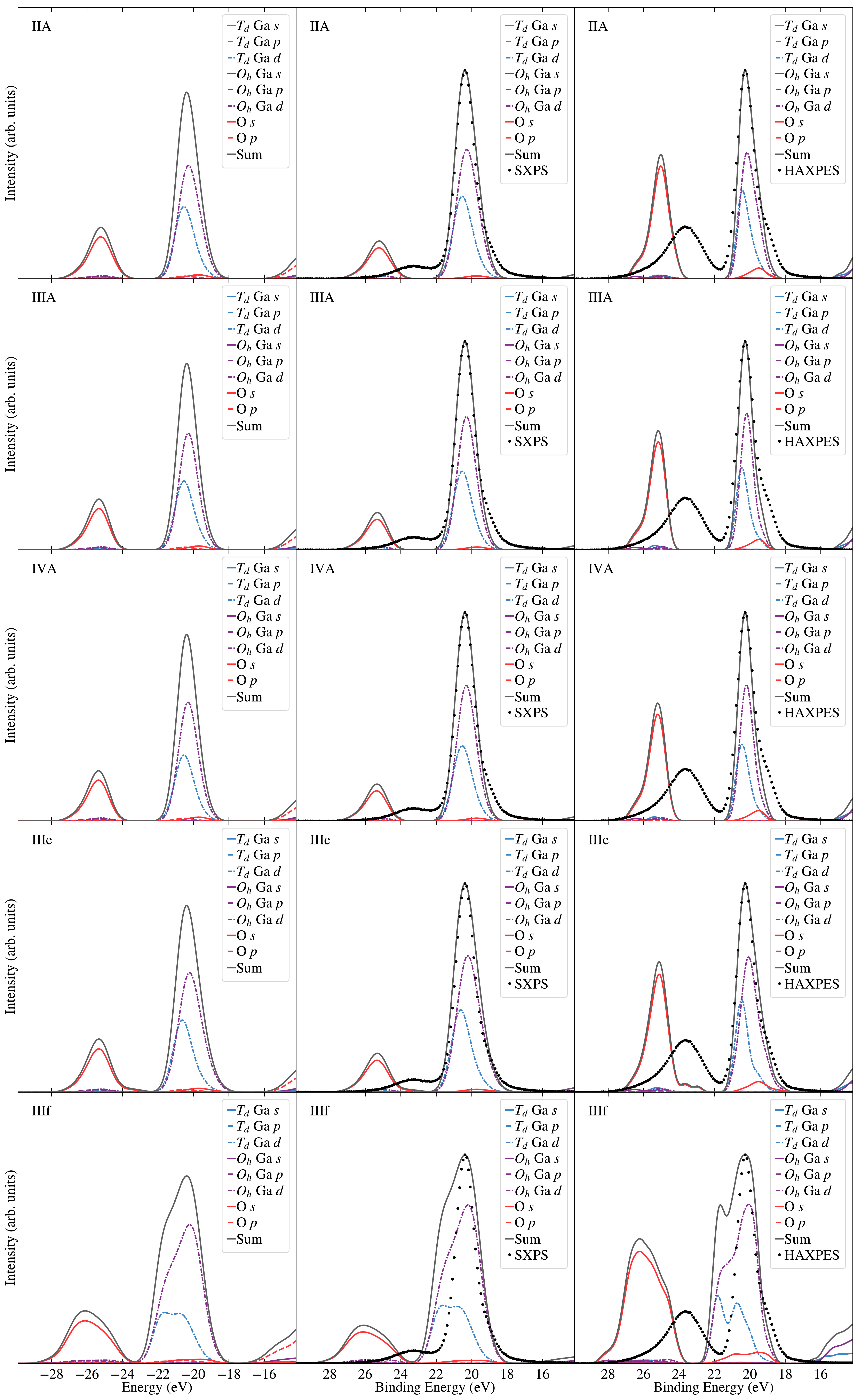}
\caption{Unweighted PDOS and comparison between the calculated and measured semicore XPS for soft and hard X-rays for select relaxed structures. Spectra have been aligned and normalised as described in the main text.
\label{fig:semicore_si1}}
\end{figure*}
 
 \clearpage
 
\begin{figure*}[htb]
\includegraphics[scale=0.33]{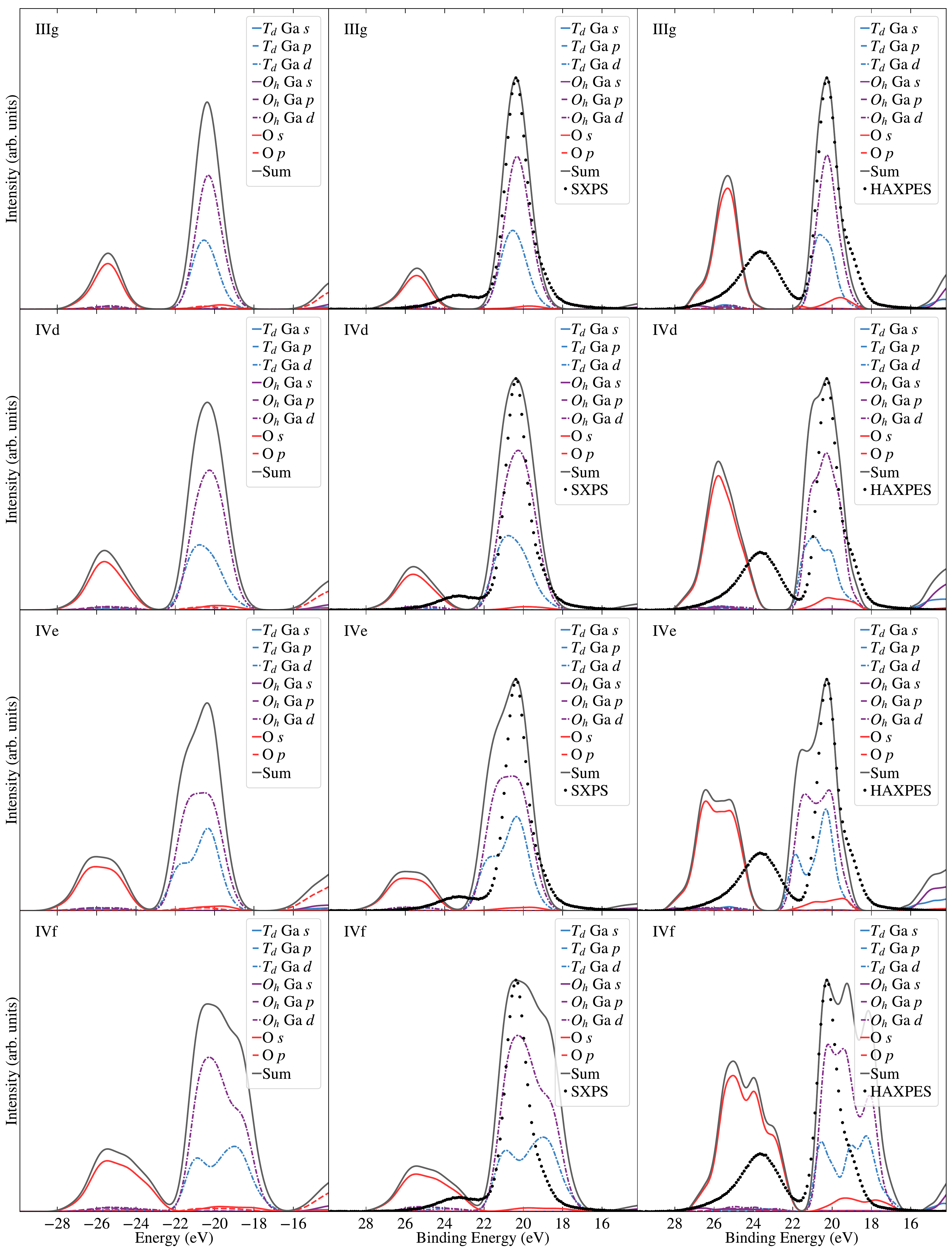}
\caption{Unweighted PDOS and comparison between the calculated and measured semicore XPS for soft and hard X-rays for select relaxed structures. Spectra have been aligned and normalised as described in the main text.
\label{fig:semicore_si2}}
\end{figure*}

\clearpage

\begin{figure*}[htb]
\includegraphics[scale=0.33]{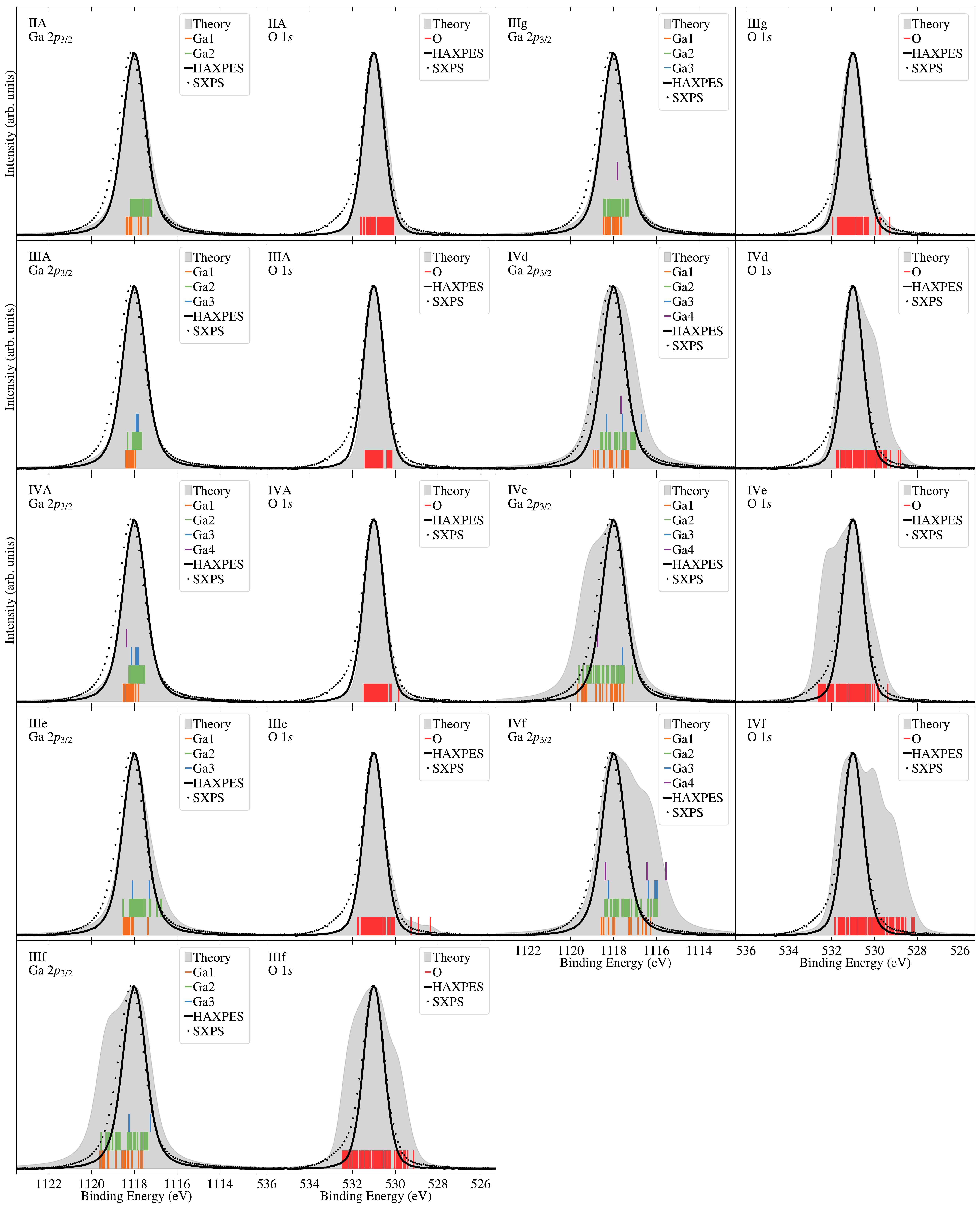}
\caption{Comparison between the calculated and measured core states for Ga~2$p_{3/2}$ and O~1$s$, for both soft and hard X-rays. The theoretical results are for the nine relaxed structures listed in Table~\ref{tab:core_bes}. The theoretical spectra have been normalised and aligned with respect to experiment.
\label{fig:core_si}}
\end{figure*}

\clearpage

\begin{table*}[htb]
\centering
\begin{threeparttable}
\caption{Spread of calculated core binding energies, $\Delta E$, for the investigated $\gamma$-\ce{Ga2O3} structures, as well as $\beta$-\ce{Ga2O3}.  The spreads are shown for distinct Ga sites, across all Ga sites, and for O.}
\label{tab:core_bes}
\begin{tabular*} {0.65\textwidth}{l @{\extracolsep{\fill}} rrrr rr }
\hline \hline
& \multicolumn{6}{c}{$\Delta E$ (eV)}  \\
\cline{2-7}\\[-2.5ex]
 & Ga1 & Ga2 & Ga3 & Ga4 & All Ga & O \\
\cline{1-1}\cline{2-2} \cline{3-3} \cline{4-4} \cline{5-5}\cline{6-6}\cline{7-7}\\[-2.5ex]
\textbf{Low Energy}\\
\textbf{2-site} & & & & &  \\
IIA & 1.00 & 0.99 & - & - & 1.18 & 1.54 \\
\textbf{3-site}\\
IIIA & 0.43 & 0.62 & 0.08 & - & 0.71 & 1.25 \\
\textbf{4-site}\\
IVA & 0.71 & 0.72 & 0.32 & 0.00 & 1.00 & 1.61 \\
\textbf{Random}\\
\textbf{3-site}\\
IIIe & 1.15 & 1.80 & 0.78 & - & 1.80 & 3.40 \\
IIIf & 1.99 & 2.18 & 0.99 & - & 2.34 & 3.31 \\
IIIg & 0.84 & 1.16 & - & 0.00 & 1.16 & 2.66\\
\textbf{4-site}\\
IVd & 1.63 & 1.61 & 1.62 & 0.00 & 2.24 & 2.99 \\
IVe & 2.15 & 2.50 & 0.00 & 0.00 & 2.55 & 3.25 \\
IVf & 2.31 & 2.43 & 2.27 & 2.83 & 3.02 & 3.69 \\
\hline
\textbf{$\beta$} & 0.00 & 0.00 & - & - & 0.25 & 0.36  \\
 \hline \hline
\end{tabular*}
\end{threeparttable}
\end{table*}

\begin{figure}[!ht]
\includegraphics[scale=0.33]{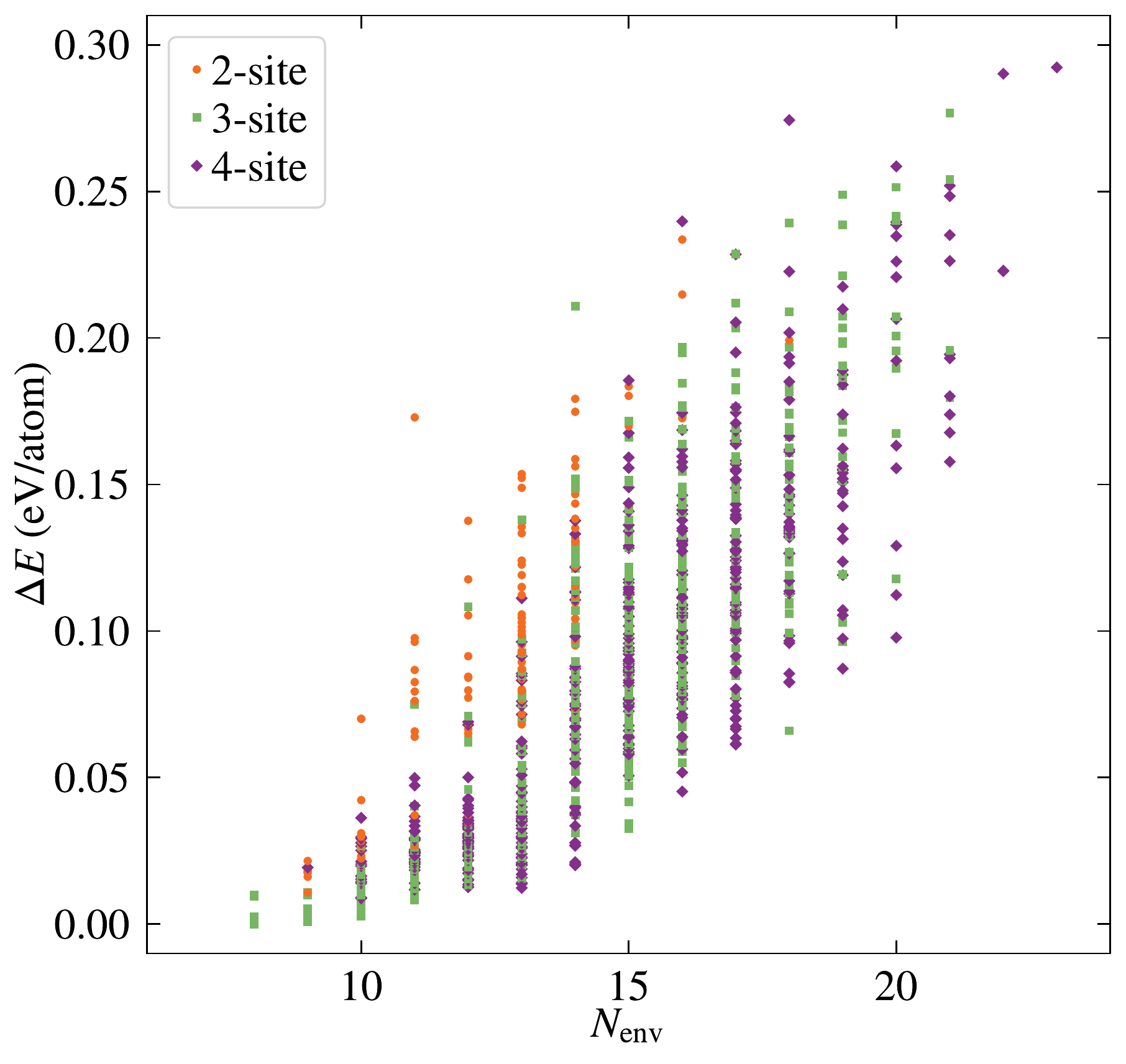}
\caption{Comparison between number of environments present in a given unrelaxed structure, $N_{\mathrm{env}}$, and its energy, $\Delta E$, relative to the lowest energy unrelaxed structure.
\label{fig:code_comp_env}}
\end{figure}

\clearpage
\begin{figure*}[!ht]
\centering
\includegraphics[scale=0.9]{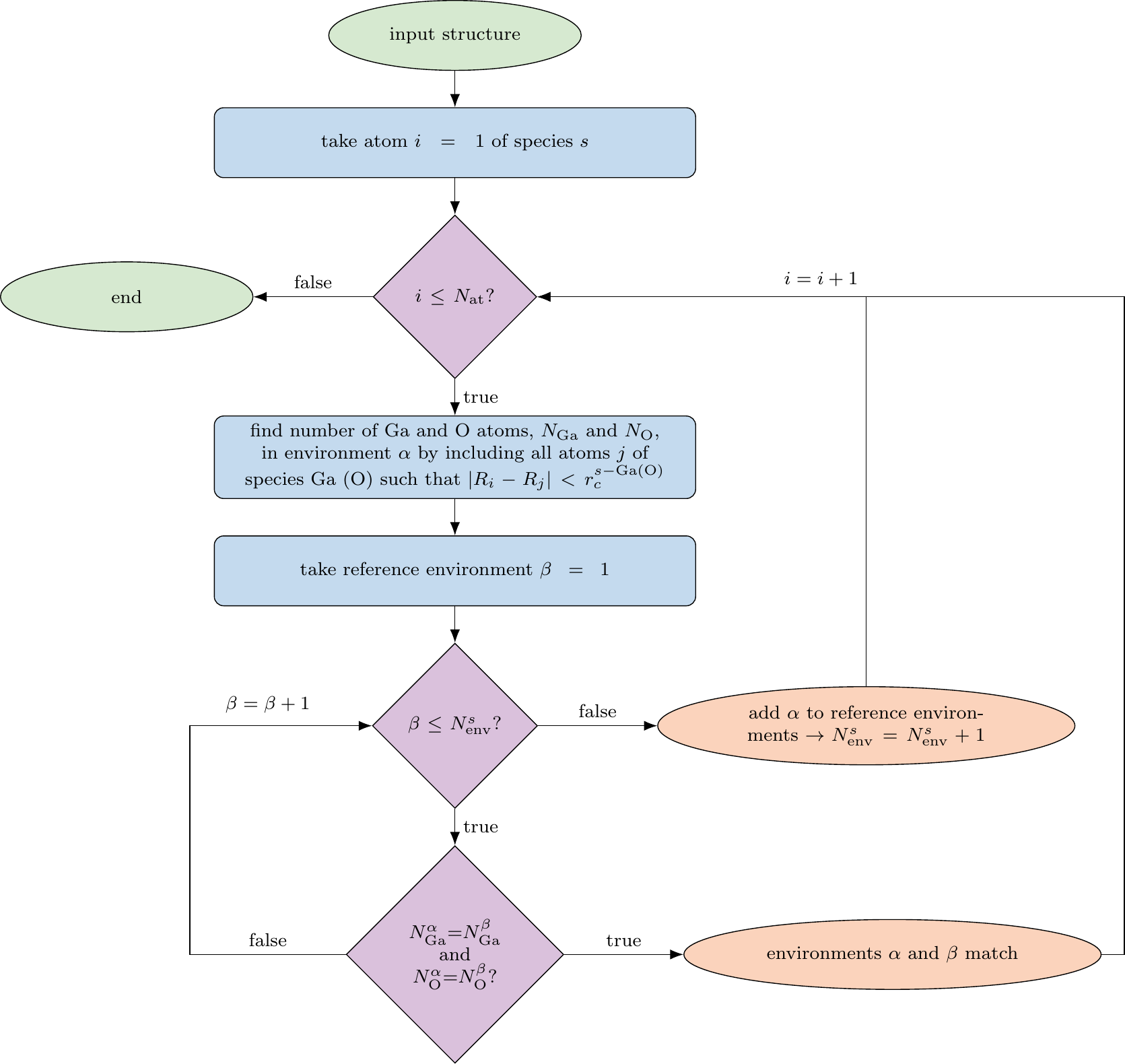}
\caption{Flowchart summarizing the process for identifying the environments present in a given structure of \ce{Ga2O3} containing $N_\mathrm{at}$ atoms, given cut-offs $r_c^{\mathrm{Ga-Ga}}$ and $r_c^{\mathrm{Ga-O}}$, starting from $N_{\mathrm{env}}^{\mathrm{Ga(O)}}$ reference environments associated with Ga (O) atoms, where $r_c^{\mathrm{O-O}}=0$. \label{fig:flowchart}}
\end{figure*}

\clearpage
\begin{figure*}[!ht]
\centering
\includegraphics[scale=0.33]{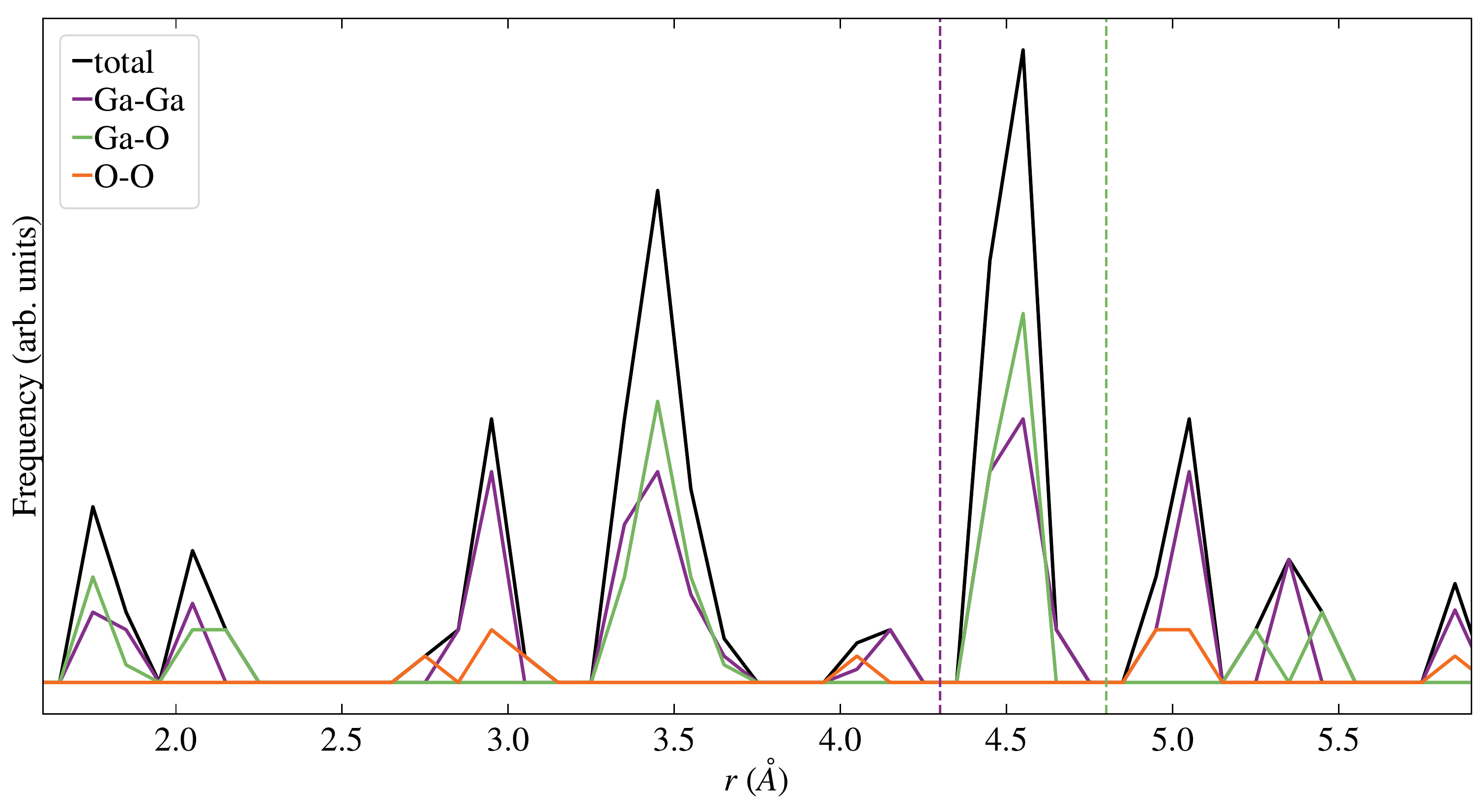}
\caption{Correlation function for the unrelaxed structure in the case where all Ga sites are fully occupied. The dashed vertical lines indicate the final selected cut-off radii, $r_c^{\mathrm{Ga-Ga}}$ and $r_c^{\mathrm{Ga-O}}$.\label{fig:correlation_all}}
\end{figure*}
\clearpage

\begin{table}[!ht]
\centering
\begin{threeparttable}
\caption{Mean absolute errors (MAE) for the validation set in meV/atom, as averaged over 5 different random splittings of the data into 534 (226) training (validation) structures, for the initial 800 DFT $1\times 1\times 3$ supercell calculations and different cut-off values, $r_c$ in \AA. Also shown is the total number of distinct environments detected, $N_{\mathrm{env}}$. In all cases $r_c^{\mathrm{O-O}}=0.0$~\AA.}
\label{tab:fitting}
\begin{tabular*} {0.7\textwidth}{r @{\extracolsep{\fill}} rrr}
\hline \hline
\multicolumn{2}{c}{$r_c$}\\
\cline{1-2} \\[-2.5ex]
Ga-Ga & Ga-O & $N_{\mathrm{env}}$ & MAE  \\
\cline{1-1}\cline{2-2} \cline{3-3} \cline{4-4}\\[-2.5ex]
0.0	&	2.4	&	9	&	20.6 \\
3.2	&	2.4	&	23	&	16.7 \\
3.8	&	2.4	&	30	&	7.1 \\
4.3	&	2.4	&	27	&	6.4 \\
4.8	&	2.4	&	29	&	13.2 \\
5.2	&	2.4	&	31	&	8.7 \\
5.6	&	2.4	&	27  &   7.4 \\
\\
0.0	& 3.8	& 11	& 13.7 \\
3.2	& 3.8	& 25	& 9.8 \\
3.8	& 3.8	& 32	& 8.8 \\
4.3	& 3.8	& 29	& 7.3 \\
4.8	& 3.8	& 31	& 12.6 \\
5.2	& 3.8	& 33	& 10.3 \\
5.6 & 3.8   & 29    & 10.0 \\
\\
0.0	& 4.8 & 12	&	9.9 \\
3.2	& 4.8 & 26  &	7.9 \\
3.8	& 4.8 & 33	&	7.0 \\
4.3	& 4.8 & 30	&	6.0 \\
4.8	& 4.8 & 32	&	8.5 \\
5.2	& 4.8 & 34  &	8.1 \\
5.6 & 4.8 & 30  &   9.1 \\
\\
0.0	& 5.6	& 12	& 11.7 \\
3.2	& 5.6	& 20	& 8.4 \\
3.8	& 5.6	& 33	& 7.9 \\
4.3	& 5.6	& 30	& 6.3 \\
4.8	& 5.6	& 32	& 9.7 \\
5.2	& 5.6	& 34	& 8.7 \\
5.6	& 5.6	& 30	& 9.7 \\

 \hline \hline
\end{tabular*}
\end{threeparttable}
\end{table}

\begin{figure}[!ht]
\centering
\includegraphics[scale=0.33]{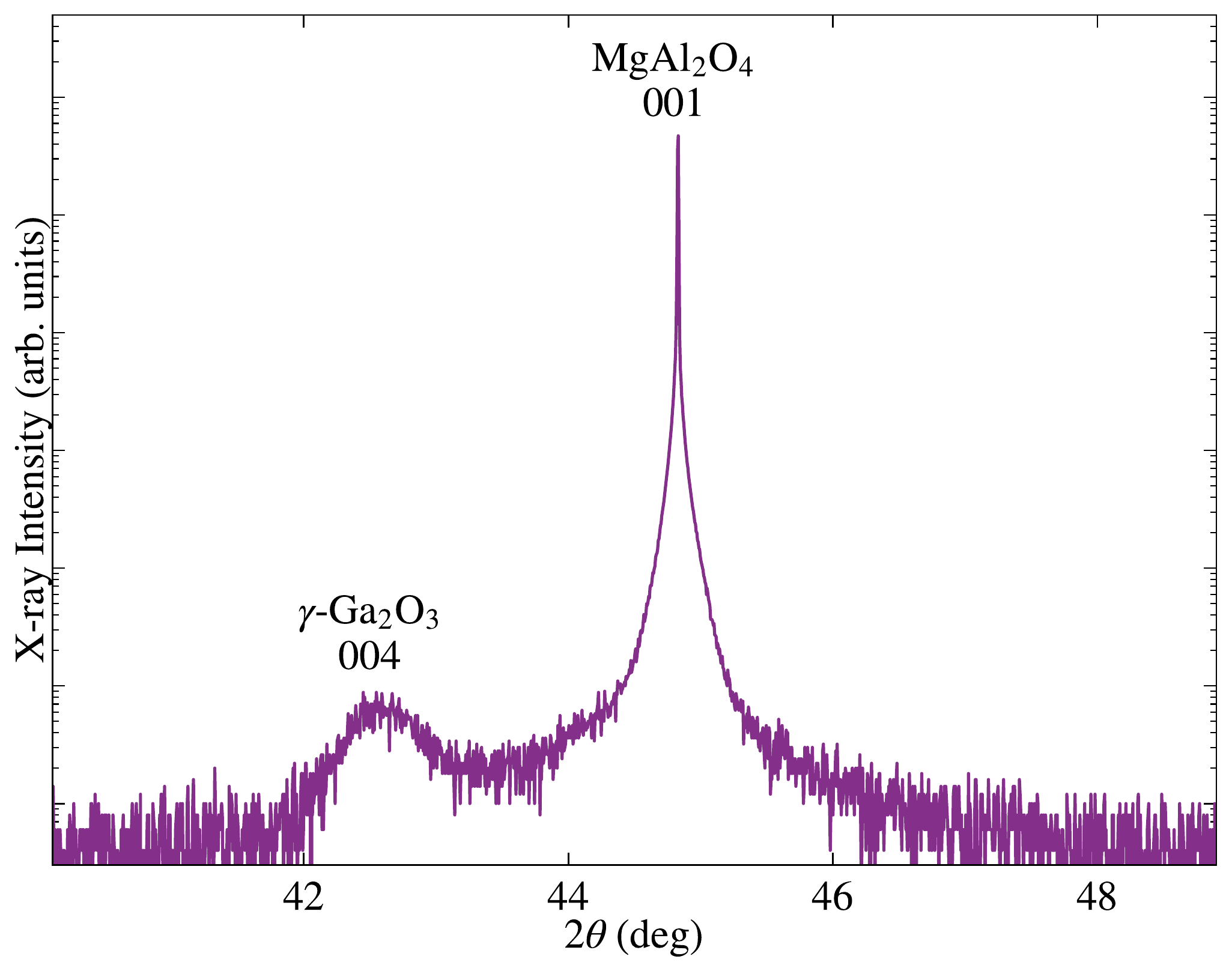}
\caption{X-ray diffraction pattern of the epitaxial $\gamma$-\ce{Ga2O3} film grown on (001) \ce{MgAl2O4} using plasma-assisted molecular beam epitaxy showing symmetric out-of-plane 2$\theta$-$\omega$ scans on a logarithmic scale.}
\label{fig:xrd}
\end{figure}

\begin{figure}[!ht]
\centering
\includegraphics[scale=0.33]{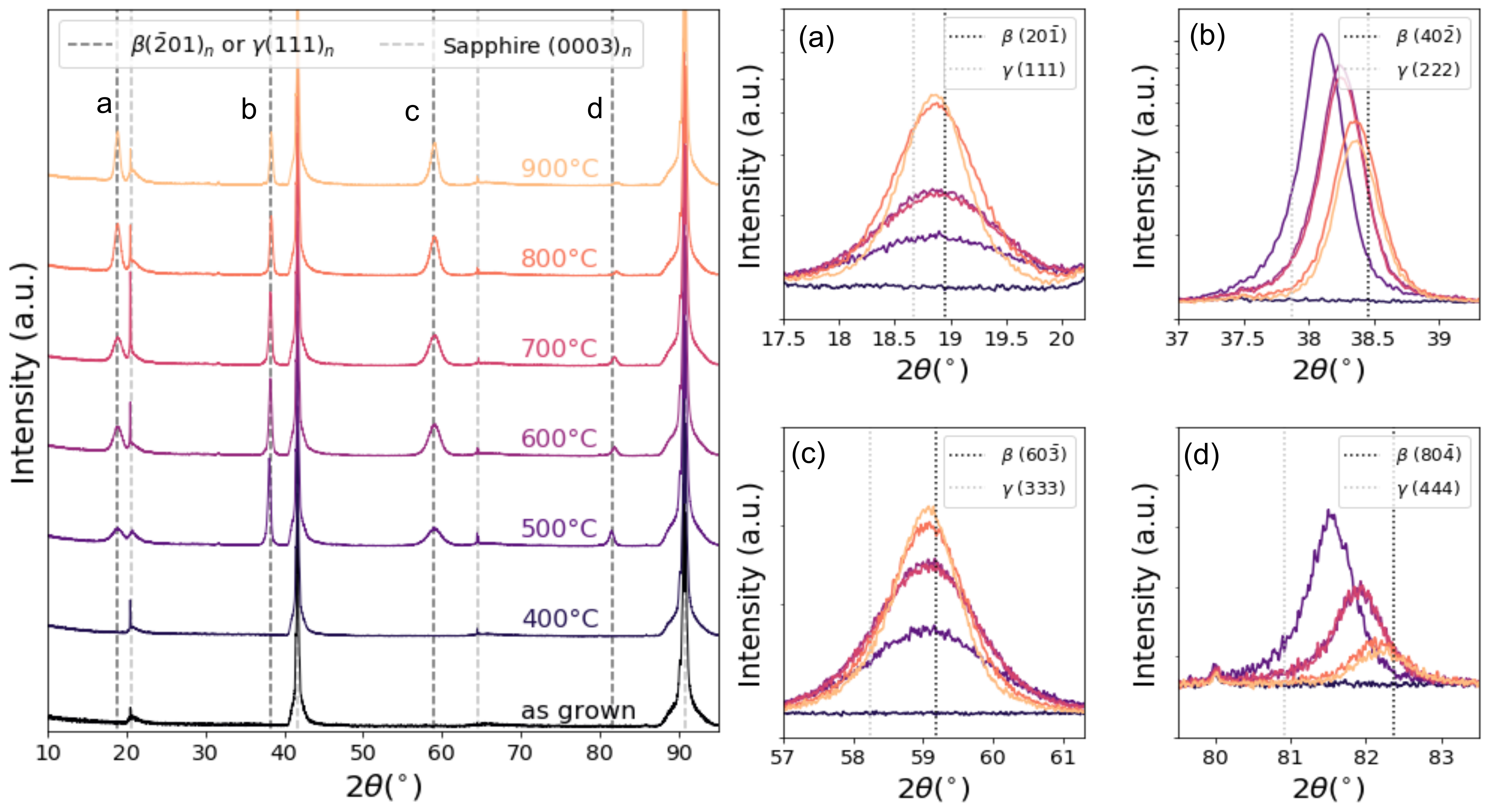}
\caption{X-ray diffraction $\omega-2\theta$ scans of as grown amorphous \ce{Ga2O3} thin films on a sapphire substrate grown by solid-phase epitaxy after annealing at varying temperatures under an \ce{O2} atmosphere. The evolution of the main four peaks marked with (a, b, c, and d) is also included. The literature values of the (111)$\gamma$ and (20$\bar{1}$)$\beta$ family of reflections are represented by dotted lines.}
\label{fig:xrd2}
\end{figure}

\clearpage
\begin{figure*}[!ht]
\includegraphics[scale=0.33]{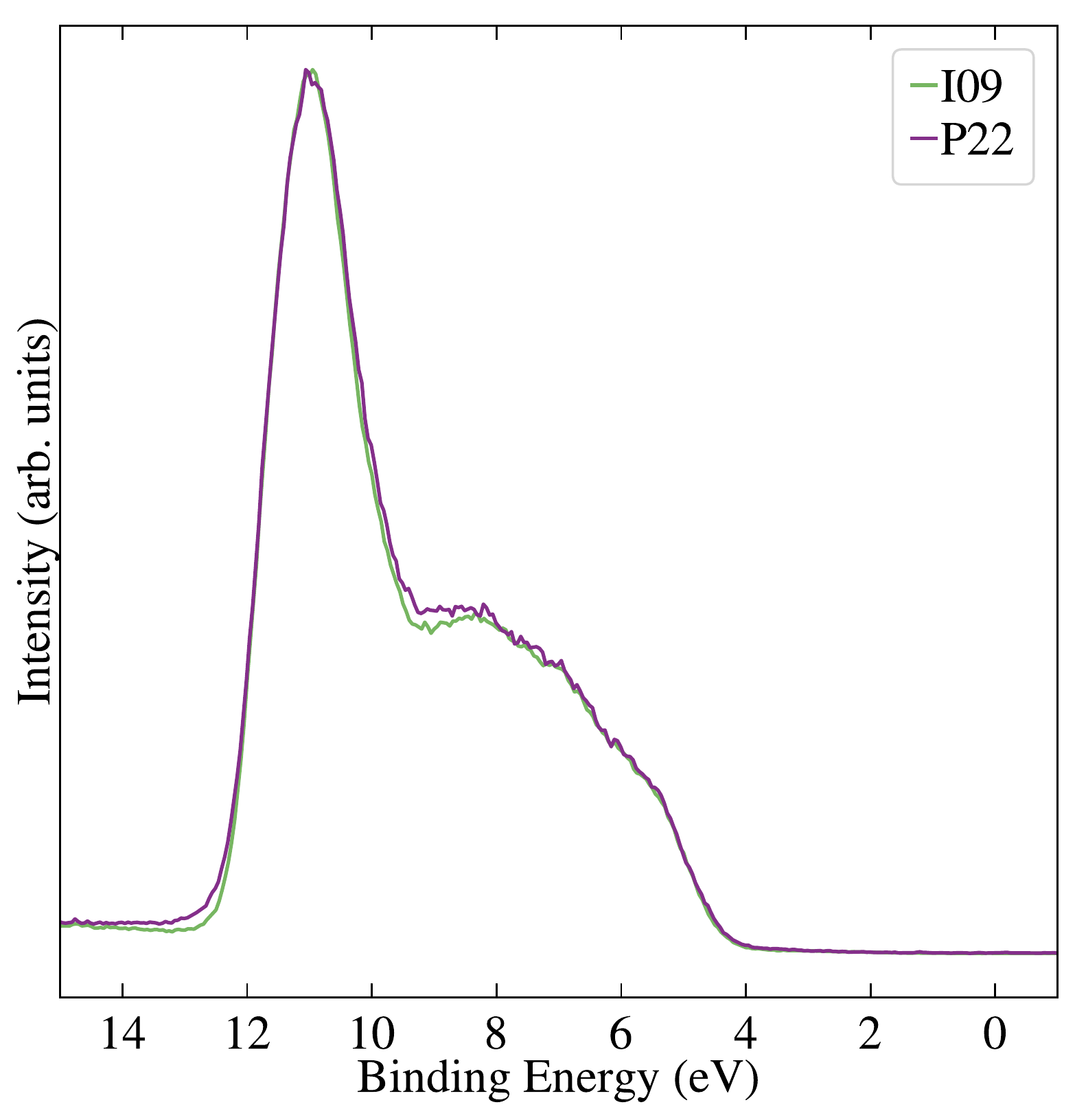}
\caption{Comparison of HAXPES valence band spectra collected at beamlines I09 at Diamond Light Source and P22 at PETRA III.
\label{fig:beamline_comp}}
\end{figure*}

In addition to the SXPS and HAXPES data collected at beamline I09 at Diamond Light Source, HAXPES measurements were conducted at the dedicated HAXPES beamline P22 of PETRA III (DESY, Hamburg). A 6~keV photon energy was used for all measurements, achieved using the 1\textsuperscript{st} harmonic beam from a Si(311) double crystal monochromator and providing an energy resolution of 260~meV (determined from the 16/84\% width of the Fermi edge of a polycrystalline gold foil). P22 is equipped with a SPECS Phoibos 225HV hemispherical analyser and grazing incidence and near-normal emission geometry were used for all measurements. The sample was affixed to the sample plate holder using conductive silver paint and to minimise charging and X-ray beam-damage attenuators were used to reduce the photon flux.\par

